\documentclass[fleqn,usenatbib]{mnras}

\usepackage{newtxtext,newtxmath}
\usepackage[T1]{fontenc}

\usepackage{graphicx}	% Including figure files
\usepackage{amsmath}	% Advanced maths commands
\usepackage[dvipsnames]{xcolor}
\usepackage[flushleft]{threeparttable}
\usepackage{verbatim}
\usepackage{array}
\usepackage{tabularx}
\usepackage{colortbl}
\usepackage{hyperref}
\usepackage{float}
\usepackage{subcaption}
\usepackage{booktabs}
\usepackage{adjustbox}
\usepackage{pdflscape}
\usepackage{booktabs}
\usepackage{inputenc}
\usepackage{soul}

\DeclareRobustCommand{\VAN}[3]{#2}
\let\VANthebibliography\thebibliography
\def\thebibliography{\DeclareRobustCommand{\VAN}[3]{##3}\VANthebibliography}

\newcommand{\kepler}{{\it Kepler}}
\newcommand{\corot}{{\it CoRoT}}
\newcommand{\TESS}{{\it TESS}}
\newcommand{\tess}{{\it TESS}}

\newcommand{\gaia}{{\it Gaia}}

\newcommand{\twomass}{{\it 2MASS}}

\newcommand{\jwst}{{\it JWST}}
\newcommand{\NGTS}{{\it NGTS}}
\newcommand{\ngts}{{\it NGTS}}

\newcommand{\HARPS}{{\it HARPS}}
\newcommand{\harps}{{\it HARPS}}
\newcommand{\Minerva}{{\it MINERVA-Australis}}
\newcommand{\minerva}{{\it MINERVA-Australis}}

\newcommand{\ishell}{{\it iSHELL}}

\newcommand{\hires}{{\it HIRES}}

\newcommand{\pfs}{{\it PFS}}
\newcommand{\PFS}{{\it PFS}}
\newcommand{\MEarth}{{\it MEarth}}
\newcommand{\MEarthSouth}{{\it MEarth-South}}
\newcommand{\astep}{{\it ASTEP}}
\newcommand{\ASTEP}{{\it ASTEP}}
\newcommand{\lco}{{\it LCOGT}}
\newcommand{\lcogt}{{\it LCOGT}}

\newcommand{\waspsouth}{{\it WASP-South}}
\newcommand{\wasp}{{\it WASP}}
\newcommand{\kelt}{{\it KELT}}
\newcommand{\hatnet}{{\it HATNet}}
\newcommand{\hatsouth}{{\it HAT-South}}
\newcommand{\cheops}{{\it CHEOPS}}

\newcommand{\keck}{{\it Keck}}

\newcommand{\gemini}{{\it Gemini}}

\newcommand{\vlt}{{\it VLT}}

\newcommand{\soar}{{\it SOAR}}

\newcommand{\TRES}{{\it TRES}}

\newcommand{\apass}{{\it APASS}}

\newcommand{\kms}{km\,s$^{-1}$}
\newcommand{\ms}{m\,s$^{-1}$}
\newcommand{\msy}{m\,s$^{-1}$\,yr$^{-1}$}
\newcommand{\masy}{mas\,yr$^{-1}$}

\newcommand{\tc}{T\textsubscript{\textit{c}}}
\newcommand{\texp}{\mbox{t$_{exp}$}}

\newcommand{\rpl}{\mbox{R$_{p}$}}
\newcommand{\mstar}{\mbox{M$_{*}$}}
\newcommand{\rstar}{\mbox{R$_{*}$}}

\newcommand{\msun}{\mbox{M$_{\odot}$}}
\newcommand{\rsun}{\mbox{R$_{\odot}$}}

\newcommand{\fsun}{\mbox{S$_{\odot}$}}
\newcommand{\rearth}{R$_{\oplus}$}
\newcommand{\mearth}{M$_{\oplus}$}
\newcommand{\gccc}{g\,cm$^{-3}$}

\newcommand{\teff}{T\textsubscript{\textit{eff}}}

\newcommand{\logg}{$\log g$}

\newcommand{\vsini}{$v \sin i_*$}

\newcommand{\feh}{[Fe/H]}
\newcommand{\mgh}{[Mg/H]}
\newcommand{\sih}{[Si/H]}

\newcommand{\Msun}{M$_{\odot}$}
\newcommand{\Rsun}{R$_{\odot}$}

\newcommand{\TStar}{TOI-836} % TOI
\newcommand{\Tstar}{TOI-836} % TOI
\newcommand{\TTIC}{TIC 440887364} % TIC
\newcommand{\THIP}{HIP 73427} % HIP
\newcommand{\TTwomass}{J15001942-2427147}
\newcommand{\TGaiaId}{6230733559097425152} % GAIA EDR3

\newcommand{\TRa}{\mbox{$15^{\rmn{h}} 00^{\rmn{m}} 19\fs16$}} % GAIA EDR3
\newcommand{\TDec}{\mbox{$-24\degr 27\arcmin 15\farcs14$}} % GAIA EDR3
\newcommand{\TPropRa}{\mbox{$-199.48\pm0.018$}} % GAIA EDR3
\newcommand{\TPropDec}{\mbox{$-27.997\pm0.017$}} % GAIA EDR3
\newcommand{\TPropTotal}{\mbox{$201.438\pm0.025$}} % GAIA EDR3
\newcommand{\TPropshort}{\mbox{$\sim200$}} % GAIA EDR3
\newcommand{\TParallax}{\mbox{$36.353\pm0.016$}} % GAIA EDR3
\newcommand{\TDistance}{\mbox{$27.504\pm0.029$}} % TICv8
\newcommand{\GaiaRV}{\mbox{$-26.603\pm0.922$}} % GAIA DR2

\newcommand{\Tloggporto}{\mbox{$4.743\pm0.105$}} % Porto, labelled dr3
\newcommand{\Teffporto}{\mbox{$4552\pm154$}} % Porto
\newcommand{\Tstarmassporto}{\mbox{$0.678_{-0.041}^{+0.049}$}} % Porto, labelled dr3
\newcommand{\Tstarmassportoshort}{\mbox{$0.68\pm0.05$}} % Porto
\newcommand{\Tstarradiusporto}{\mbox{$0.666\pm0.010$}} % Porto
\newcommand{\Tstarradiusportoshort}{\mbox{$0.67\pm0.01$}} % Porto
\newcommand{\Tstarvsiniporto}{\mbox{$1.86\pm0.50$}} % Porto 1
\newcommand{\Tstarfehporto}{\mbox{$-0.284\pm-0.067$}} % Porto
\newcommand{\Tstarfehshort}{\mbox{$-0.28$}} % Porto
\newcommand{\Tstarmghporto}{\mbox{$-0.23\pm0.17$}}
\newcommand{\Tstarsihporto}{\mbox{$-0.29\pm0.20$}}
\newcommand{\Tstarageporto}{\mbox{$5.4^{+6.3}_{-5.0}$}}

\newcommand{\Tstarmassexo}{\mbox{$0.678_{-0.041}^{+0.049}$}} % exoplanet
\newcommand{\Tstarradiusexo}{\mbox{$0.665\pm0.010$}} % exoplanet
\newcommand{\Tstardensityexo}{\mbox{$3.294^{+0.079}_{-0.092}$}}
\newcommand{\Tstarperiodexo}{\mbox{$21.987\pm0.097$}} % exoplanet
\newcommand{\Tstarperiodexoshort}{\mbox{$21.99\pm0.097$}}
\newcommand{\ldu}{\mbox{$0.039\pm0.235$}}
\newcommand{\ldv}{\mbox{$0.023\pm0.335$}}

\newcommand{\TBmag}{\mbox{$11.138\pm0.028$}} % TICv8
\newcommand{\TGaiaMagBP}{\mbox{$10.126\pm0.003$}} % GAIA EDR3
\newcommand{\TGaiaMagRP}{\mbox{$8.587\pm0.004$}} % GAIA EDR3
\newcommand{\TGMag}{\mbox{$9.407\pm0.0003$}} % GAIA EDR3
\newcommand{\TJmag}{\mbox{$7.580\pm0.023$}} % TICv8
\newcommand{\THmag}{\mbox{$6.983\pm0.040$}} % TICv8
\newcommand{\TKmag}{\mbox{$6.804\pm0.018$}} % TICv8
\newcommand{\TVmag}{\mbox{$9.920\pm0.030$}} % TICv8
\newcommand{\TTESSmag}{\mbox{$8.649\pm0.006$}} % TICv8

\newcommand{\Tplanetb}{TOI-836\,b}
\newcommand{\Tplanetc}{TOI-836\,c}
\newcommand{\Tdrift}{\mbox{$-7.95\pm2.14$}} %DACE
\newcommand{\Tperiodb}{\mbox{$3.81673\pm0.00001$}} % exoplanet
\newcommand{\Tperiodbshort}{\mbox{$3.82$}} % exoplanet
\newcommand{\Tperiodc}{\mbox{$8.59545\pm0.00001$}} % exoplanet
\newcommand{\Tperiodcshort}{\mbox{$8.60$}} % exoplanet
\newcommand{\TTDurfullb}{\mbox{$1.805^{+0.222}_{-0.007}$}} % exoplanet
\newcommand{\TTDurcutb}{\mbox{$1.6823^{+0.0012}_{-0.3292}$}} % exoplanet
\newcommand{\TTDurfullc}{\mbox{$2.486^{+0.161}_{-0.192}$}} % exoplanet
\newcommand{\TTDurcutc}{\mbox{$2.256^{+0.144}_{-0.432}$}} % exoplanet
\newcommand{\Tcb}{\mbox{$1599.9953\pm0.0019$}} % exoplanet
\newcommand{\Tcc}{\mbox{$1599.7623\pm0.0008$}} % exoplanet
\newcommand{\TMassb}{\mbox{$4.53^{+0.92}_{-0.86}$}} % exoplanet
\newcommand{\TMassbshort}{\mbox{$4.5\pm0.9$}} % exoplanet
\newcommand{\TMassc}{\mbox{$9.6^{+2.7}_{-2.5}$}} % exoplanet
\newcommand{\TMasscshort}{\mbox{$9.6\pm{2.6}$}} % exoplanet
\newcommand{\TRadiusb}{\mbox{$1.704\pm0.067$}} % exoplanet
\newcommand{\TRadiusbshort}{\mbox{$1.70\pm0.07$}} % exoplanet
\newcommand{\TRadiusc}{\mbox{$2.587\pm0.088$}} % exoplanet
\newcommand{\TRadiuscshort}{\mbox{$2.59\pm0.09$}} % exoplanet
\newcommand{\Tdensityb}{\mbox{$5.02^{+0.36}_{-0.44}$}} % exoplanet
\newcommand{\Tdensitybshort}{\mbox{$5.02^{+0.36}_{-0.44}$}} % exoplanet
\newcommand{\Tdensityc}{\mbox{$3.06^{+0.48}_{-0.54}$}} % exoplanet
\newcommand{\Tdensitycshort}{\mbox{$3.06^{+0.47}_{-0.54}$}} % exoplanet
\newcommand{\Trorb}{\mbox{$0.0235\pm0.0013$}} % exoplanet
\newcommand{\Trorc}{\mbox{$0.0357\pm0.0018$}} % exoplanet
\newcommand{\Timpactb}{\mbox{$0.58\pm0.11$}} % exoplanet
\newcommand{\Timpactc}{\mbox{$0.53\pm0.13$}} % exoplanet
\newcommand{\Tkb}{\mbox{$2.38\pm0.35$}} % exoplanet
\newcommand{\Tkc}{\mbox{$3.86\pm0.85$}} % exoplanet
\newcommand{\Tincb}{\mbox{$87.57\pm{0.44}$}} % exoplanet
\newcommand{\Tincc}{\mbox{$88.7\pm{1.5}$}} % exoplanet
\newcommand{\Taub}{\mbox{$0.04220\pm0.00093$}} % exoplanet
\newcommand{\Tauc}{\mbox{$0.0750\pm0.0016$}} % exoplanet
\newcommand{\Teqb}{\mbox{$871\pm36$}}
\newcommand{\Teqc}{\mbox{$665\pm27$}}
\newcommand{\Teccb}{\mbox{$0.053\pm0.042$}}
\newcommand{\Teccc}{\mbox{$0.078\pm0.056$}}
\newcommand{\Tomegab}{\mbox{$9\pm92$}}
\newcommand{\Tomegac}{\mbox{$-28\pm113$}}
\newcommand{\TSMb}{\mbox{$65.7\pm5.8$}}
\newcommand{\TSMbshort}{\mbox{$65.7$}}
\newcommand{\TSMc}{\mbox{$82.4\pm5.8$}}
\newcommand{\TSMcshort}{\mbox{$82.4$}}
\newcommand{\Tfluxb}{\mbox{$78.838\pm0.015$}}
\newcommand{\Tfluxc}{\mbox{$26.707\pm0.003$}}

\newcommand{\Pconfirmed}{$4935$}

\newcommand{\depthbexofop}{$580$}
\newcommand{\depthcexofop}{$1140$}

\newcommand{\Mcoreb}{\mbox{$0.12^{+0.16}_{-0.11}$}}
\newcommand{\Mwaterb}{\mbox{$0.18^{+0.25}_{-0.16}$}}
\newcommand{\Mgasb}{\mbox{$-8.33^{+3.95}_{-3.30}$}}
\newcommand{\Fecoreb}{\mbox{$0.90^{+0.09}_{-0.08}$}}
\newcommand{\Simantleb}{\mbox{$0.41^{+0.08}_{-0.07}$}}
\newcommand{\Mgmantleb}{\mbox{$0.45^{+0.15}_{-0.17}$}}

\newcommand{\Mcorec}{\mbox{$0.10^{+0.15}_{-0.09}$}}
\newcommand{\Mwaterc}{\mbox{$0.33^{+0.15}_{-0.28}$}}
\newcommand{\Mgasc}{\mbox{$-1.99^{+0.93}_{-6.77}$}}
\newcommand{\Fecorec}{\mbox{$0.90^{+0.09}_{-0.08}$}}
\newcommand{\Simantlec}{\mbox{$0.41^{+0.08}_{-0.07}$}}
\newcommand{\Mgmantlec}{\mbox{$0.44^{+0.15}_{-0.17}$}}

\title[\TStar]{\TStar:  A super-Earth and mini-Neptune transiting a nearby K-dwarf}

\author[F. Hawthorn et al.]{\parbox{\textwidth}{\Large
Faith~Hawthorn$^{1,2}$,
Daniel~Bayliss$^{1,2}$,
Thomas~G.~Wilson$^{3}$,
Andrea~Bonfanti$^{4}$,
Vardan~Adibekyan$^{5,6}$,
Yann~Alibert$^{7}$,
S\'ergio~G.~Sousa$^{5,6}$,
Karen~A.~Collins$^{8}$,
Edward~M.~Bryant$^{1,2}$,
Ares~Osborn$^{1,2}$
% ...
}
\vspace{0.2cm}
\\
\parbox{\textwidth}{
The authors' affiliations are shown in Appendix \ref{sec:affiliations}.\\
*E-mail: faith.hawthorn@warwick.ac.uk}\vspace{-0.3cm}}

\date{Accepted XXX. Received YYY; in original form ZZZ}

\pubyear{2022}

\begin{document}
\label{firstpage}
\pagerange{\pageref{firstpage}--\pageref{lastpage}}
\maketitle

%%%%%%%%%%%%%%%%%%%%%%%%%%%%%%%%%%%%%%%%%%%%%%%%%
%%%%%%%%%%%%%%%%%%% ABSTRACT %%%%%%%%%%%%%%%%%%%%
%%%%%%%%%%%%%%%%%%%%%%%%%%%%%%%%%%%%%%%%%%%%%%%%%

\begin{abstract}
We present the discovery of two exoplanets transiting \TStar\ (TIC 440887364) using data from \tess\ Sector 11 and Sector 38.  \TStar\ is a bright (T = 8.5\,mag), high proper motion (\TPropshort\,\masy), low metallicity (\feh$\approx$\Tstarfehshort) K-dwarf with a mass of \Tstarmassportoshort\,\msun\ and a radius of \Tstarradiusportoshort\,\rsun. We obtain photometric follow-up observations with a variety of facilities, and we use these data-sets to determine that the inner planet, \Tplanetb, is a \TRadiusbshort\,\rearth\ super-Earth in a \Tperiodbshort\,day orbit, placing it directly within the so-called `radius valley'.  The outer planet, \Tplanetc, is a \TRadiuscshort\,\rearth\ mini-Neptune in an \Tperiodcshort\,day orbit.  Radial velocity measurements reveal that \Tplanetb\ has a mass of \TMassbshort\,\mearth, while \Tplanetc\ has a mass of \TMasscshort\,\mearth. Photometric observations show Transit Timing Variations (TTVs) on the order of 20\,minutes for \Tplanetc, although there are no detectable TTVs for \Tplanetb. The TTVs of planet \Tplanetc\ may be caused by an undetected exterior planet.
\end{abstract}


%%%%%%%%%%%%%%%%%%%%%%%%%%%%%%%%%%%%%%%%%%%%%%%%%%
%%%%%%%%%%%%%%%%% BODY OF PAPER %%%%%%%%%%%%%%%%%%
%%%%%%%%%%%%%%%%%%%%%%%%%%%%%%%%%%%%%%%%%%%%%%%%%%

\section{Introduction} \label{sec:intro}
Since the groundbreaking discovery of 51 Pegasi\,b \citep{MayorQueloz1995}, the field of exoplanet research has grown to now include an impressive \Pconfirmed\footnote{\url{https://exoplanetarchive.ipac.caltech.edu} as of 2022 February 22, \citep{akeson2013}} discoveries using a variety of detection methods. Transit photometry and radial velocity spectroscopy continue to be the most fruitful methods of exoplanet discovery, and combined they also allow us to determine the fundamental properties of exoplanets, including their mass, radius, bulk density, and possible composition. Ground-based transit photometry surveys such as \hatnet\ \citep{bakos2004}, \wasp\ \citep{pollacco2006}, \kelt\ \citep{pepper2007}, \hatsouth\ \citep{bakos2013}, and \ngts\ \citep{Wheatley2018} among others have greatly added to the population of known transiting exoplanets.

The advent of space-based transit surveys such as \corot\ \citep{corot2009}, \kepler\ \citep{kepler2010}, \textit{K2} \citep{howell2014}, and \tess\ \citep{Ricker:2015} has allowed us to extend the range of detectable exoplanets down to the regimes of Neptune and super-Earth radii.  In this paper we present the discovery of two such exoplanets found from \TESS\ photometry to be transiting the bright star \TStar.  This system was included in the Magellan \pfs\ survey paper \citet{magellanteske}.

The general conclusion from a number of studies is that Kepler compact planetary systems are flat, with the inclination dispersion on the order of a few degrees \citep{lissauer2011kepler, tremainedong, figueira2012, johansen2012, fangmargot2012, fabrycky2014}. The discovery of such multi-planet systems \citep[eg;][]{Wilson2022} confers significant advantages over those stars where only a single exoplanet is detected.  Firstly, the statistical likelihood that the transits are astrophysical false positives is greatly reduced \citep{Lissauer2012}.  Secondly, the dynamical interactions between the planets can result in observable transit timing variations (TTVs), which in some cases may reveal the presence of non-transiting planets \citep[eg;][]{nesvorny2014}. Thirdly, the comparative properties of the planets can reveal possible formation and migration pathways.

One particularly interesting aspect of small-radius multi-planet systems is looking at how they might allow us to study the origin and characteristics of the radius valley seen at around \rpl\, $\approx$ 2.0\,\rearth\ in the exoplanet population \citep{Fulton:2017, Owen2013}. In the case of the \TStar\ system, we find that \Tplanetb\ lies within the radius valley itself, and \Tplanetc\ lies close to the peak on the right hand side. The radius valley is valid for all systems, however multi-planet systems such as this may give us significant insights into formation mechanisms through comparative planetology.
% that the inner planet (\Tplanetb) sits interior to the radius gap while the outer planet (\Tplanetc) lies exterior to the radius gap.
% \textcolor{red}{planets straddle radius gap}

This paper is structured as follows: we present our transit photometry, radial velocity and imaging observations of the \TStar\ system in Section \ref{sec:obs}, our global modelling methods, associated computational implementations and results in Section~\ref{sec:methods}. Finally we present our discussion and conclusion of these results in Sections \ref{sec:disc} and \ref{sec:conc} respectively.

\section{Observations} \label{sec:obs}
\begin{center}
\begin{table}
    \centering
    \caption{Catalog stellar parameters of \TStar.}
    \label{tab:star_props}
    \begin{threeparttable}
    \begin{tabularx}{0.96\columnwidth}{ l  p{0.3\linewidth} X }
    \toprule
    \textbf{Property} & \textbf{Value} & \textbf{Source} \\
    \hline
    \textbf{Identifiers}                & & \\
    TIC ID      & \TTIC              & TICv8 \\
    HIP ID      & \THIP                 &   \\
    2MASS ID    & \TTwomass      & \twomass \\
    Gaia ID     & \TGaiaId   & \gaia\ EDR3 \\
    \hline
    \textbf{Astrometric properties}     & & \\
    R.A. (J2015.5)  & \TRa           & \gaia\ EDR3 \\
    Dec (J2015.5)   & \TDec          & \gaia\ EDR3 \\
    Parallax (mas)  & \TParallax      & \gaia\ EDR3 \\
    Distance (pc)   & \TDistance      & \\
    $\mu_{\rm{R.A.}}$ (mas yr$^{-1}$)       & \TPropRa   & \gaia\ EDR3 \\
    $\mu_{\rm{Dec}}$ (mas yr$^{-1}$)        & \TPropDec  & \gaia\ EDR3 \\
    $\mu_{\rm{Total}}$ (mas yr$^{-1}$)      & \TPropTotal  & \gaia\ EDR3 \\
    RV\textsubscript{sys} (\kms) &\GaiaRV & \gaia\ DR2 \\
    \hline
    % \textbf{Key parameters} & & \\
    % \mstar\ (\msun)  &\Tstarmass &TICv8 \\
    % \rstar\ (\rsun)  &\Tstarradius    &TICv8 \\
    % \lstar\ (\lsun) &\Tstarlum  &TICv8 \\
    % \teff\ (K)   &\Teff  &TICv8 \\
    % \hline
    \textbf{Photometric properties} & & \\
    TESS (mag)  & \TTESSmag    & TICv8 \\
    B (mag)     & \TBmag     & \apass \\
    V (mag)     & \TVmag      & \apass \\
    G (mag)     & \TGMag  & \gaia\ EDR3 \\
    J (mag)     & \TJmag      & \twomass \\
    H (mag)     & \THmag      & \twomass \\
    K (mag)     & \TKmag      & \twomass \\
    Gaia BP (mag)     & \TGaiaMagBP  & \gaia\ EDR3 \\
    Gaia RP (mag)   & \TGaiaMagRP   & \gaia\ EDR3 \\
    \bottomrule
    \end{tabularx}
    \begin{tablenotes}
    \item Sources: TICv8 \citep{Stassun2019}, \twomass\ \citep{Skrutskie2006}, \gaia\ Early Data Release 3 \citep{GAIA_EDR3}, \apass\ \citep{apass}
    \end{tablenotes}
    \end{threeparttable}
\end{table}
\end{center}

%%%%%%%%%%%%%%%%%%%%%%%%%%%%%%%%%%%%%%%%%%%%%%%%%%

\subsection{TESS discovery photometry}
\label{sec:tessphot}

The transit signatures of \Tplanetb\ and \Tplanetc\ were originally identified by the TESS Science Processing Operations Center \citep{Jenkins:2016} using an adaptive matched filter \citep{jenkins2002, jenkins2010, jenkins2020} to search the Sector 11 light curve on 2019 June 5. The transit signatures were fitted with an initial limb-darkened transit model \citep{li2019}, and passed all the diagnostic tests performed and reported in the Data Validation reports \citep{twicken2018}. The \TESS\ Science Office reviewed the Data Validation reports and issued an alert for TOI-836 on 2019 June 17. Subsequent searches of the combined light curves from sectors 11 and 38 located the source of the transit events to within 3.73\,$\pm$\,2.5\,\arcsec\ and 0.98\,$\pm$\,1.5\,\arcsec\ of the host star for \Tplanetb\ and \Tplanetc, respectively. Note that the difference image centroiding results complement the high resolution imaging results presented in Section~\ref{sec:imaging}.

\begin{figure}
    \centering
    \includegraphics[width=0.5\textwidth]{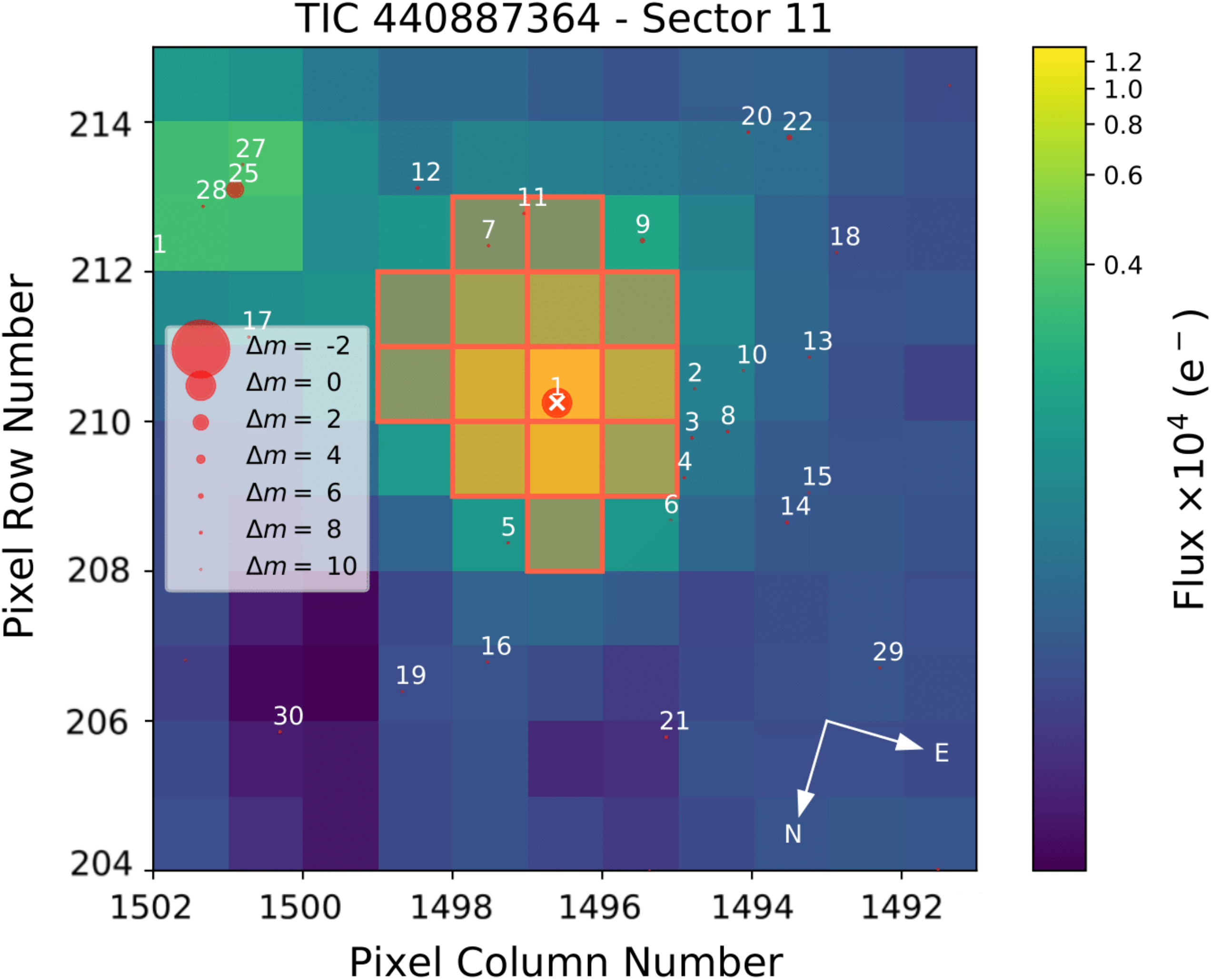}
    \caption{Target Pixel File (TPF) from \tess\ centered on \TStar\ from the \gaia\ catalog, with \gaia\ DR2 sources indicated by red circles with scaled magnitudes, where the numbers indicate ranked distance from the target represented by a white cross. The aperture mask is outlined in red.}
    \label{fig:tpfgaia}
\end{figure}

\TStar\ was first identified as a \TESS\ Object of Interest \citep[TOI;][]{toi2021} in \tess\ Sector 11, Camera 1, CCD 3 from 2019 April 22 to 2019 May 21.
Stellar identifiers, astrometric properties and photometric properties for \TStar\ are listed in Table~\ref{tab:star_props}. Figure~\ref{fig:tpfgaia} shows the Target Pixel File (TPF) from \tess\ created in \texttt{tpfplotter}\footnote{\url{https://github.com/jlillo/tpfplotter}} \citep{tpfplotter}, centred on \TStar\ (indicated by a white cross), with the \gaia\ DR2 catalog data for sources overplotted in red along with scaled magnitudes and the aperture mask for photometry extraction.

\begin{figure*}
    \begin{subfigure}{\textwidth}
        \centering
        \includegraphics[width=\textwidth]{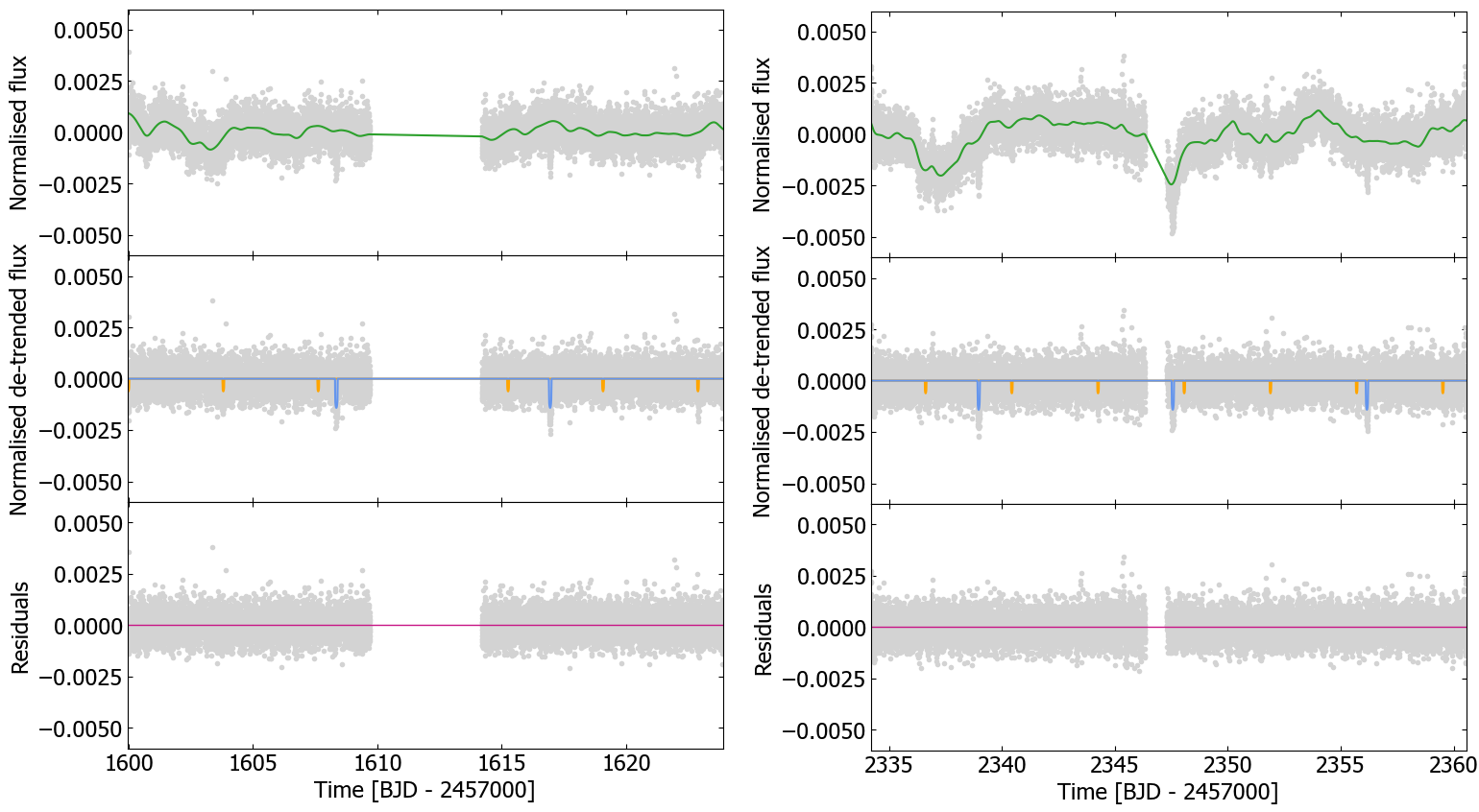}
    \end{subfigure}
    \caption{\textbf{Top left panel:} \tess\ PDC-SAP light curve from Sector 11 with the GP model plotted in green. \textbf{Middle left panel:} \tess\ PDC-SAP light curve data minus the GP model, with transits plotted for \Tplanetc\ (blue line) and \Tplanetb\ (orange line). \textbf{Bottom left panel:} Residuals between the best fit model and the \tess\ datapoints. \textbf{Top right panel:} \tess\ PDC-SAP light curve from Sector 38 with the GP model plotted in green. \textbf{Middle right panel:} \tess\ PDC-SAP light curve data minus the GP model, with transits plotted for \Tplanetc\ (blue line) and \Tplanetb\ (orange line). \textbf{Bottom right panel:} Residuals between the best fit model and the \tess\ datapoints.}
    \label{fig:TESSlightcurve}
    \begin{subfigure}{\textwidth}
        \centering
        \includegraphics[width=\textwidth]{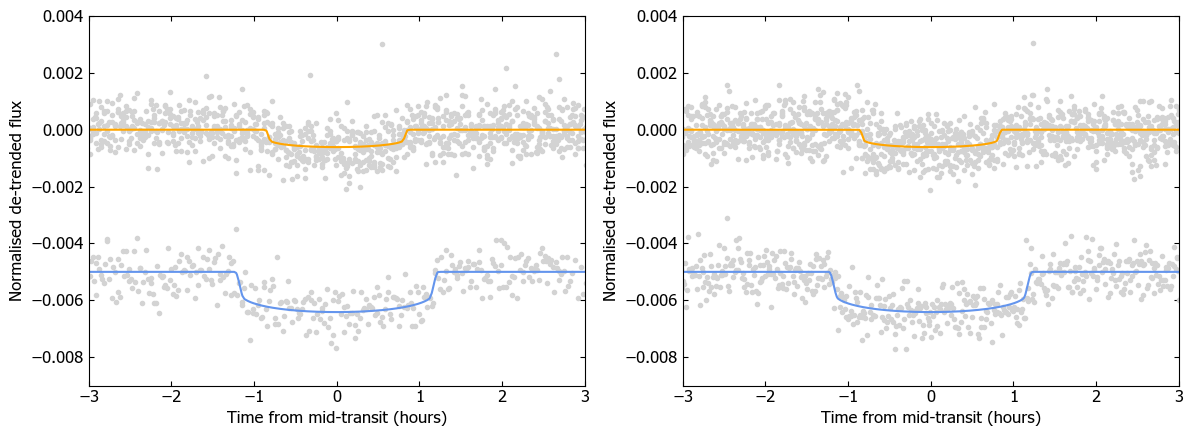}
    \end{subfigure}
    \caption{\textbf{Left panel:} \tess\ PDC-SAP light curve from Sector 11 minus the GP model, phase-folded to a period corresponding to that of \Tplanetb\ with the transit model shown in orange and phase-folded to a period corresponding to that of \Tplanetc\ with the transit model shown in blue. The data for \Tplanetc\ has been offset by -0.005 for clarity. \textbf{Right panel:} \tess\ PDC-SAP light curve from Sector 38 minus the GP model, phase folded and offset for each planet analogously to that of Sector 11.}
    \label{fig:TESSphasefold}
\end{figure*}

\TStar\ showed transit events from two exoplanet candidates, designated TOI-836.01 (\Tplanetc; SNR\,=\,21) and TOI-836.02 (\Tplanetb; SNR\,=\,17), identified from the \TESS\ light-curves. In Sector 11, \Tplanetb\ shows five transit events and one partial (egress only) transit, while \Tplanetc\ shows two transit events. One transit event of \Tplanetb\ would have occurred in the gap during which the satellite downloads data. See Table~\ref{tab:photobs} and the left-hand panel of Figure~\ref{fig:TESSlightcurve}.

\TStar\ was observed again in the third year of \tess\ operations during Sector 38, Camera 1, CCD 4 from 2021 April 28 to 2021 May 26. Seven transit events were observed for \Tplanetb, and three for \Tplanetc. See Table~\ref{tab:photobs} and right-hand panel of Figure~\ref{fig:TESSlightcurve}.

The transits of \Tplanetb\ indicate an orbital period of \Tperiodbshort\,days. The transit depth was \depthbexofop\,ppm, implying the planet candidate is a potential hot super-Earth.  For \Tplanetc\ the orbital period is \Tperiodcshort\,days, and the transit depth is \depthcexofop\,ppm, implying the candidate is potentially sub-Neptune in size.

For this work we use the Presearch Data Conditioning Simple Aperture Photometry (PDC-SAP) light-curve produced by the SPOC pipeline. The PDC-SAP light-curves have non-astrophysical trends removed from the raw Simple Aperture Photometry (SAP) light-curves using the PDC algorithm \citep{Stumpe2012,Stumpe2014,Smith2012}. The PDC-SAP light-curves for \TStar\ were retrieved from the Mikulski Archive for Space Telescopes (\href{https://mast.stsci.edu}{MAST}) portal and used in our joint model in Section \ref{sec:methods}.

To mitigate for the effects of stellar variability on the transit lightcurves in the Sector 11 and Sector 38 \tess\ data, we apply a Gaussian Process (GP) model using the \texttt{PyMC3} and \texttt{celerite} packages. We constrain this GP model for each sector using three hyperparameters as priors set up with log(\textit{s2}) (a jitter term describing the excess white noise, \citealt{exoplanet:pymc3}) and log(\textit{Sw4}) as normal distributions with a mean equal to the variance of the flux of each sector and a standard deviation of 0.1 for Sector 11 and 0.05 for Sector 38 (this is done to prevent overfitting of the GP); and the same is applied to log(\textit{w0}). log(\textit{Sw4}) and log(\textit{w0}) both represent terms that describe the non-periodic variability of the light-curves \citep{exoplanet:pymc3}. %; and \textit{Q} is set to 1/$\sqrt{2}$ for both sectors.
These hyperparameter setups are identical to those described for TOI-431 in \citet{ares} and informed by the \texttt{exoplanet} and \texttt{PyMC3} documentation. These hyperparameters are then incorporated into the SHOTerm kernel within the \texttt{exoplanet} framework, representing a stochastically-driven simple harmonic oscillator \citep{exoplanetdanforeman}. The GP model is then subtracted from the PDC-SAP flux to recover a flattened light curve from which transit models of \Tplanetb\ and \Tplanetc\ can be drawn. The effect of this can be seen in the first and second panels of Figure \ref{fig:TESSlightcurve} for Sector 11 and Sector 38 of \tess\ respectively.   We also plot the phase-folded \tess\ data for \Tplanetb\ and \Tplanetc\ in Figure \ref{fig:TESSphasefold} for both sectors.

For all follow-up photometry, we convert each time system to TBJD (\tess\ Barycentric Julian Date, BJD - 2457000) for consistency, and normalise each lightcurve by dividing by the median of the out-of-transit flux datapoints and subtracting the mean of the out-of-transit flux. The transits themselves are then modelled using a quadratic limb-darkened Keplerian orbit (with coefficients \textit{u\textsubscript{1}} and \textit{u\textsubscript{2}}) according to \citet{exoplanet:kipping13}, with parameters including stellar radius (\rstar) and mass (\mstar) in Solar units, planetary orbital period (\textit{P}) in days, transit ephemeris (T\textsubscript{\textit{c}}) in TBJD, impact parameter (\textit{b}), eccentricity (\textit{e}) and argument of periastron (\textit{$\omega$}) defined for each of \Tplanetb\ and \Tplanetc\ with priors informed by our spectral analysis and catalog data (see Appendices~\ref{tab:priorsstar}, \ref{tab:priorsb} and \ref{tab:priorsc} for details of the priors used). Transit models for each set of photometry time-series data are then created using the \texttt{starry} package within \texttt{exoplanet}, along with their corresponding planetary radii (\rpl), time of the data (\textit{t}) and exposure times for each instrument \texp.

%%%%%%%%%%%%%%%%%%%%%%%%%%%%%%%%%%%%%%%%%%%%%%%%%%
\subsection{CHEOPS photometry}
\label{sec:cheopsphot}

\begin{figure}
    \centering
    \includegraphics[width=\columnwidth]{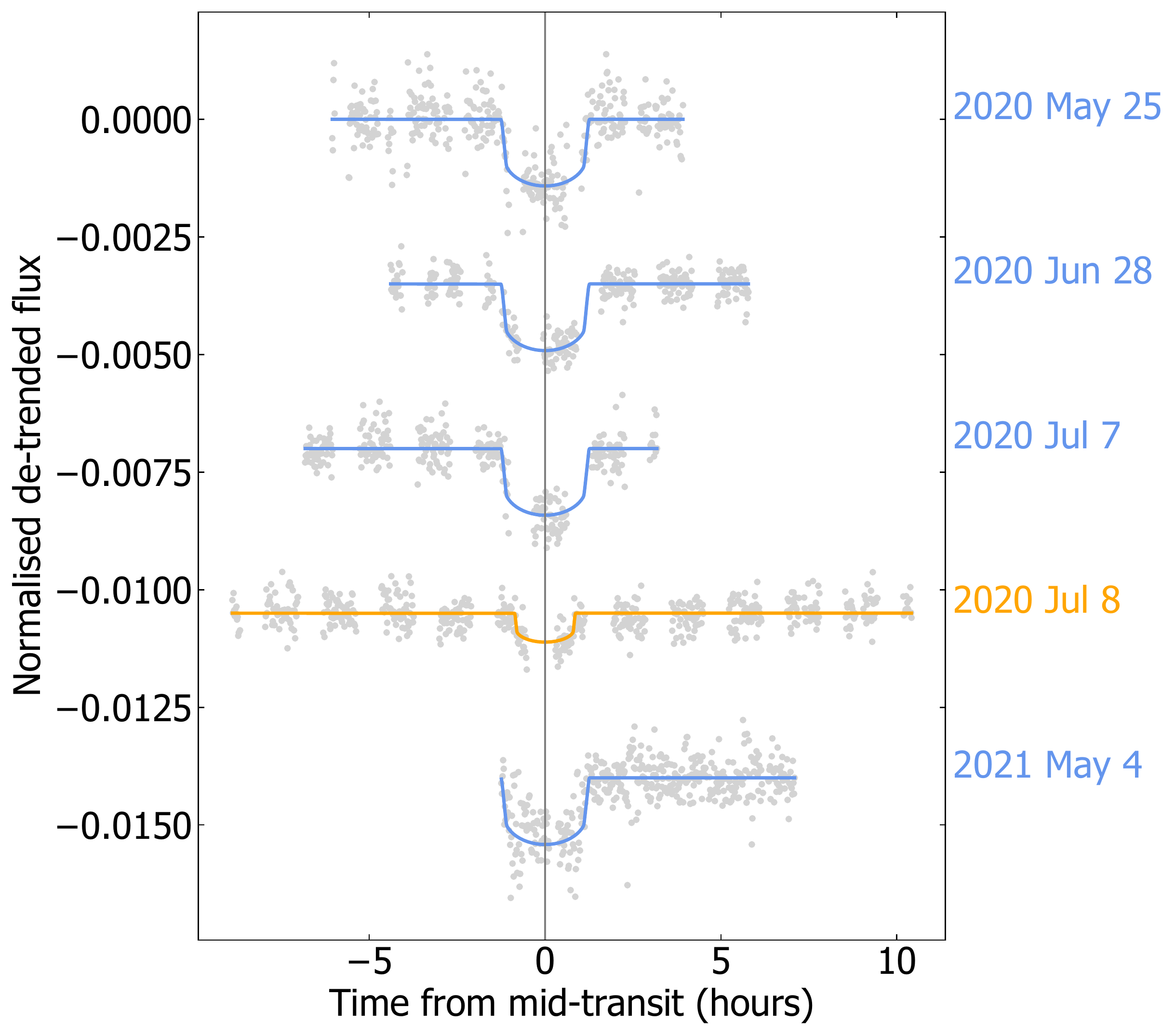}
    \caption{Light curves of \Tplanetb\ and \Tplanetc\ taken by the \cheops\ satellite as detailed in Table~\ref{tab:photobs}, plotted with our best fit \texttt{exoplanet} models for \Tplanetb\ in orange and \Tplanetc\ in blue, and offset for clarity.}
    \label{fig:cheopslcs}
\end{figure}

%\textcolor{green}{Tom Wilson}
%It is very difficult to obtain follow-up photometry for transiting exoplanet candidates with shallow transit depths from the ground. \textcolor{red}{the following subsections give 5 ground based instruments that you used for photometry! You don't need to justify the use of Cheops here over ground based photometry I think.} 
The transit depths for \Tplanetb\ and \Tplanetc\ are \depthbexofop\,ppm and \depthcexofop\,ppm respectively, making them challenging for photometric follow-up efforts.  The \cheops\ mission is able to reach a precision of 15\,ppm per 6\,h for a star with V\,=\,9\,mag \citep{benz2021}, and \cheops\ is therefore in a unique position to confirm and characterise shallow transit discoveries from \tess, as has been shown in recent publications \citep{Bonfanti2021,Delrez2021,leleu2021}.

In order to better determine the planet radii and orbital ephemerides, and check for any TTVs, we observed \TStar\ with \cheops\ spacecraft between 2020 May 25 and 2021 May 4, as a part of the Guaranteed Time Observing programme, yielding a total of 57.81\,h on target. Five observations of \TStar\ were taken by the \cheops\ satellite, resulting in the recovery of four transits of \Tplanetc, and one transit of \Tplanetb. For all visits, we use an exposure time of 60\,s. See details set out in Table~\ref{tab:photobs}.

The \cheops\ spacecraft is in a low-Earth orbit and thus parts of the observations are unobtainable because the telescope passes through the South Atlantic Anomaly (SAA), and as the amount of stray-light entering the telescope becomes higher than the accepted threshold, our observations are interrupted by Earth occultations. These effects that occur on orbital timescales ($\sim$98.77\,min) result in onboard rejections of images and manifest in a decrease in observational efficiency, corresponding to 72\%, 55\%, 56\%, 54\%, \& 96\% per visit, as can be seen in Figure~\ref{fig:cheopslcs}.

For all visits, the data were automatically processed using the \cheops\ data reduction pipeline (DRP v13; \citealt{Hoyer2020}), that conducts image calibration, such as bias, gain, non-linearity, dark current, and flat fielding corrections, and performs rectifications of environmental and instrumental effects, for example cosmic-ray hits, smearing trails, and background variations. Aperture photometry is subsequently done on the corrected images using a set of standard apertures; $R$ = 22.5\arcsec\ (RINF), 25.0\arcsec\ (DEFAULT), and 30.0\arcsec\ (RSUP), and an additional aperture that aims to optimise the radius based on contamination level and instrumental noise (ROPT). For the \cheops\ observations of \TStar, this radius is either 29.0 or 29.5\arcsec. The DRP also computes a contamination estimate of background sources, as detailed in section~6.1 of \cite{Hoyer2020}, that is subtracted from the light curves. 

Due to the orbit of \cheops\ and thus the rotating field of view, \cheops\ data include short-term, non-astrophysical flux trends due to nearby contaminants, background variations, or changes in instrumental environment that vary on the timescale of the orbit of \cheops. Whilst previous works have used linear decorrelation with instrumental basis vectors \citep{Bonfanti2021,Delrez2021,leleu2021} or Gaussian process regression \citep{Lendl2020}, a recent study has shown that a novel PSF detrending method can also remove these roll angle trends \citep{Wilson2022}. In brief, this method assesses PSF shape changes over a visit by conducting a principal component analysis on the autocorrelation function of the \cheops\ subarray images, as it was found that a myriad of causes of systematic variation within \cheops\ data affects the PSF shape. A leave-one-out-cross-validation \citep{loocv} is used to select the most prominent components that are subsequently used to decorrelate the light curve produced by aperture photometry. We apply this method to the TOI-836 \cheops\ observations with fluxes obtained with the DEFAULT aperture. The decorrelated CHEOPS data are presented in Table~\ref{tab:cheopsphot}, along with the resulting light-curves in Figure~\ref{fig:cheopslcs}.

\subsection{Ground-based Follow-up Photometry}

\begin{center}
\begin{table*}
    \centering
    \caption{Photometric observations of \TStar.}
    \label{tab:photobs}
    \begin{tabularx}{0.92\textwidth}{ l c c c c p{0.1\textwidth} p{0.07\textwidth} p{0.12\textwidth} }
    \toprule
    \textbf{Instrument} &\textbf{Aperture} &\textbf{Filter} &\textbf{Exposure time (s)} & \textbf{No. of images}    & \textbf{UT night} & \textbf{Planet}    & \textbf{Epoch no.} \\
    \hline
    \tess   & 0.105\,m   & \tess\textsuperscript{1}   & 120   & 19527   & 2019 Apr 22 - 2019 May 20   & \Tplanetb\ \Tplanetc & Epochs\,\,1-7 Epochs\,\,1-2\\
    \MEarthSouth    & 0.4\,m $\times$ 7   & RG715 & 32   & 3054    & 2019 Jul 4    & \Tplanetc & Epoch\,\,8 \\
    \lcogt-SSO  & 1.0\,m & \textit{Y} & 40   & 232 & 2020 Feb 29 & \Tplanetc    & Epoch\,\,36 \\
    \lcogt-CTIO\textsuperscript{A}    & 1.0\,m    & \textit{Y} & 100   & 138   & 2020 Mar 8   & \Tplanetb   & Epoch\,\,83 \\
    \lcogt-SSO\textsuperscript{B}  & 1.0\,m    & \textit{Y}    & 100   & 109   & 2020 Mar 20 & \Tplanetb & Epoch\,\,86 \\
    \lcogt-SSO     & 1.0\,m  & \textit{z\textsubscript{s}} & 30   & 341    & 2020 Apr 12   & \Tplanetc & Epoch\,\,41\\
    \lcogt-SSO  & 1.0\,m    & \textit{Y}    & 100   & 260   & 2020 May 4  & \Tplanetb & Epoch\,\,98 \\
    \lcogt-SAAO\textsuperscript{C}    & 1.0\,m  & \textit{z\textsubscript{s}}   & 30   & 327 & 2020 May 16  & \Tplanetc    & Epoch\,\,45 \\
    \cheops & 0.32\,m & \cheops\textsuperscript{2} & 60    & 398   & 2020 May 25   & \Tplanetc & Epoch\,\,46 \\
    \cheops & 0.32\,m & \cheops\textsuperscript{2} & 60    & 319   & 2020 Jun 28   & \Tplanetc & Epoch\,\,50 \\
    \cheops & 0.32\,m & \cheops\textsuperscript{2} & 60    & 318   & 2020 Jul 7   & \Tplanetc  & Epoch\,\,51 \\
    \cheops & 0.32\,m & \cheops\textsuperscript{2} & 60    & 574   & 2020 Jul 8   & \Tplanetb  & Epoch\,\,115 \\
    \lcogt-SSO  & 1.0\,m  & \textit{z\textsubscript{s}}   & 30    & 345    & 2021 Apr 8    & \Tplanetc & Epoch\,\,83 \\
    \astep  & 0.4\,m  & \textit{R\textsubscript{c}} & 25    & 370    & 2021 Apr 8    & \Tplanetc\ (egress)   &Epoch\,\,83 \\
    \ngts   & 0.2\,m $\times$ 3   & \ngts\textsuperscript{3}  & 10     & 5405   & 2021 Apr 16   & \Tplanetc & Epoch\,\,84 \\
    \lcogt-CTIO   & 1.0\,m  & \textit{z\textsubscript{s}}   & 30    & 382 & 2021 Apr 16  & \Tplanetc    & Epoch\,\,84 \\
    \tess   & 0.105\,m & \tess\textsuperscript{1} & 120    & 19226   & 2021 Apr 29 - 2021 May 26   & \Tplanetb\ \Tplanetc & Epochs\,\,194-200\,\, Epochs\,\,86-88\\
    \cheops & 0.32\,m & \cheops\textsuperscript{2} & 60   & 431    & 2021 May 4   & \Tplanetc  & Epoch 86 \\
    \lcogt-CTIO    & 1.0\,m   & \textit{z\textsubscript{s}}    & 30 & 300  & 2021 Jun 24  & \Tplanetc    & Epoch\,\,92 \\
    \bottomrule
    \end{tabularx}
    \begin{tabularx}{0.92\textwidth}{ l p{0.07\textwidth} p{0.12\textwidth} }
    % \item Sources: \tess\ \citep{Ricker:2015}, \cheops\ \citep{benz2021}, \MEarthSouth\ \citep{mearth2015, mearthnorth2015}, \ngts\ \citep{Wheatley2013, Wheatley2018},\\
    % \astep\ \citep{astep2010}, \lcogt\ \citep{lcogt2013}\\
        \textsuperscript{1}\tess\ custom 600--1000 nm
        \hspace{5 mm} \textsuperscript{2}\cheops\ custom 350--1100 nm \hspace{5 mm} \textsuperscript{3}\ngts\ custom 520--890 nm
    \end{tabularx}
    \begin{tabularx}{0.92\textwidth}{ l p{0.07\textwidth} p{0.12\textwidth} }
        \textsuperscript{A}CTIO - Cerro Tololo Inter-American Observatory \hspace{3 mm} \textsuperscript{B}SSO - Siding Spring Observatory \hspace{3 mm} \textsuperscript{C}SAAO - South Africa Astronomical Observatory
    \end{tabularx}
    % \begin{tablenotes*}
    %     \item Sources: \tess\ \citep{Ricker:2015}, \cheops\ \citep{benz2021}, \MEarthSouth\ \citep{mearth2015, mearthnorth2015}, \ngts\ \citep{Wheatley2013, Wheatley2018}, \astep\ \citep{astep2010}, \lcogt\ \citep{lcogt2013}
    %     \item \textsuperscript{1}\tess\ custom filter at 600--1000 nm \hspace{5 mm} \textsuperscript{2}\cheops\ custom filter at 350--1100 nm \hspace{5 mm} \textsuperscript{3}\ngts\ custom filter at 520--890 nm
    % \end{tablenotes*}
\end{table*}
\end{center}

\begin{center}
\begin{table}
    \centering
    \caption{\cheops\ photometric data for \Tstar. This table is available in its entirety online.}
    \label{tab:cheopsphot}
    \begin{tabular}{l c c c c }  
%    \begin{tabularx}{\columnwidth}{ p{0.115\textwidth} p{0.12\textwidth} X }
    \toprule
    \textbf{Time (BJD}   & \textbf{Normalised flux} & \textbf{Flux uncertainty} \\
    \textbf{-2457000}) & & \\
    \hline
    1994.88704 &0.99981 &0.00025\\
    1994.88773 &0.99955 &0.00026\\
    1994.88843 &1.00105 &0.00027\\
    1994.88912 &1.00140 &0.00030\\
    1994.88982 &1.00033 &0.00035\\
    1994.90649 &0.99897 &0.00027\\
    1994.90718 &0.99896 &0.00026\\
    1994.90788 &1.00011 &0.00025\\
    1994.90857 &1.00045 &0.00025 \\
    \multicolumn{1}{c}{...} & \multicolumn{1}{c}{...} & \multicolumn{1}{c}{...} \\
    \bottomrule
    \end{tabular}
    %\end{tabularx}
\end{table}
\end{center}

%%%%%%%%%%%%%%%%%%%%%%%%%%%%%%%%%%%%%%%%%%%%%%%%%%

\subsubsection{MEarth-South photometry}
\label{sec:mearthphot}
% \textcolor{green}{Jonathan Irwin}
A transit of \Tplanetc\ was observed using the \MEarthSouth\
telescope array \citep{irwin2015} at Cerro Tololo
Inter-American Observatory (CTIO), Chile on 2019 July 3-4.  Seven
telescopes were operated defocused to a half-flux diameter of 12
pixels (10.1\,\arcsec, given the pixel scale of 0.84\,\arcsec/pix), and an exposure time of 32\,s, observing continuously starting from twilight until the target set below 2 airmasses.  Observations were made using an RG715 filter.  A meridian flip occurred during the transit and has been taken into account in the analysis by allowing for a separate magnitude zero-point on either side of the meridian to remove any residual flat fielding error.

Data were reduced following standard procedures for \MEarthSouth\ data  (e.g. \citealt{irwin2007,irwin2015}) with a photometric extraction aperture of radius 17 pixels (14.3\,\arcsec). To account for residual colour-dependent atmospheric extinction the transit model included linear decorrelation against airmass. The edge of the photometric aperture is slightly contaminated by fainter sources, the most significant being TIC 440887361, but we estimate that this source is approximately 10.6 \TESS\ magnitudes fainter than the target star, so the resulting dilution of the measured transit depth should be negligible. The \MEarthSouth\ light curve is shown in Figure~\ref{fig:followuplcs} and used in the joint modeling in Section~\ref{sec:jointmodel}.

%%%%%%%%%%%%%%%%%%%%%%%%%%%%%%%%%%%%%%%%%%%%%%%%%%

\subsubsection{ASTEP photometry}
\label{sec:astepphot}

\ASTEP\ (Antarctic Search for Transiting ExoPlanets) is a 40\,cm Newtonian telescope designed to perform high precision photometry under the extreme conditions of the Antarctic winter \citep{Fressin2005a, Daban2010, Abe2013, Guillot2015, Mekarnia2016}. It is installed at the French-Italian Concordia station at Dome\,C, Antarctica (75$^{\circ}\,$06' S, 123$^{\circ}\,$21' E) on a summit of the high Antarctic plateau, at an altitude of 3233\,m, 1100\,km inland. Dome\,C is an ideal location for time-series observations thanks to the 4-month continuous night during the Antarctic winter and favourable weather conditions \citep{Crouzet2010, Crouzet2018}. 
\ASTEP\ is equipped with a FLI Proline KAF 16801 E $\rm 4096 \times 4096$ pixel CCD camera observing in an $\rm R_c$ band-pass, the field of view is $1^{\circ} \times 1^{\circ}$ and the pixel size is 0.9 arcsec/pixel.

We observed \TStar\ on 2021 April 8, during 5 hours between BJD 2459313.20 and 2459313.41, and we detected the second half of the transit of \Tplanetc. We scheduled the observation using a custom scheduling tool that sends queries to the {\tt TESS Transit Finder}. We set the exposure time to 25\,s, the cadence was 50\,s, and we collected 370 frames. The median Full Width Half Maximum (FWHM) was 4.06\,\arcsec\ and the airmass varied between 1.57 and 1.94. The details of the \ASTEP\ observations are set out in Table~\ref{tab:photobs}. We performed differential aperture photometry using a custom data reduction pipeline based on the pipeline described in \citet{Mekarnia2016} and adapted to \tess\ follow-up.  We used an aperture radius of 10 pixels (9.3\,\arcsec) and 8 comparison stars. The light curve RMS is 1.43\,ppt and decreases to 1.2\,ppt after binning the light curve with a bin size of 3 points, for a predicted transit depth of 1.38\,ppt. The transit appears clearly and is on target.  The \ASTEP\ light curve is shown in Figure~\ref{fig:followuplcs} and used in the joint modelling in Section~\ref{sec:jointmodel}. The \ASTEP\ telescope is now being upgraded with two new cameras that will observe simultaneously in two colors and will provide a much better throughput \citep{Crouzet2020}.

%%%%%%%%%%%%%%%%%%%%%%%%%%%%%%%%%%%%%%%%%%%%%%%%%%%%%%%%%%%%%%%%%%%%%%%%%%%%%

\subsubsection{NGTS photometry} \label{sec:ngtsphot}

We monitored a full transit of \Tplanetc\ on the night of 2021 April 16 using three of the \ngts\ \citep[Next Generation Transit Survey;][]{Wheatley2018} telescopes at the ESO Paranal Observatory, Chile. The observations were performed using the \ngts\ multi-telescope observing method described in \citet{bryant:wasp166b:2020} and \citet{smith2020}. \NGTS\ consists of an array of 0.2\,m robotic telescopes, each with a wide field-of-view of 8 square degrees. A custom \ngts\ filter of 520--890\,nm is used, and images are taken using Andor iKon-L 936 cameras, which deliver a plate-scale of $5\,\arcsec$\,pix$^{-1}$. We use an exposure time of 10\,s, and with readout time this translates to a cadence of approximately 13\,s.  The details of the \NGTS\ observations are set out in Table~\ref{tab:photobs}. 

\begin{figure}
    \centering
    \includegraphics[width=\columnwidth]{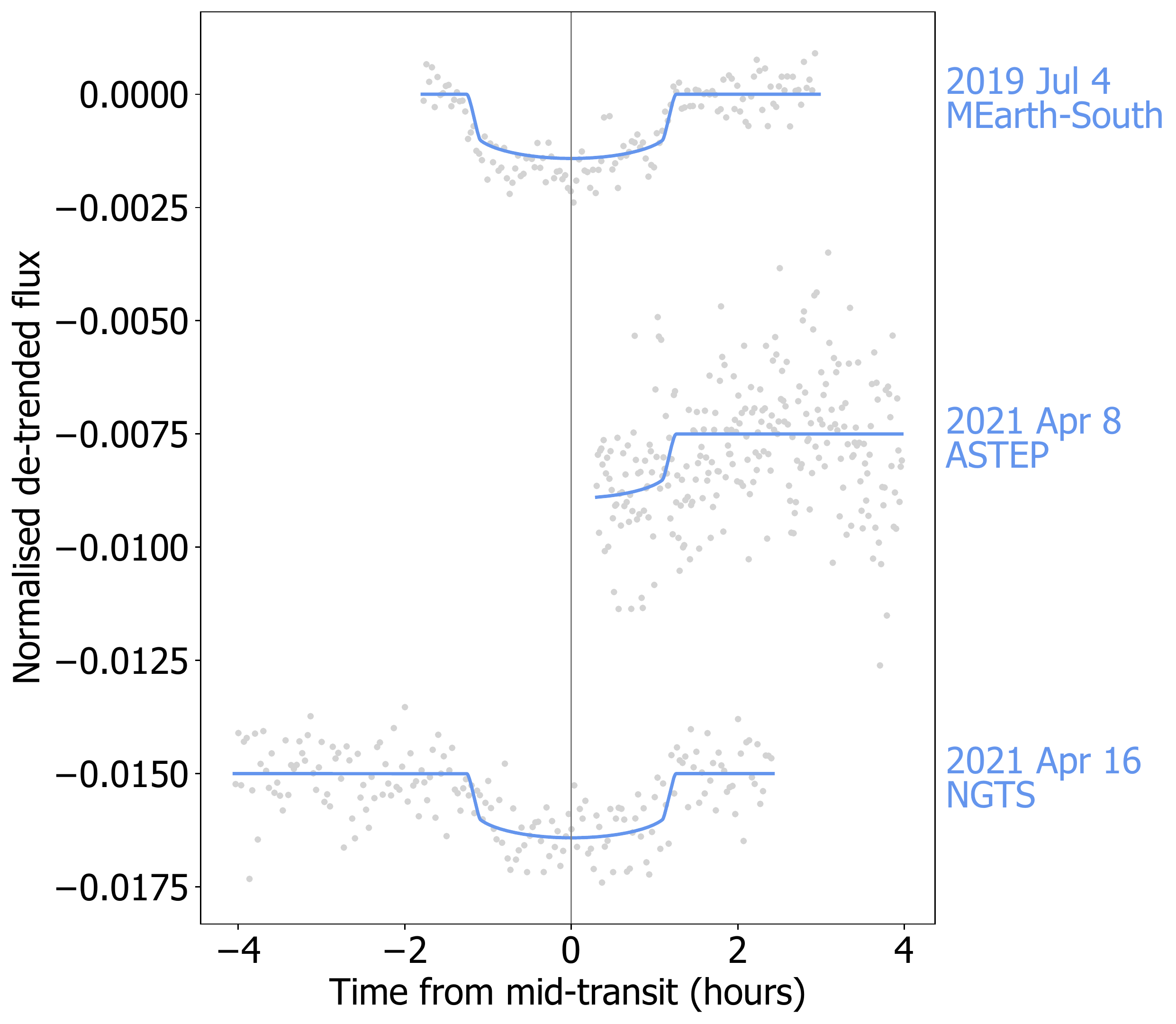}
    \caption{Lightcurves of \Tplanetc\ taken by the \MEarthSouth, \ngts\ and \astep\ facilities as detailed in Table~\ref{tab:photobs}, plotted with our best fit \texttt{exoplanet} models and offset for clarity.}
    \label{fig:followuplcs}
\end{figure}

%We detrended the flux using a linear fit to the out-of-transit flux.
The \ngts\ image reduction was performed using an adapted version of the standard \ngts\ pipeline \citep{Wheatley2018}, which has been updated to perform aperture photometry for a single star. Comparison stars which are isolated and similar to \Tstar\ in brightness and CCD position were automatically identified by the pipeline using \textit{Gaia} DR2 \citep{GAIA_DR2}. The resultant flux from each telescope was detrended independently against airmass, and the photometry from the three telescopes is combined into a single light curve file, which is publicly available from the ExoFOP-TESS website\footnote{\url{https://exofop.ipac.caltech.edu/tess/}}. The \NGTS\ light curve is shown in Figure~\ref{fig:followuplcs} and used in the joint modeling in Section~\ref{sec:jointmodel}.

%%%%%%%%%%%%%%%%%%%%%%%%%%%%%%%%%%%%%%%%%%%%%%%%%%

\subsubsection{LCO photometry}
\label{sec:lcophot}
\begin{figure}
    \centering
    \includegraphics[width=\columnwidth]{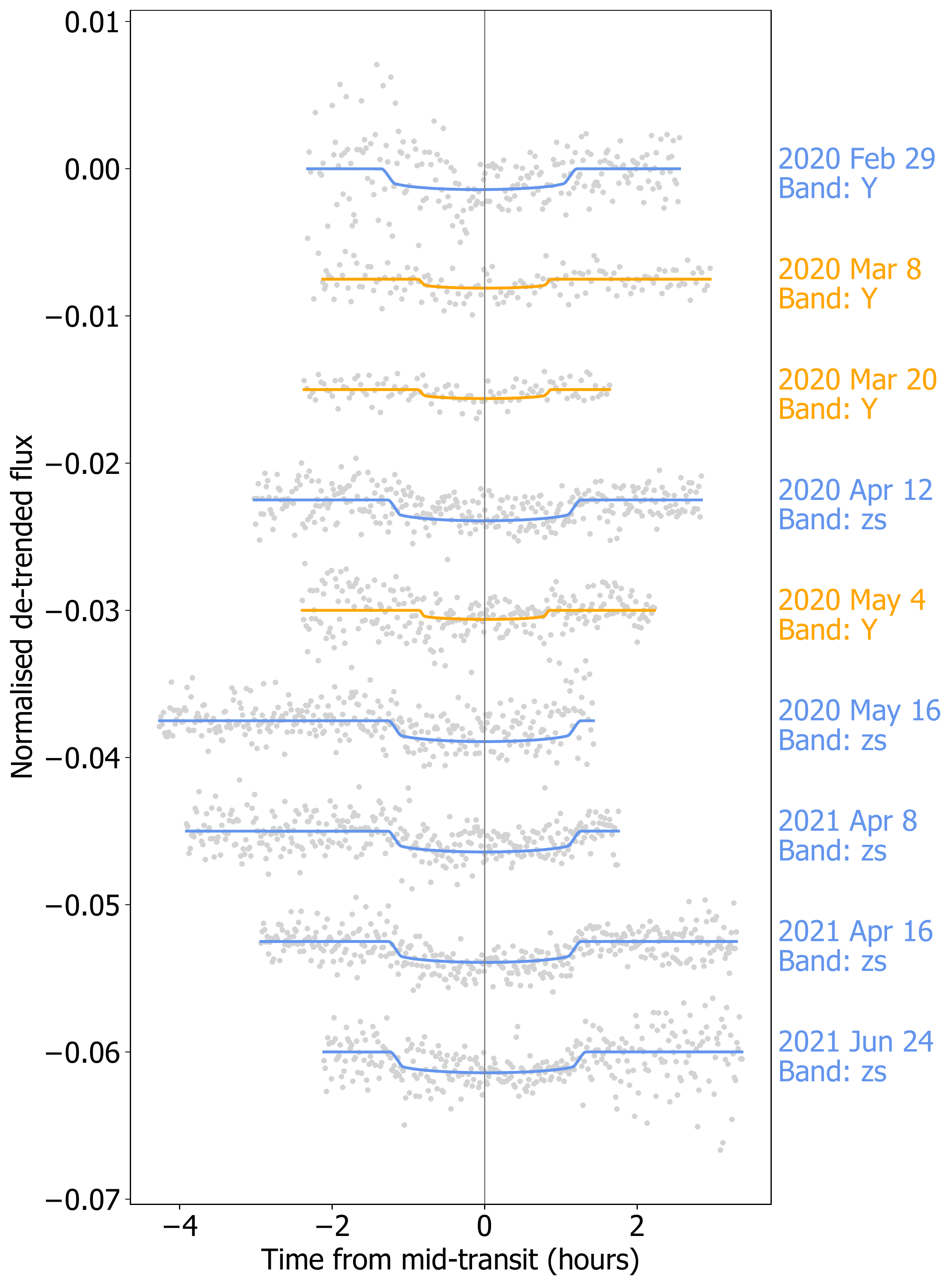}
    \caption{Light curves of \Tplanetb\ and \Tplanetc\ taken by the \lcogt\ network as detailed in Table~\ref{tab:photobs}, plotted with our best fit \texttt{exoplanet} models for \Tplanetb\ in orange and \Tplanetc\ in blue, and offset for clarity.}
    \label{fig:lcolcs}
\end{figure}

% \textcolor{green}{Karen Collins}
We observed three full transits of \Tplanetb\ and six full transits of \Tplanetc\ from the Las Cumbres Observatory Global Telescope \citep[\lcogt;][]{lcogt2013} 1.0\,m network. The details of the \lcogt\ observations are set out in Table~\ref{tab:photobs}. We used the {\tt TESS Transit Finder}, which is a customized version of the {\tt Tapir} software package \citep{Jensen:2013}, to schedule our transit observations. The telescopes are equipped with $4096\times4096$ SINISTRO cameras having an image scale of 0.389\arcsec\ per pixel, resulting in a $26\arcmin\times26\arcmin$ field of view. The images were calibrated by the standard \lcogt\ {\tt BANZAI} pipeline \citep{McCully:2018}, and photometric data were extracted using {\tt AstroImageJ} \citep{Collins:2017}. The \lcogt\ light curves are shown in Figure~\ref{fig:lcolcs} for \Tplanetb\ and \Tplanetc, and used in the joint modelling in Section~\ref{sec:jointmodel}.

%%%%%%%%%%%%%%%%%%%%%%%%%%%%%%%%%%%%%%%%%%%%%%%%%%

\subsubsection{WASP-South photometry}
\label{sec:waspphot}
\begin{figure}
    \centering
    \includegraphics[width=0.5\textwidth]{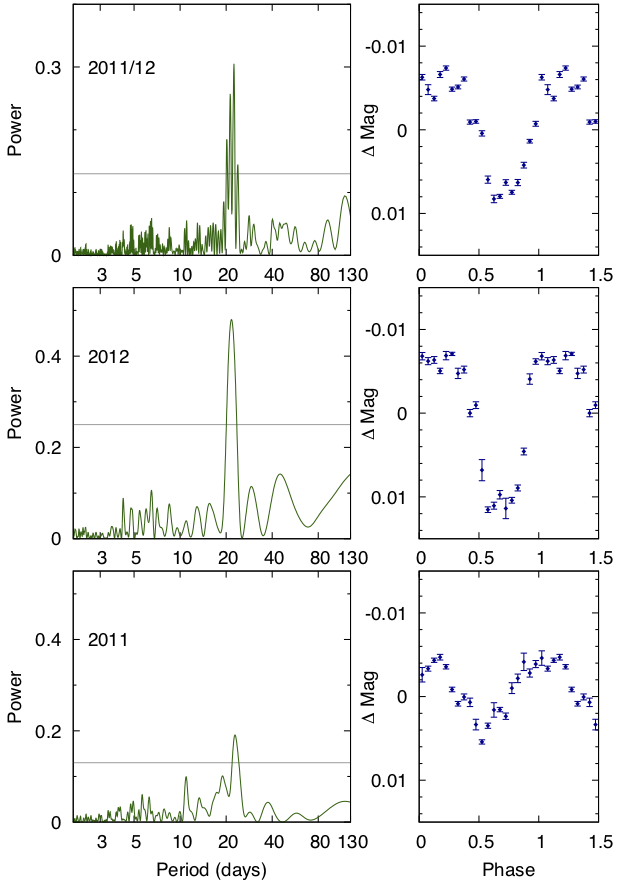}
    \caption{\textbf{Left panels:} Periodograms of the \waspsouth\ lightcurves of \Tstar\ from 2011 and 2012, and for 2011 \&\ 2012 combined. The horizontal line is the estimated 1\%-likelihood false-alarm level. \textbf{Right panels:} \waspsouth\ photometry data, phase-folded to the best stellar rotation period estimate.}
    \label{fig:waspperiodogram}
\end{figure}

% \textcolor{green}{Coel Hellier}
The \waspsouth\ array of 8 wide-field cameras was the Southern station of the \wasp\ transit-search project \citep{pollacco2006}. \waspsouth\ observed the field of TOI-836 repeatedly over the years 2006 to 2014, observing with a broad-band filter, and accumulating a total of 93,000 photometric data points. While the precision of these observations is not sufficient to detect the transits, the long-duration monitoring is ideal for detecting photometric activity due to star spots. We thus searched the data for a rotational modulation using the methods discussed in \citet{2011PASP..123..547M}. We find a persistent periodicity with a period of 22.0 $\pm$ 0.1\,days, where the uncertainty estimate makes allowance for phase changes caused by changing star-spot patterns. The amplitude varies from 3 to 8\,mmag and the false-alarm probability in each season's dataset is typically $<$\,1\%. In Figure~\ref{fig:waspperiodogram} we show periodograms from two seasons of data, together with the resulting modulation profile from folding the data.

The 22\,day period is consistent with activity seen in the \tess\ data, particularly in Sector 38 data (see Figure~\ref{fig:TESSlightcurve}). We therefore adopt this as the likely spin period of the star and use it to inform our joint modelling in Section~\ref{sec:jointmodel}.

%%%%%%%%%%%%%%%%%%%%%%%%%%%%%%%%%%%%%%%%%%%%%%%%%%%

\subsection{Follow-up Spectroscopy}
In order to determine the stellar parameters and measure radial velocity variations, a number of spectrographs were used to observe \Tstar.  Two reconnaissance spectra were taken on 2019 July 1 and 2021 May 28 with the Tillinghast Reflector Echelle Spectrograph (\TRES) \citep{tres} on the 1.5\,m telescope at the Fred Lawrence Whipple Observatory (FLWO). The spectra were used to derive stellar parameters using the Stellar Parameter Classification (SPC) tool \citep{spc2012, spc2014}. These spectra indicated that \Tstar \ is a K-dwarf with a low \vsini\ that would be amenable to high-precision radial velocity follow-up.  In this section we describe these high-precision radial velocity data, which are obtained using the \harps\ and \pfs\ spectrographs. We also obtain 11 spectra from the \hires\ spectrograph \citep{hires}, taken from 2009 April 6 to 2013 February 3, which we use to examine long-term radial velocity trends. The iSHELL radial velocities were taken at 2.3 microns, and as we do not implement a chromatic RV analysis as in \citet{cale2021}, we exclude them from our analysis. Additional radial velocity data from \minerva\ also exist, but the lower precision of these data mean that we omit them from our analysis.

\subsubsection{HARPS radial velocity observations}
\label{sec:harpsrv}
%\textcolor{green}{Dave Armstrong}
\harps\ \citep[High Accuracy Radial velocity Planet Searcher;][]{HARPS} is an Echelle spectrograph mounted on the ESO 3.6\,m telescope situated at La Silla Observatory, Chile. A total of 52 spectra of \Tstar\ were obtained with \harps\ as part of the \textit{NCORES} program (PI D. Armstrong, 1102.C-0249). 15 of these spectra were obtained  from 2020 March 16 to 2020 March 23 (7 nights), followed by a further 37 spectra from 2021 January 22 to 2021 March 2 (39 nights). These data were obtained in \harps\ High-Accuracy Mode with a 1\arcsec\ diameter fibre, standard resolution of R$\sim$115,000, and exposure times of approximately 1500\,s. Raw data were reduced according to the standard \harps\ data reduction software detailed in \citet{lovis2007}. The data table for these observations can be found in Table~\ref{tab:harpsobs}, which we use in our joint modelling (Section~\ref{sec:jointmodel}). The \HARPS\ data are marked with an asterisk in Table~\ref{tab:rvfu}.

\begin{center}
\begin{table*}
    \centering
    \caption{\HARPS\ spectroscopic data for \Tstar. This table is available in its entirety online.}
    \label{tab:harpsobs}
    \begin{tabular}{c c c c c c c}  
%    \begin{tabularx}{\columnwidth}{ p{0.115\textwidth} p{0.12\textwidth} X }
    \toprule
    \textbf{Time (BJD} & \textbf{RV} & \textbf{RV error} & \textbf{FWHM} & \textbf{Bisector} & \textbf{Contrast} & \textbf{S-index\textsubscript{MW}} \\
    \textbf{-2457000)} & \textbf{(\ms)} & \textbf{(\ms)} & \textbf{(\ms)} & \textbf{(\ms)} &  & \\
    \hline
1924.744232 & -26270.62 & 1.20 & 6479.82 & 59.29 & 42.086199 & 1.118916 \\
1924.847515 & -26272.89 & 1.13 & 6477.87 & 58.02 & 42.082108 & 1.088405 \\
1925.765286 & -26277.15 & 1.33 & 6483.37 & 54.98 & 42.104065 & 1.099795 \\
1925.897310 & -26278.60 & 1.42 & 6484.65 & 62.33 & 42.063377 & 1.035016 \\
1926.748165 & -26279.33 & 1.23 & 6481.65 & 63.03 & 42.111069 & 1.073716 \\
1926.891093 & -26276.88 & 1.25 & 6474.28 & 65.77 & 42.150971 & 1.039492 \\
1927.807982 & -26280.90 & 1.66 & 6472.36 & 61.35 & 42.201152 & 1.068344 \\
1927.885303 & -26283.22 & 1.24 & 6470.19 & 62.19 & 42.177954 & 1.035070 \\
1928.764641 & -26288.22 & 1.24 & 6465.28 & 65.38 & 42.164275 & 1.058810 \\
1928.890901 & -26289.86 & 1.37 & 6466.36 & 65.65 & 42.174431 & 1.042093 \\
...         & ...       & ...  & ...     & ...   & ...       & ... \\
\bottomrule
    %\end{tabularx}
    \end{tabular}
\end{table*}
\end{center}

\begin{center}
\begin{table}
    \centering
    \caption{Radial velocity follow-up details for \Tstar. Observations used in the joint model are marked with an asterisk.}
     \label{tab:rvfu}
    \begin{tabularx}{0.865\columnwidth}{ p{0.27\linewidth} p{0.13\linewidth} p{0.11\linewidth} X }
    \toprule
    \textbf{Facility} & \textbf{Telescope aperture} & \textbf{No. of spectra} & \textbf{Resolution} \\
    \hline
        \harps\ * &3.6\,m &52  &115000 \\
        \hires &10.0\,m &11 &60000 \\
        \pfs\ * &6.5\,m &30 &130000 \\
        \ishell &3.0\,m &10 &70000 \\
        \minerva &0.7\,m $\times$ 6 &27 &75000 \\
    \bottomrule
    \end{tabularx}
    \begin{tablenotes}
    \item Sources: \harps\ \citep{HARPS}, \hires\ \citep{hires}, \pfs\ \citep{pfs2006}, \ishell\ \citep{ishell2012}, \Minerva\ \citep{minerva2018, addison2019, addison2021}
    \end{tablenotes}
\end{table}
\end{center}

%%%%%%%%%%%%%%%%%%%%%%%%%%%%%%%%%%%%%%%%%%%%%%%%%%

\subsubsection{PFS radial velocity observations}
\label{sec:pfsrv}
% \textcolor{green}{Johanna Teske}
The Planet Finder Spectrograph (\PFS) \citep{pfs2006, crane2008, crane2010} is a high resolution optical Echelle spectrograph mounted on the 6.5\,m Magellan II Telescope at Las Campanas Observatory, Chile. \PFS\ is calibrated via an iodine-cell, and raw data are reduced to 1D spectra and relative radial velocities extracted using a custom pipeline based on \citet{butler1996}. The spectrograph was upgraded in 2018, and now operates with a default slit width of 0.3\,\arcsec, which delivers a resolving power of R$\sim$130,000.

\Tstar\ was observed as part of the Magellan-\TESS\ Survey \citep{magellanteske} between 2019 July 10 to 2020 March 17. Exposure times were approximately 900-1200\,s per individual observation, and usually two observations were taken per night (separated by $\sim$2 hours) and binned together.  In total, 38 binned radial velocities were published in Teske et al. for \Tstar, and these are set out in table~4 of \citet{magellanteske}. We use the PFS radial velocities in our joint modelling (Section~\ref{sec:jointmodel}). The \pfs\ data are marked with an asterisk in Table~\ref{tab:rvfu}.

\begin{figure}
\centering
\includegraphics[width=\columnwidth]{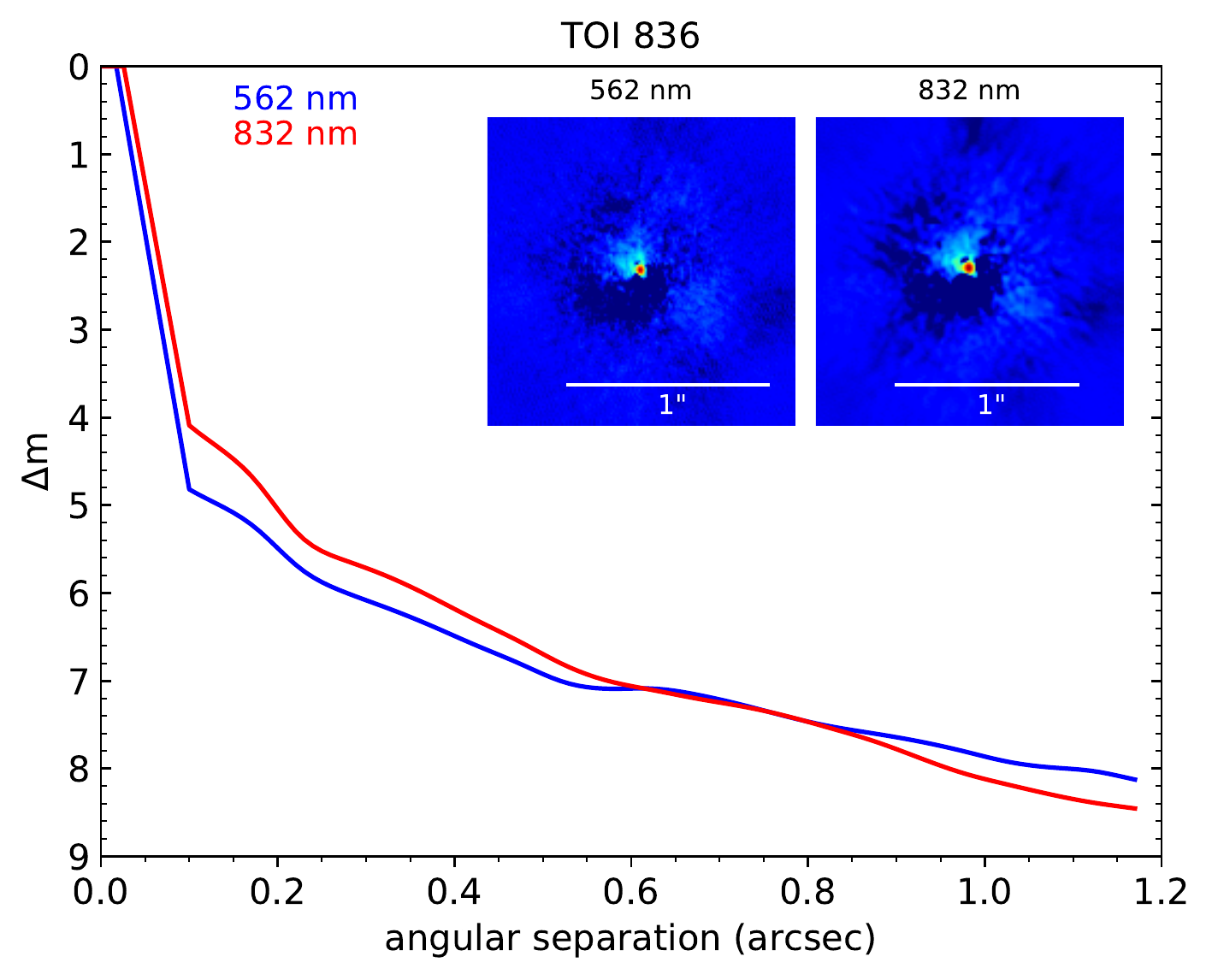}
\caption{Reconstructed images and speckle sensitivity curves of \TStar\ taken on 2020 March 13 using Zorro on the \gemini-South 8.0\,m telescope at Cerro Pach\'{o}n, Chile, in each of the two bandpasses. No close companions are visible brighter than a contrast of 5 mag for separations between 0.2 and 1.2 \arcsec. Other direct imaging data also place similar constraints on the presence of close companions.}
\label{fig:gemini}
\end{figure}

% \begin{figure}
%     \centering
%     \includegraphics[width=0.48\textwidth]{3.Results/Figures/HARPS_PFS_RV_phase_TOI-836.png}
%     \caption{\textbf{Top panel:}  \harps\ (purple circles) and \pfs\ (orange triangles) RV data, phase-folded to a period corresponding to that of \Tplanetb\ with the RV model shown in orange. \textbf{Bottom panel:} \harps\ and \pfs\ data, phase-folded to a period corresponding to that of \Tplanetc\ with the RV model shown in blue.}
%     \label{fig:harpsrvphase}
% \end{figure}

% \begin{figure}
%     \centering
%     \includegraphics[width=\columnwidth]{3.Results/Figures/dace1.png}
%     \caption{Radial velocity data of \Tstar\ from the \hires\ instrument on the \keck\ telescope from 2009 April 6 to 2013 February 3, and fit with a linear trend represented by the solid grey line.}
%     \label{fig:hiresdace}
% \end{figure}

%%%%%%%%%%%%%%%%%%%%%%%%%%%%%%%%%%%%%%%%%%%%%%%%%%

\subsubsection{HIRES radial velocity observations}
\label{sec:hiresrv}
\hires\ \citep[High Resolution Echelle Spectrometer;][]{hires} is an R$\sim$60,000 resolving power spectrograph mounted on the 10\,m Keck Telescope at Mauna Kea Observatory, Hawaii.  Like \pfs, \hires\ also operates with an iodine-cell wavelength calibration, and data are reduced using a custom pipeline based on \citet{butler1996}.

\Tstar\ was observed as part of the Lick-Carnegie Exoplanet Survey \citep{butler2017} between 2009 April 6 to 2013 February 3.  In total, 11 observations were made over this four year time period, with a typical exposure time of approximately 500\,sec.  These data are set out in table~1 of \citet{butler2017}.  The observations were made prior to the discovery of the transiting planets \Tplanetb\ and \Tplanetc.  The low cadence of these observations, coupled with the stellar activity of \Tstar, means that we decided not to use them in our GP-based joint model of Section~\ref{sec:jointmodel} - however they do enable us to study any long-term radial velocity trends for the system (see Section~\ref{sec:longterm}).

%%%%%%%%%%%%%%%%%%%%%%%%%%%%%%%%%%%%%%%%%%%%%%%%%%

\subsection{Imaging}
\label{sec:imaging}
% \textcolor{green}{Steve Howell}
The large size of the TESS pixels (21$\arcsec$) necessitates a careful study of neighbouring regions in order to determine if there are stars blended in to the TESS photometric data. In such cases, planet transits can be mimicked by other stellar configurations (e.g., \citealt{lillobox12, howell11, lillobox14, furlan17}).  \gaia\ shows \Tstar\ to be a relatively isolated star, with no neighbours with $\Delta\,T_{mag}\,<\,6$  in the photometric aperture to within its sensitivity limits (see Figure~\ref{fig:tpfgaia}).  To probe regions very close to \Tstar\ (<\,1.5\,\arcsec), where \gaia\ is known to be incomplete, we use direct imaging from large ground-based telescopes.

% know that there are no stars within    Such contaminating light sources will affect the observed transit depths, and in an extreme case could even be a blended eclipsing binary star that causes the apparent transit signal itself. \textcolor{red}{Table \ref{tab:imaging}}

\Tstar\ was imaged by multiple telescopes and instruments in order to check for close companions.  This imaging includes \gemini-Zorro and \gemini-'Alopeke \citep{scott2021}, \vlt-NaCo \citep{rousset2003}, \keck-2-NIRC2 \citep{ciardi2015} and \soar-HRCam \citep{ziegler2020}. These imaging data are publicly available from the ExoFOP-TESS website\footnote{\url{https://exofop.ipac.caltech.edu/tess/}}.  The conclusion from all of these imaging data is that \Tstar\ has no close companions outside a separation of 0.2\,\arcsec.

As an example of this direct imaging data, Figure~\ref{fig:gemini} shows the reconstructed images and speckle sensitivity curves from the observation taken using the Zorro instrument \citep{scott2021} on \gemini-South at Cerro Pach\'{o}n Observatory, Chile. This imaging was taken on 2020 March 13 in two simultaneous passbands (562\,nm and 832\,nm), and like all the direct imaging, shows that \Tstar\ is an isolated star to within the 5\,$\sigma$ contrast limits.

\section{Methods and Results} \label{sec:methods}
%%%%%%%%%%%%%%%%%%%%%%%%%%%%%%%%%%%%%%%%%%%%%%%%%%%%%%%%%%%%%%%%%%%%%%%%%%%%%%%%%%%%%%%%%%%%%%%%%%%%%%%%%%%%%%%%%%

\subsection{Stellar analysis} \label{sec:stellaranalysis}
%\textcolor{green}{Nuno Santos, Elisa Delgado Mena, Vardan Adibekyan, Sergio Sousa, text reviwed by Sergio and Vardan}
% Obtaining stellar parameters of \Tstar\ is a necessity if we are to derive further values for \Tplanetb\ and \Tplanetc\ such as the ages, masses and radii. 
To determine the stellar parameters for \Tstar, we co-add the 52 \harps\ spectra (Section~\ref{sec:harpsrv}) into a single combined spectrum with a signal-to-noise of {$\sim$400} at 550\,nm. We use the method described in \citet{sousa2014} and \citet{Santos:2013} in order to derive the stellar atmospheric parameters including a trigonometric surface gravity \logg, effective temperature \teff\ and metallicity \feh. This method measures the equivalent widths of iron lines in the combined \harps\ spectrum via the \texttt{ARES} v2 code \citep{sousa2015}. The abundances are then estimated using the \texttt{MOOG} code \citep{Sneden:1973} for radiative transfer, which includes a grid of model atmospheres from \citet{Kurucz:1993}, and we find the best set of spectroscopic parameters by assuming equilibriums of ionization and excitation. Following the same methodology as described in \citet[][]{sousa2021}, we use the \gaia\ EDR3 parallax and estimate the trigonometric surface gravity. This spectral analysis shows that \Tstar\ is a K-dwarf with a \logg\,=\,\Tloggporto\,dex and a \teff\,=\,\Teffporto\,K.  We find a metallicity of \feh\,=\,\Tstarfehporto\,dex and a \vsini\,=\,\Tstarvsiniporto\,\kms. %, and a mass of \mstar\,=\,\Tstarmassporto\,\msun.
%We find values of \mstar\,=\,\Tstarmassporto\,\msun\ and \rstar\,=\,\Tstarradiusporto\,\rsun\ using the param web-interface \citep[][]{daSilva2006}.

To obtain the radius of TOI-836, we use a Markov-Chain Monte Carlo (MCMC) modified infrared flux method (IRFM; \citealt{Blackwell1977,Schanche2020}). This is done by building spectral energy distributions (SEDs) from \textsc{atlas} Catalogue stellar atmospheric models \citep{Castelli2003} and stellar parameters derived via our spectral analysis, and calculating synthetic fluxes by integrating the SEDs over bandpasses of interest after attenuation to account for extinction. These fluxes are compared to observed broadband photometry retrieved from the most recent data releases for the following bandpasses; {\it Gaia} \textit{G}, \textit{G$_{\rm BP}$}, and \textit{G$_{\rm RP}$} \citep{GaiaCollaboration2021}, 2MASS \textit{J}, \textit{H}, and \textit{K} \citep{Skrutskie2006}, and {\it WISE} \textit{W1} and \textit{W2} \citep{wise} to calculate the apparent bolometric flux, and hence the stellar angular diameter and effective temperature. By converting the angular diameter to the stellar radius using the offset-corrected \gaia\ EDR3 parallax \citep{Lindegren2021}, we obtain \rstar\,=\,\Tstarradiusporto\,\rsun.

Starting from the basic input set given by (\teff, [Fe/H], \rstar), we then derived the isochronal mass \mstar\ and age $t_*$. To provide robust estimates, we employed two different evolutionary models, namely \texttt{PARSEC}\footnote{\textit{PA}dova and T\textit{R}ieste \textit{S}tellar \textit{E}volutionary \textit{C}ode: \url{http://stev.oapd.inaf.it/cgi-bin/cmd}} v1.2S \citep{marigo17} and \texttt{CLES} \citep[Code Liègeois d'Évolution Stellaire,][]{scuflaire08}. In detail, we derived a first pair of mass and age values using the isochrone placement technique \citep{bonfanti15,bonfanti16}, which we applied to pre-computed tables of \texttt{PARSEC} tracks and isochrones. Besides the basic input set, we further inputted the $v\sin{i_*}$ value to improve the convergence of the interpolating routine as detailed in \citet{bonfanti16}.
A second pair of mass and age estimates was instead retrieved through the \texttt{CLES} code, which generates the best stellar evolutionary track that reproduces the basic input set following the Levenberg-Marquadt minimisation scheme \citep{salmon21}.
After carefully checking the mutual consistency of the two respective pairs of values through the $\chi^2$-based criterion outlined in \citet{Bonfanti2021}, we finally merged the two output mass and age distributions and we obtained \mstar\,=\,\Tstarmassporto\,\msun\, and $t_*=$\Tstarageporto\,Gyr. We use these values of the stellar mass and radius as priors within our \texttt{exoplanet} modelling (described in Section~\ref{sec:jointmodel}), which are then fit for in the code to produce the final values seen in Table~\ref{tab:star_props_results}.

% The low temperature of the star also makes derivation of \vsini\ difficult, \textcolor{red}{so we estimate a value of \vsini=\,=\,\Tstarvsiniporto}. Through synthesis we find a value of \vsini\,=\,\Tstarvsinielisa, however it should be noted that a \vsini\ below 3\,\kms\ is degenerate with macroturbulent velocity \vmac, which itself is a broadening factor of spectral lines. \vmac\ is based on calibrations, however the \teff\ used is outside the calibration range, so we assign it a value of 2\,\kms\ such that \vmac\ decreases as \teff\ decreases.

Further to this, we derive stellar abundances using the curve-of-growth analysis method in local thermodynamic equilibrium, as employed in \citet{adibekyan2012, adibekyan2015}. We are unable to derive reliable values for the abundances of C and O because the lines for those elements become very weak and blended with other species for cool dwarf stars, as it is in the case of \Tstar\ (see eg. \citealt{delgadomena2021}). The values of \mgh\ and \sih\ are \Tstarmghporto\ dex and \Tstarsihporto\ dex respectively. These are typical values for a thin-disk star, which agrees with our calculated Galactic space velocity components and thin-disk membership probability as described in the next paragraph. There is no evidence in the stellar spectrum (such as a strong Li line) to suggest that \Tstar\ is a young star. The full set of results from our spectral analysis are set out in Table~\ref{tab:star_props_results}.

Following the formulation of \citet{johnson1987}, and using the values of proper motion and parallax from \gaia\ EDR3 (see Table \ref{tab:star_props}) and a radial velocity from \gaia\ DR2 of \GaiaRV\,\kms\ \citep{GAIA_DR2}, we calculate the values and uncertainties for $U$, $V$ and $W$, the heliocentric velocity components of the Galactic space velocities, in the direction of the galactic centre, rotation, and pole respectively, in Table \ref{tab:star_props_results}. We should note that we do not subtract the Solar motion and compute the $U$, $V$ and $W$ values in the right-handed system. %\textcolor{red}{please fill in sources column, TW: Done}

We also use the approach of \citet{reddy2006} in a Monte Carlo fashion with 100,000 samples to determine the probability that \Tstar\ is in a given kinematic Galactic family, using a weighted average of the results obtained using the velocity dispersion standards of \citet{bensby2003, bensby2014}, \citet{reddy2006}, and \citet{chen2021}. We find a Galactic thin disk membership probability for \Tstar\ of 98.9\%, thick disk membership probability of 1.1\%, and halo membership probability of 0\%. This agrees well with the Galactic eccentricity of \Tstar\ of 0.08, and the high Galactic Z-component of the angular momentum of $Z$ $\approx$\,1770\,kpc\,\kms. We compute these values using the {\tt galpy} package after a Galactic orbit determination using the  \gaia\ EDR3 position, proper motions, and parallax, and \gaia\ DR2 radial velocity integrating over 5\,Gyr, as well as the typical values for \mgh\ and \sih\ from stellar analysis.

%%%%%%%%%%%%%%%%%%%%%%%%%%%%%%%%%%%%%%%%%%%%%%%%%%%%%%%%%%%%%%%%%%%%%%%%%%%%%%%%%%%%%%%%%%%%%%%%%%%%%%%%%%%%%%%%%%

\subsection{Exoplanet data analysis} \label{sec:jointmodel}

We model the photometric and spectroscopic data presented in Section~\ref{sec:obs} using the \texttt{exoplanet} package \citep{exoplanetdanforeman, exoplanetnew}, which incorporates \texttt{starry} \citep{starry}, \texttt{celerite} \citep{celerite} and \texttt{PyMC3} \citep{pymc3} within its framework. We have selected the high-quality follow-up light curves, which includes all observations from \tess\ and \cheops\ as our space-based photometry, one observation from \ngts, nine observations from \lcogt, one observation from \astep, and one observation from \MEarth\ as our ground-based photometry sample (see Table~\ref{tab:photobs}). Our radial velocity modelling of short-term trends is comprised of data from \harps\ and \pfs.

%%%%%%%%%%%%%%%%%%%%%%%%%%%%%%%%%%%%%%%%%%%%%%%%%%%%%%%%%%%%%%%%%%%%%%%%%%%%%%%%%%%%%%%%%%%%%%%%%%%%%%%%%%%%%%%%%%

\subsubsection{Transit Timing Variations} \label{sec:ttvs1}

In order to account for perceived transit timing variations (TTVs) on \Tplanetc\ in 2020 (year 2 of observation), we introduce an offset parameter \tc. This offset parameter is calculated by fitting each detrended, normalised dataset using the \texttt{EXOFAST} modelling tool \citep{Exofast, Exofastv2}. The offset parameter represents the value of the central transit time found in \texttt{EXOFAST}, and $\delta$\tc\, is the difference from the expected transit ephemeris. The corresponding $\delta$\tc\, for each transit can be found in Table \ref{tab:offset}. We omit offset parameters for the transits of \Tplanetb\ taken by \lcogt, as these observations are not of sufficient precision to allow for suitably accurate determination of the offset parameter. We omit offset parameters for the \lcogt\ transit of \Tplanetc\ on 2020 February 29 and the \astep\ transit on 2021 April 8 for these same reasons. We also choose to omit transits of both planets in the \tess\ light-curves that occur very close to the start and end of sectors and close to the data download gap, as they are likely to be highly affected by systematics which may affect transit timings.

\begin{figure}
    \centering
    \includegraphics[width=0.48\textwidth]{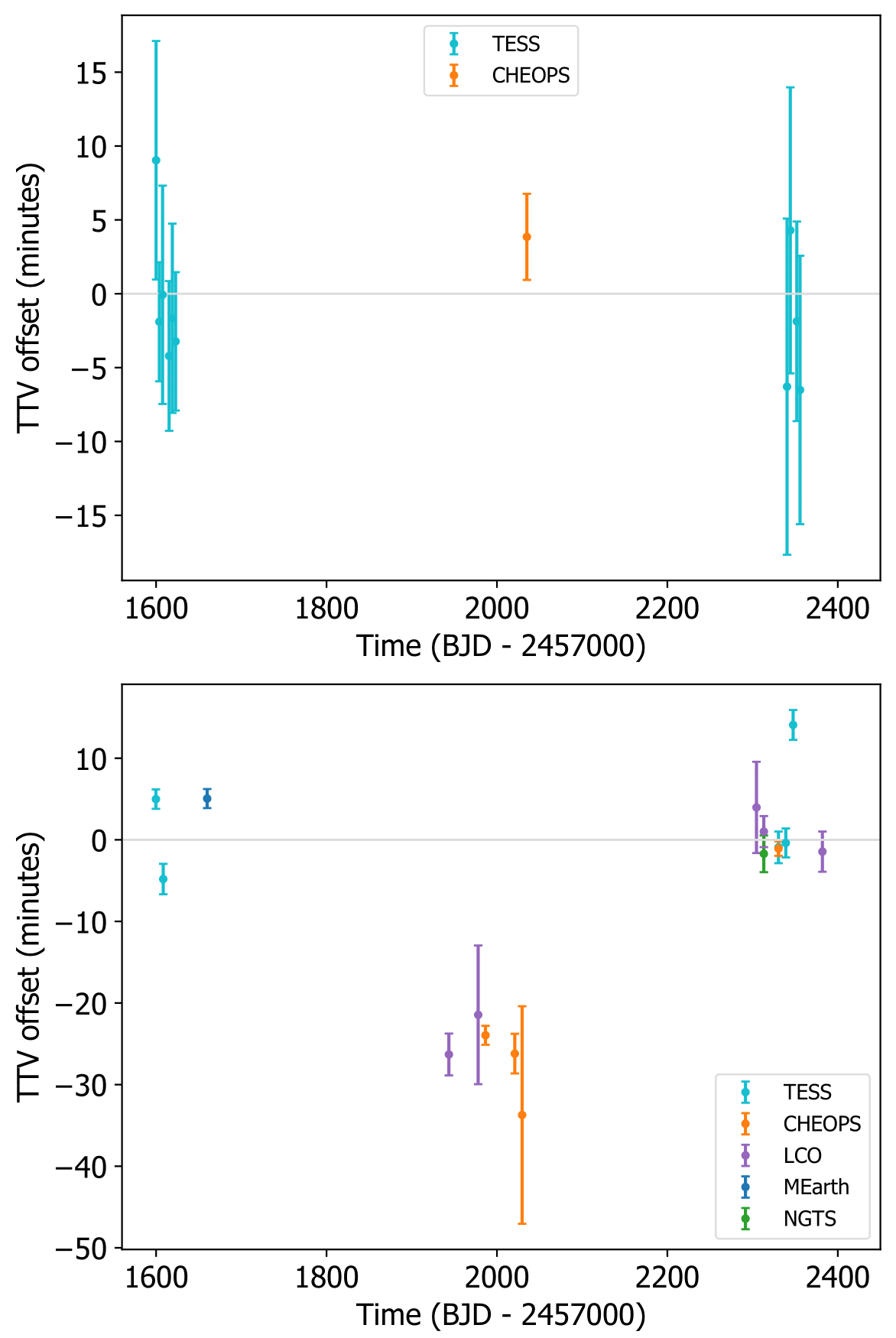}
    \caption{\textbf{Top panel:} Transit Timing Variations (TTVs) for each transit of  \Tplanetb\ from the following photometry sources: \tess\ in turquoise and \cheops\ in yellow. \textbf{Bottom panel:} Transit Timing Variations (TTVs) for each transit of  \Tplanetc\ from the following photometry sources: \tess\ in turquoise, \cheops\ in yellow, \lcogt\ in purple, \MEarthSouth\ in blue and \ngts\ in green.}
    \label{fig:ttv}
\end{figure}

We plot the resulting offset for the central transit time \tc\ for each of \Tplanetb\ and \Tplanetc\ in Figure~\ref{fig:ttv}. We note that there appear to be no  significant TTVs in the observed transits of \Tplanetb, however in \Tplanetc\ we detect an offset within the \tc\ values ranging from approximately 20 to 30 minutes. The presence of these TTVs is supported by observations from both the space-based \cheops\ satellite and multiple ground-based facilities. These TTV measurements alone are not enough to be able to put meaningful constraints on the mass of \Tplanetc, but with further TTV monitoring it may be possible.
%\textcolor{green}{Ed Bryant - any more ideas?}

%\begin{equation} \label{eq:ttv}
%\centering
%    T\textsubscript{off} = T\textsubscript{calc} - %(T\textsubscript{c} + nP)
%\end{equation}

%%%%%%%%%%%%%%%%%%%%%%%%%%%%%%%%%%%%%%%%%%%%%%%%%%%%%%%%%%%%%%%%%%%%%%%%%%%%%%%%%%%%%%%%%%%%%%%%%%%%%%%%%%%%%%%%%%

\subsubsection{Radial velocity (RV)} \label{sec:radialvelocity}

\begin{figure*}
    \centering
    \includegraphics[width=\textwidth]{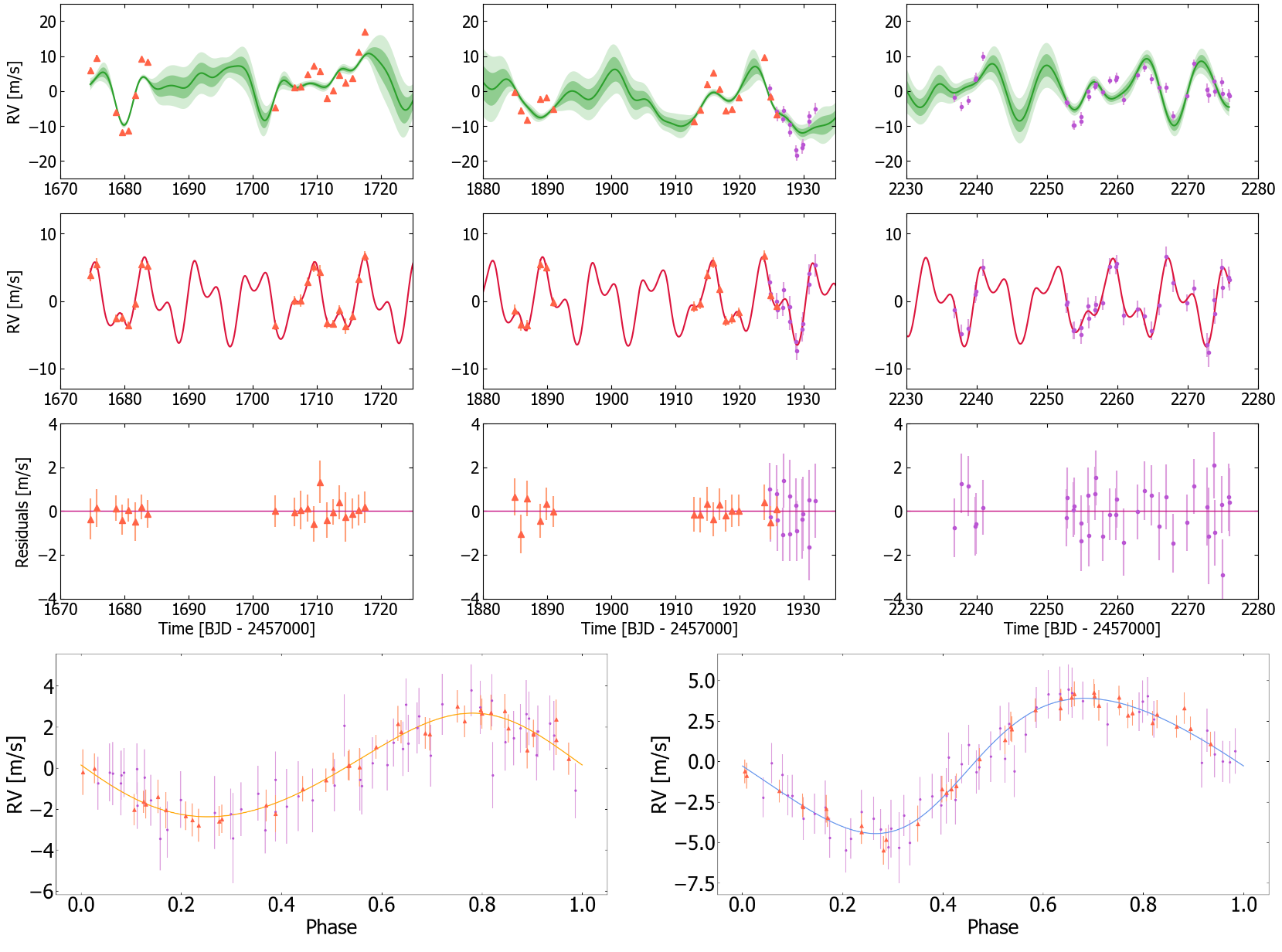}
    \caption{\textbf{Top panels:} \harps\ (purple circles) and \pfs\ (orange triangles) RV data with formal uncertainties with the GP model plotted as a solid green line, with 1 and 2 standard deviations in lighter shades. \textbf{Second panels:} Combined RV models of the two planets, with the GP subtracted, with \harps\ and \pfs\ RV datapoints. \textbf{Third panels:} Residuals for \harps\ and \pfs\ datapoints relative to a baseline RV of 0 \ms. \textbf{Fourth panels (left):} \harps\ (purple circles) and \pfs\ (orange triangles) RV data, phase-folded to a period corresponding to that of \Tplanetb\ with the RV model shown in orange.\ \textbf{Fourth panels (right):}  \harps\ and \pfs\ data, phase-folded to a period corresponding to that of \Tplanetc\ with the RV model shown in blue.}
    \label{fig:harpspfsrv}
\end{figure*}

We model the radial velocity of \Tstar\ using the \harps\ and \pfs\ data simultaneously, seen in Figure~\ref{fig:harpspfsrv}. We analysed these radial velocity data with various models, including linear and quadratic drift and a third planet. None of these were able to account for the large scatter in the radial velocity measurements, and therefore we find it necessary to apply a GP model for both of our chosen datasets in order to account for stellar variability. We apply a quasi-periodic kernel (commonly used in works with similar goals, such as \citealt{ares}), as implemented in \texttt{celerite}. We assign a prior probability distribution for the rotation period as a normal distribution centered around 22 days, with a standard deviation of 0.1 days, based on the results from the \waspsouth\ periodogram.
In completion, our kernel is a combination of two available kernels in the \texttt{PyMC3} package\footnote{\url{https://docs.pymc.io/api/gp/cov.html}} \citep{pymc3} - the \texttt{Periodic} and \texttt{ExpQuad} kernels are multiplied to create the final quasi-periodic kernel. As part of this analysis, we define a set of GP hyperparameters which are fit concurrently for both sets of radial velocity data: $\eta$ representing the GP amplitude, the stellar rotation period \textit{P}, the smoothing parameter \textit{l\textsubscript{P}} and the timescale of active region evolution \textit{l\textsubscript{E}}. This has been shown to successfully model stellar activity in eg. \citet{grunblatt2015}, \citet{santerne2018} and \citet{ares}.

% \begin{equation} \label{eq:periodic}
%     k(x,x') = \eta^2 \exp \left(-\frac{\sin^2(\pi |x-x'| \frac{1}{T})}{2l_P^2} \right),
% \end{equation}

% \begin{equation} \label{eq:expquad}
%     k(x,x') = \eta^2 \exp \left(-\frac{(x-x')^2}{2l_E^2} \right),
% \end{equation}

% \begin{equation} \label{eq:rvkernel}
%     k(x,x') = \eta^2 \exp \left(-\frac{\sin^2(\pi |x-x'| \frac{1}{T})}{2l_P^2} - \frac{(x-x')^2}{2l_E^2} \right).
% \end{equation}

When modelling the \harps\ and \pfs\ data, we utilise \texttt{exoplanet} to find values for the radial velocity semi-amplitude \textit{K} with priors from 0 to 10 \,\ms. We also fit for values for the offsets as a normal distribution centered around the mean of the radial velocity of each dataset. We also fit for jitter terms centered around the minimum radial velocity error multiplied by 2, which represent other variability not accounted for in the \harps\ and \pfs\ formal uncertainties, and the application of the GP model to the data.

% \textit{n} which follows Equation \ref{eq:rvnoise}, where \textit{j} denotes the jitter term, and \textit{V\textsubscript{e}} the radial velocity error.

% \begin{equation} \label{eq:rvnoise}
%     n = e^j + V_e^2
% \end{equation}

%These parameters are combined by adding the predicted values for radial velocity (obtained by finding RV values at each \harps\ and \pfs\ timestamp in the \texttt{exoplanet} modelling code) to the offset values and subtracting these from the actual observed values, and providing the GP with the noise term described above.
Modelled planetary reflex motions are subtracted from the radial velocities at each timestamp before being passed to the GP kernel, and we use the same time system for both the \harps\ and \pfs\ data sets (BJD - 2457000). The prior distributions for each of the parameters used in the code can be found in Appendices~\ref{tab:priorsstar}, \ref{tab:priorsb} and \ref{tab:priorsc} for the host star \Tstar, and the planets \Tplanetb\ and \Tplanetc\ respectively.

%%%%%%%%%%%%%%%%%%%%%%%%%%%%%%%%%%%%%%%%%%%%%%%%%%%%%%%%%%%%%%%%%%%%%%%%%%%%%%%%%%%%%%%%%%%%%%%%%%%%%%%%%%%%%%%%%%

\subsubsection{Joint fitting} \label{sec:jointfit}

To bring the two observational methods together, we utilise the \texttt{exoplanet} package to fit for our initial values from the maximum log probability, which are then passed into the \texttt{PyMC3} sampler as a starting point in a No U-Turn Sampler (NUTS) variant of the Hamilton Monte Carlo (HMC) algorithm \citep{mcmcnuts}. We set our run to have a burn-in of 4000 samples, 4000 steps and 10 chains, giving our modelling significant opportunity to explore the parameter spaces.

As a result of our joint fitting of transit and radial velocity data, we find that \Tplanetb\ is a super-Earth planet with a radius of \TRadiusbshort\,\rearth\ and mass of \TMassbshort\,\mearth, on a period of \Tperiodbshort\,days, and \Tplanetc\ is a sub-Neptune planet with a radius of \TRadiuscshort\,\rearth\ and mass of \TMasscshort\,\mearth\, on a period of \Tperiodcshort\,days. From this we can infer a bulk density of \Tdensitybshort\,\gccc\ for \Tplanetb, and \Tdensitycshort\,\gccc\ for \Tplanetc. A full set of parameters for \Tstar\ can be found in Table \ref{tab:star_props_results}, and parameters for each planet can be found in Table \ref{tab:planet_props}.

\begin{center}
\begin{table}
    \centering
    \caption{Stellar parameters of \TStar.}
    \label{tab:star_props_results}
    \begin{threeparttable}
    \begin{tabularx}{0.96\columnwidth}{ l  l X }
    \toprule
    \textbf{Property (unit)} & \textbf{Value} & \textbf{Source} \\
    \hline
    % \textbf{Key parameters} & & \\
    Mass (\Msun)    & \Tstarmassexo & \texttt{exoplanet} \\
    Radius (\Rsun)  & \Tstarradiusexo   & \texttt{exoplanet} \\
    Density (\gccc) & \Tstardensityexo  & \texttt{exoplanet} \\
    P\textsubscript{\textit{rot}} (days)   & \Tstarperiodexo   & \texttt{exoplanet} \\
    LD coefficient \textit{u\textsubscript{1}}  & \ldu  & \texttt{exoplanet} \\
    LD coefficient \textit{u\textsubscript{2}}  & \ldv  & \texttt{exoplanet} \\
    \logg\ &\Tloggporto  &\texttt{ARES + MOOG + \gaia} \\
    \teff\ (K)   &\Teffporto  &\texttt{ARES + MOOG} \\
    \vsini\ (\kms) &\Tstarvsiniporto &\texttt{ARES + MOOG}\\
    Age (Gyr)   &\Tstarageporto &Isochrones \\
    % V\textsubscript{\textit{tur}} (\kms)    & \Tstarvturbporto  &Porto \\
% \hline
%     \textbf{Chemical abundance} &\textbf{Value (dex)} &\textbf{Source} \\
    \hline
    \multicolumn{3}{l}{\textbf{Stellar abundances}} \\
    \feh\,(dex) &\Tstarfehporto &\texttt{ARES + MOOG} \\
    \mgh\,(dex)   &\Tstarmghporto &\texttt{ARES + MOOG} \\
    \sih\,(dex)    &\Tstarsihporto &\texttt{ARES + MOOG} \\
    \hline
    \multicolumn{3}{l}{\textbf{Galactic space velocity components}} \\
    $U$ (\kms)   & -35.6 $\pm$ 0.7 & \gaia\ EDR3  \\
    $V$ (\kms)  & -10.7 $\pm$ 0.3 & \gaia\ EDR3 \\
    $W$ (\kms)  & -3.50 $\pm$ 0.5 & \gaia\ EDR3 \\
    \bottomrule
    \end{tabularx}
    \begin{tablenotes}
    \item Sources: \texttt{exoplanet} \citep{exoplanetdanforeman, exoplanetnew}, \texttt{ARES} \citep{sousa2015}, \texttt{MOOG} \citep{Sneden:1973, Kurucz:1993}, \gaia\ \citep{GAIA_EDR3}
    \end{tablenotes}
    \end{threeparttable}
\end{table}
\end{center}

\begin{center}
\begin{table}
    \centering
    \caption{Timing offsets for observations of \Tplanetb\ and \Tplanetc.}
    \label{tab:offset}
    \begin{tabularx}{0.43\textwidth}{ p{0.21\linewidth} p{0.20\linewidth} l X }
    \toprule
    \textbf{Facility}   & \textbf{UT night} & \textbf{$\delta$T\textsubscript{c} (days)}   & \textbf{$\delta$T\textsubscript{c} error (days)} \\
    \hline
    \textbf{\Tplanetb}   &   &   & \\
    \TESS\ S11    & ---  & 0.009757  & 0.005609 \\
    \TESS\ S11    & ---  & 0.002165  & 0.002800  \\
    \TESS\ S11    & ---  & 0.003431  & 0.005132 \\
    \TESS\ S11    & ---  & 0.000558  & 0.003520 \\
    \TESS\ S11    & ---  & 0.002330  & 0.004450 \\
    \TESS\ S11    & ---  & 0.001242  & 0.003252 \\
    \lcogt-CTIO   & 2020 Mar 8   & ---    & --- \\
    \lcogt-SSO   & 2020 Mar 20   & ---    & --- \\
    \lcogt-SSO   & 2020 May 4   & ---    & --- \\
    \cheops & 2020 Jul 8  & 0.0061575  & 0.002024  \\
    \TESS\ S38  & ---  & --- & --- \\
    \TESS\ S38  & ---  & -0.000887 & 0.007903 \\
    \TESS\ S38  & ---  & 0.006464  & 0.006724 \\
    \TESS\ S38  & ---  & ---  & --- \\
    \TESS\ S38  & ---  & 0.0021850  & 0.004691 \\
    \TESS\ S38  & ---  & -0.001041 & 0.006310 \\
    \TESS\ S38  & ---  & --- & --- \\
    \hline
    \textbf{\Tplanetc}  &   &   & \\
    \TESS\ S11   & --- & 0.0034651   & 0.000826 \\
    \TESS\ S11   & ---  & 0.0033399  & 0.001295 \\
    \MEarthSouth    & 2019 Jul 4    & 0.0035104  & 0.000811 \\
    \lcogt-SSO   & 2020 Feb 29   & ---    & --- \\
    \lcogt-SSO  & 2020 Apr 12   & 0.0182677  & 0.001780 \\
    \lcogt-SAAO   & 2020 May 16   & -0.0148950  & 0.005903  \\
    \cheops & 2020 May 25    & -0.0166364   & 0.000806 \\
    \cheops & 2020 Jun 28   & -0.0181972  & 0.001690 \\
    \cheops & 2020 Jul 7    & -0.0234211   & 0.000923 \\
    \lcogt-SSO   & 2021 Apr 8    & 0.0027583   & 0.003884 \\
    \astep  & 2021 Apr 8    & ---    & --- \\
    \ngts   & 2021 Apr 16   & -0.0011893  & 0.001562 \\
    \lcogt-CTIO   & 2021 Apr 16   & 0.0007001    & 0.001320 \\
    \lcogt-CTIO   & 2021 Jun 24   & -0.0010068  & 0.001712  \\
    \cheops & 2021 May 4    & 0.0007432  & 0.000622 \\
    \TESS\ S38   & --- & -0.0006412   & 0.001347 \\
    \TESS\ S38  & ---  & -0.0002651  & 0.001231 \\
    \TESS\ S38   & ---  & 0.0097779  & 0.001272 \\
    \bottomrule
    \end{tabularx}
    \begin{tablenotes}
        \item Sources: \lcogt\ \citep{lcogt2013}, \cheops\ \citep{benz2021}, \astep\ \citep{Daban2010}, \ngts\ \citep{Wheatley2018}, \MEarthSouth\ \citep{mearth2015}, \tess\ \citep{Ricker:2015}
    \end{tablenotes}
\end{table}
\end{center}

\begin{center}
\begin{table}
    \centering
    \caption{Parameters of \Tplanetb\, and \Tplanetc.}
    \label{tab:planet_props}
    \begin{tabular}{lll}
    \toprule
    \textbf{Property} & \multicolumn{2}{c}{\textbf{Value}} \\ \hline
     & \multicolumn{1}{l}{\textbf{\Tplanetb}} & \multicolumn{1}{l}{\textbf{\Tplanetc}} \\
    Identifier & TOI-836.02 & TOI-836.01 \\
    Period (days) & \Tperiodb & \Tperiodc \\
    Mass (\mearth) & \TMassb & \TMassc \\
    Radius (\rearth) & \TRadiusb & \TRadiusc \\
    Density (gccc) & \Tdensityb & \Tdensityc \\
    R\textsubscript{\textit{p}}/R\textsubscript{*} & \Trorb & \Trorc \\
    \tc\ (TBJD) & \Tcb & \Tcc \\
    T1-T4 duration (hours) & \TTDurfullb & \TTDurfullc \\
    T2-T3 duration (hours) & \TTDurcutb & \TTDurcutc \\
    Impact parameter & \Timpactb & \Timpactc \\
    \textit{K} (\ms) & \Tkb & \Tkc \\
    Inclination ($^{\circ}$) & \Tincb & \Tincc \\
    Semi-major axis (AU) & \Taub & \Tauc \\
    Temperature T\textsubscript{\textit{eq}} (K) * & \Teqb & \Teqc \\
    Insolation flux (\fsun) & \Tfluxb & \Tfluxc \\
    Eccentricity & \Teccb & \Teccc \\
    Argument of periastron ($^{\circ}$) & \Tomegab & \Tomegac \\
    TSM & \TSMb & \TSMc \\
    \bottomrule
    \end{tabular}
\end{table}
\end{center}

% \begin{figure}
%     \centering
%     \includegraphics[width=\columnwidth]{3.Results/Figures/CHEOPS_stacked_lightcurve_TOI-836.pdf}
%     \caption{\hl{Light curves of} \Tplanetb\ and \Tplanetc\ taken by the \cheops\ satellite as detailed in Table~\ref{tab:photobs}, plotted with our best fit \texttt{exoplanet} models for \Tplanetb\ in orange and \Tplanetc\ in blue, and offset for clarity.}
%     \label{fig:cheopslcs}
% \end{figure}

% \begin{figure}
%     \centering
%     \includegraphics[width=\columnwidth]{3.Results/Figures/LCO_stacked_lightcurves_TOI-836.pdf}
%     \caption{\hl{Light curves of} \Tplanetb\ and \Tplanetc\ taken by the \lcogt\ network as detailed in Table~\ref{tab:photobs}, plotted with our best fit \texttt{exoplanet} models for \Tplanetb\ in orange and \Tplanetc\ in blue, and offset for clarity.}
%     \label{fig:lcolcs}
% \end{figure}

% \begin{figure}
%     \centering
%     \includegraphics[width=\columnwidth]{3.Results/Figures/followup_stacked_lightcurve_TOI-836.pdf}
%     \caption{Lightcurves of \Tplanetc\ taken by the \MEarthSouth, \ngts\ and \astep\ facilities as detailed in Table~\ref{tab:photobs}, plotted with our best fit \texttt{exoplanet} models and offset for clarity.}
%     \label{fig:followuplcs}
% \end{figure}

\begin{figure}
    \centering
    \includegraphics[width=\columnwidth]{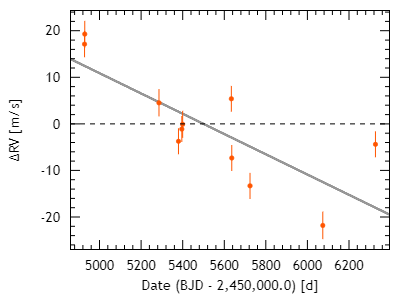}
    \caption{Radial velocity data of \Tstar\ from the \hires\ instrument on the \keck\ telescope from 2009 April 6 to 2013 February 3, and fit with a linear trend represented by the solid grey line.}
    \label{fig:hiresdace}
\end{figure}

%%%%%%%%%%%%%%%%%%%%%%%%%%%%%%%%%%%%%%%%%%%%%%%%%%%%%%%%%%%%%%%%%%%%%%%%%%%%

\subsubsection{Long-term trends} \label{sec:longterm}

In addition to our short-term radial velocity analysis with data from \harps\ and \pfs, we also make use of \hires\ data to constrain longer-term trends. We fit the data for a linear drift, and find a drift value of \Tdrift\,\msy. The fit is shown in Figure \ref{fig:hiresdace}.  The \hires\ data is sparsely sampled over a duration of approximately four years.  Therefore it is not possible to remove the stellar activity signal in the manner we did for the \harps\ and \pfs\ data, and so the marginally detected linear trend may not be real, and we do not use this trend when fitting the radial velocities in Section~\ref{sec:radialvelocity}.  However the \hires\ data is able to rule out any radial velocity drift above the level of the stellar activity signal ($\sim$\,10\,\ms) over a four year time period. 

% \begin{figure}
%     \centering
%     \includegraphics[width=\columnwidth]{3.Results/Figures/Hires_linear_trend.pdf}
%     \caption{Linear drift fit in the \hires\ data from the \texttt{DACE} platform.}
%     \label{fig:hiresdace}
% \end{figure}

% \section{Results} \label{sec:results}
% \input 3.Results/results.tex

\section{Discussion} \label{sec:disc}
\begin{figure}
    \centering
    \includegraphics[width=0.48\textwidth]{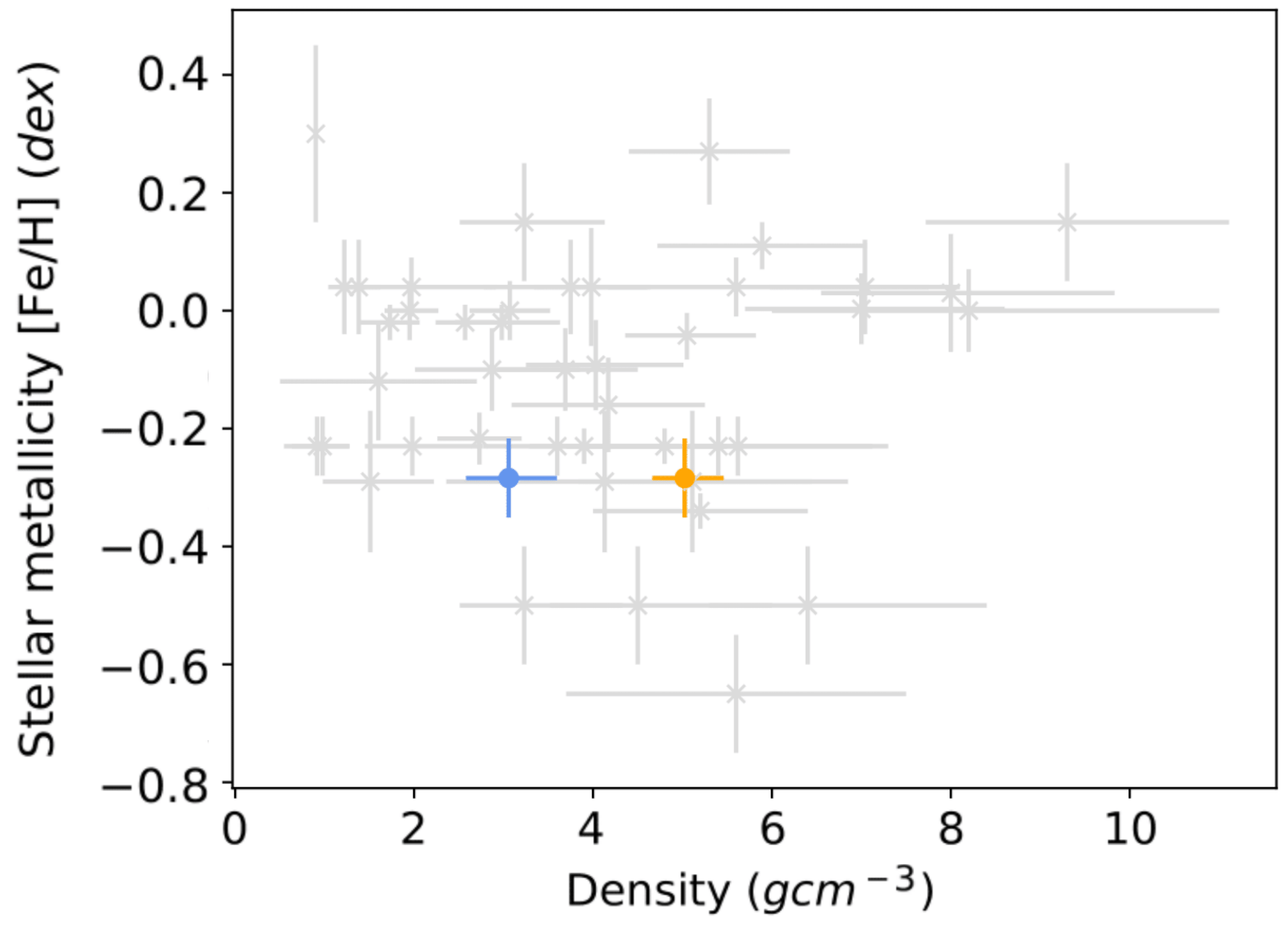}
    \caption{
    Bulk densities of \Tplanetb\ (orange) and \Tplanetc\ (blue) plotted against the stellar metallicity of \Tstar, along with a sample of planets orbiting K-dwarfs with R$<$4\,\rearth\ and $\rho$\,$<$15\,\gccc.
    }
    \label{fig:metallicity}
\end{figure}

In addition to the results from our joint modelling, we find that \Tstar\ has a relatively low metallicity of \feh\,=\,\Tstarfehporto\,dex. %\citet{buchhave2012} finds that planets with a radius of \rpl\,$<$\,4\,\rearth\ are not correlated with the metallicity of the host star, and therefore, it is not unusual that \Tplanetb\ and \Tplanetc\ orbit a metal-poor star. 
As was found in \citet{adibekyan2021}, there is a strong trend between host stellar metallicity and the iron component for low-mass exoplanets. This can be interpreted as systems that formed from metal-rich proto-stellar/planetary disks have stars with metal-rich photospheres and planets with large metallic cores. This is supported by the recent study of \citet{Wilson2022} that found a correlation between sub-Neptune planet densities and stellar metallicities across all stellar types that implies that sub-Neptunes around metal-rich stars have larger metallic cores that can retain a larger atmosphere and hence appear less dense. This effect has also been observed in radius valley trends with metallicity \citep{chen2022}. As \Tstar\ has a low-metallicity we reproduce Fig.~15 of \citet{Wilson2022} and plot the bulk densities of the two planets against the stellar metallicity in Figure~\ref{fig:metallicity}, alongside a sample of planets orbiting K-dwarfs with a radius of $<$4\,\rearth\ and a density of $<$15\,\gccc\ from the NASA Exoplanet Archive. This sample of all well-characterised super-Earths and sub-Neptunes around K-dwarfs supports previous findings and strengthens the evidence that stellar composition affects planetary internal structure.

%%%%%%%%%%%%%%%%%%%%%%%%%%%%%%%%%%%%%%%%%%%%%%%%%%%%%%%%%%%%%%%%%%%%%%%%%%%%%%%%%%%%%%%%%%%%%%%%%%%%%%%%

\subsection{Positions of the planets on the mass-radius (M-R) diagram} \label{results:MR}

\begin{figure}
    \centering
    \includegraphics[width=0.48\textwidth]{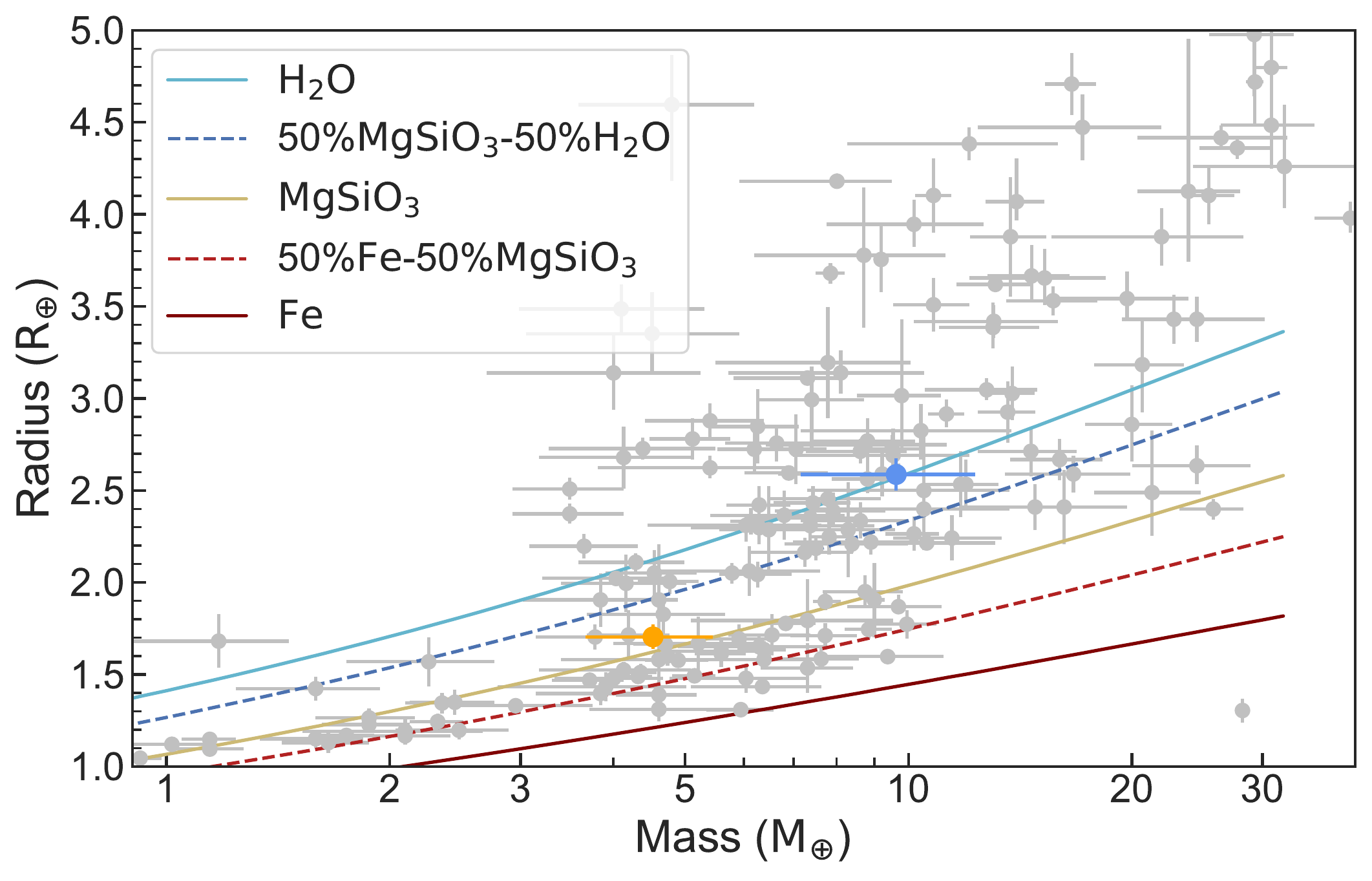}
    \caption{
    Mass-radius diagram plotted for \Tplanetb\ in orange and \Tplanetc\, with exoplanets from the \textit{TEPCAT} catalog \citep{tepcat} in grey and composition models from \citet{zeng}.
    }
    \label{fig:massradius}
\end{figure}

We plot \Tplanetb\ and \Tplanetc\ on the mass-radius (M-R) diagram in Figure~\ref{fig:massradius}, using \texttt{fancy-massradius-plot}\footnote{\url{https://github.com/oscaribv/fancy-massradius-plot}}, alongside a sample of exoplanets from the \textit{TEPCAT} catalog \citep{tepcat}. It can be seen that \Tplanetb\ sits directly between the MgSiO\textsubscript{3} and 50\%\,Fe--50\%\,MgSiO\textsubscript{3} planetary composition models from \citet{zeng}, and \Tplanetc\ sits on the H\textsubscript{2}O track. The masses and radii of \Tplanetb\ and \Tplanetc, along with their bulk densities, are consistent with the previously-determined populations of super-Earths and mini-Neptunes.

%%%%%%%%%%%%%%%%%%%%%%%%%%%%%%%%%%%%%%%%%%%%%%%%%%%%%%%%%%%%%%%%%%%%%%%%%%%%%%%%%%%%%%%%%%%%%

\subsection{Internal structure modelling} \label{sec:intstructure}

%\textcolor{green}{To discuss and rewrite with Yann Alibert}

Using the planetary and stellar parameters derived above, we used a Bayesian analysis to infer the internal structure of both planets. 
The method we use is presented in detail in \citet{leleu2021}; we just recall here the main elements. 
The Bayesian analysis relies on two parts. The first one is the forward models which allows computing the planetary radius as a function of internal structure parameters, here the mass of the solid Fe/Si core, the fraction of Fe in the core, the mass of the silicate mantle and its composition (Si, Mg and Fe molar ratios), the mass of the water layer, the mass of the gas envelope (composed in this model of pure H/He), the equilibrium temperature of the planet, and its age. The second part is the Bayesian inference itself.

The details of the forward model are given in \citet{leleu2021}, we just emphasize the fact that the gaseous (H/He) part of the planet does not influence, in our model, the `non-gas' part of the planet (core, mantle and water layer). The radius of the non-gas part is not influenced by the potential compression and thermal isolation effect from the gas envelope. The molar ratio of Fe, Si and Mg in the refractory parts of the two planets (core and mantle) are assumed to be identical and similar to the one of the star. Note, however, that \citet{adibekyan2021} recently showed that the stellar and planetary abundances may not be always correlated in a one-to-one relation. The water and gas mass ratio, on the other hand, are not required to be similar between the two planets. In terms of priors, we assume that the core, mantle and water mass fraction (relative to the non-gas part) are uniform (subject to the constraint that they add up to one), whereas the mass fraction of the H/He layer is assumed to be uniform in log. We point out the fact that considering, instead, a uniform prior for the H/He gas layer would translate to more gas-rich planets, and consequently less water-rich planets. 

The resulting internal structure of both planets presented are summarized in Table~\ref{tab:intstructure}. \Tplanetb\, is likely to contain a very small fraction of gas, and could have a non-negligible mass of water (although the solution with no water is also compatible with the data). \Tplanetc, on the other hand, has a much smaller density and is likely to contain more gas and/or water. We finally recall that the derived internal structure results from a Bayesian analysis, and that the distributions are of statistical nature and depend somewhat on the assumed priors. % removed references to corner plots

The structure of \Tplanetb\ is somewhat analogous to that of TOI-1235\,b \citep{cloutier2020}, despite the difference in the host star's spectral type, and the rocky composition of the planet may support a thermally-driven or core-powered mass loss scenario rather than a gas-poor formation scenario. \Tplanetc\ on the other hand is a little more ambiguous, but given its insolation flux of \Tfluxc\,\fsun\ and radius of \TRadiuscshort\,\rearth, we expect a non-negligible fraction of its mass to be in gaseous form.

These two planets may also support the concept of intra-system uniformity reported by \citet{milholland2017} and \citet{milholland2021}, as the two planets lie close together within the mass-radius space than if two planets were to be drawn at random from the entire distribution of exoplanets according to their radii.

% \begin{figure}
%      \centering
%      \begin{subfigure}[b]{\columnwidth}
%          \centering
%          \includegraphics[width=\columnwidth]{4.Discussion/corner_small_b.png}
%      \end{subfigure}
%      \hfill
%      \begin{subfigure}[b]{\columnwidth}
%          \centering
%          \includegraphics[width=\columnwidth]{4.Discussion/corner_small_c.png}
%      \end{subfigure}
%         \caption{Corner plot showing the results on the interior composition models of \Tplanetb\, (top) and \Tplanetc\, (bottom). The vertical dashed lines and the `error bars' given at the top of each columns represent the 5~\% and 95~\% percentiles.}
%         \label{fig:corner_IS}
% \end{figure}

\begin{center}
\begin{table}
    \centering
    \caption{Interior structure properties of \Tplanetb\ and \Tplanetc.}
    \label{tab:intstructure}
    \begin{tabularx}{0.67\columnwidth}{ l  l  X }
    \toprule
    \textbf{Property (unit)} & \multicolumn{2}{c}{\textbf{Values}} \\
    \hline
       & \textbf{\Tplanetb}  &\textbf{\Tplanetc}   \\
    M\textsubscript{\textit{core}}/M\textsubscript{\textit{total}}    & \Mcoreb  & \Mcorec \\
    M\textsubscript{\textit{water}}/M\textsubscript{\textit{total}}  & \Mwaterb & \Mwaterc   \\
    log(M\textsubscript{\textit{gas}})   & \Mgasb   & \Mgasc   \\
    Fe\textsubscript{\textit{core}} & \Fecoreb  & \Fecorec   \\
    Si\textsubscript{\textit{mantle}} & \Simantleb  & \Simantlec   \\
    Mg\textsubscript{\textit{mantle}} & \Mgmantleb  & \Mgmantlec   \\
    \bottomrule
    \end{tabularx}
\end{table}
\end{center}

%%%%%%%%%%%%%%%%%%%%%%%%%%%%%%%%%%%%%%%%%%%%%%%%%%%%%%%%%%%%%%%%%%%%%%%%%%%%%%%%%%%%%%%%%%%%

\subsection{Positions of the planets compared to the radius valley} \label{results:radiusvalley}

The radius valley is a bimodal distribution of planetary radii that separates super-Earths and sub-Neptunes either side of R\textsubscript{\textit{p}}\,$\approx$\,2\,\rearth\ \citep{vaneylen2017, Fulton:2017}, from $\approx$\,1.3\,\rearth\ and $\approx$\,2.6\,\rearth, respectively. The radius valley is important to examine on the basis of its implications for the formation and evolution of terrestrial planets \citep{Giacalone2022}. Some commonalities can be found within the population of super-Earths on the left side of the valley, consisting of atmosphere-stripped rocky cores, and the population of mini-Neptunes on the right hand side, consisting of rocky cores that have retained their atmospheres \citep{VanEylen2021}. Many possibilities for the origin of the radius valley have been speculated, including the theory that terrestrial planets lose their atmospheres through photoevaporation \citep{Owen2013, Jin2018, VanEylen2021}, mass loss due to core temperatures \citep{Ginzburg2016}, and the impacts of planetesimals \citep{Schlichting2015}.

\begin{figure}
    \centering
    \includegraphics[width=0.48\textwidth]{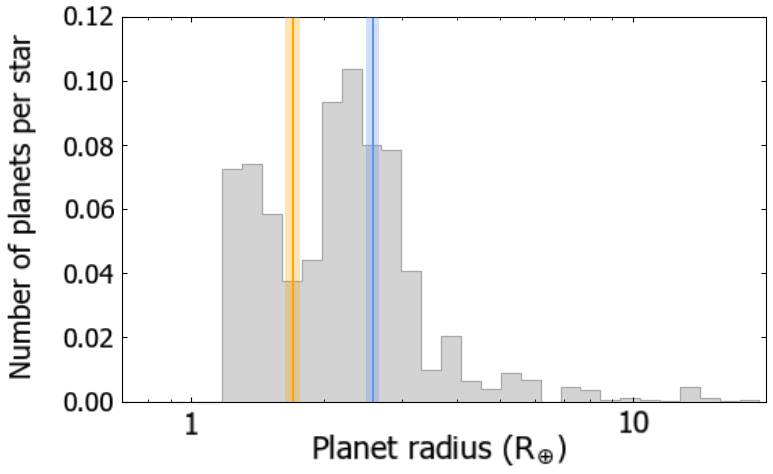}
    \caption{Histogram of confirmed planets with periods less than 100 days, using data from \citet{Fulton:2018} represented in grey, overplotted with the radii of \Tplanetb\ in orange and \Tplanetc\ in blue, including 1$\sigma$ standard deviations according to Table~\ref{tab:planet_props}.}
    \label{fig:radiusvalley}
\end{figure}

\begin{figure}
    \centering
    \includegraphics[width=0.4\textwidth]{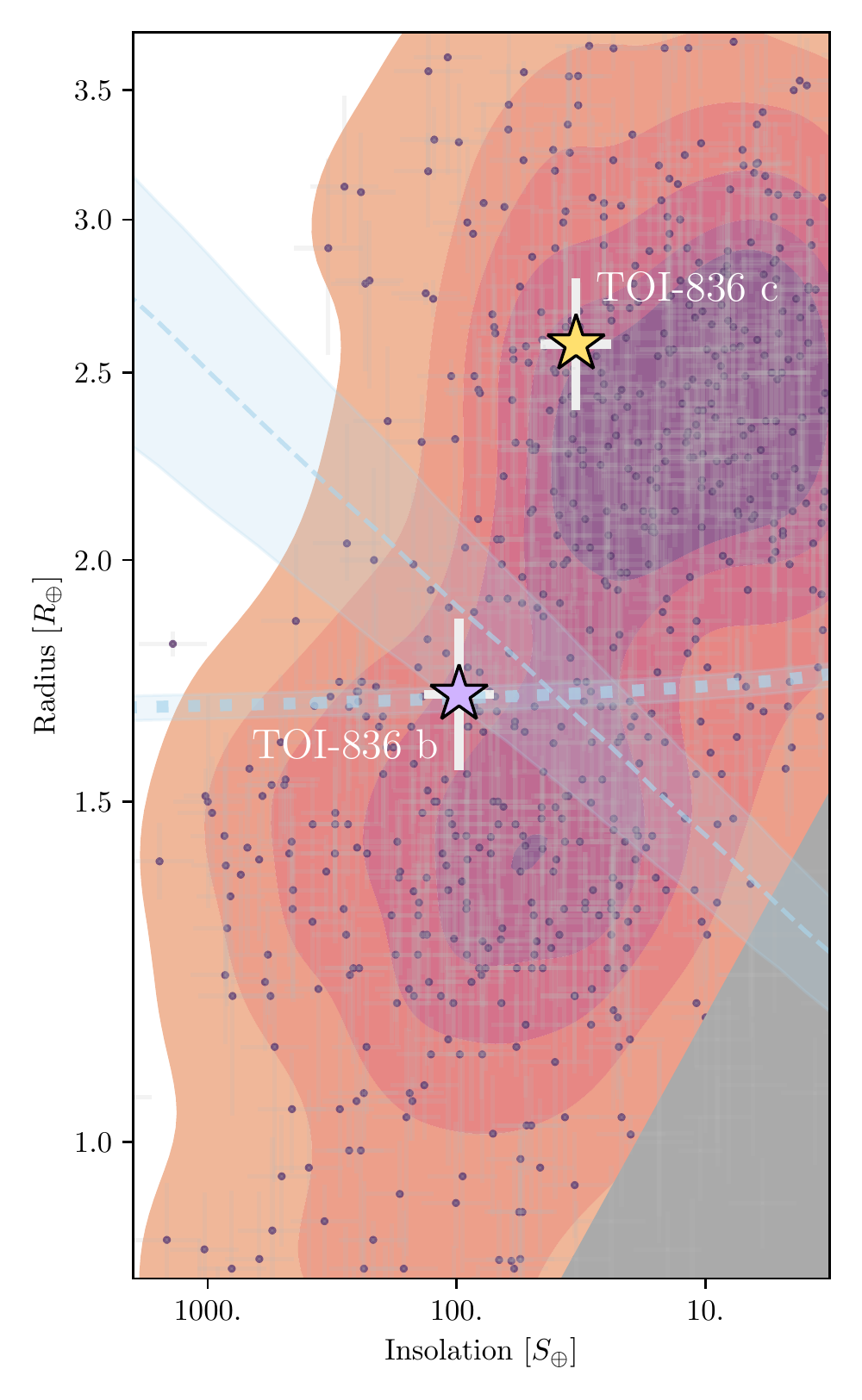}
    \caption{\Tplanetb\ and \Tplanetc\ (filled stars) as a function of planetary radius and insolation, compared with the population of exoplanets. Colours represent a kernel density estimation (KDE) applied to small (\rpl\,$<$\,4\,\rearth), transiting planets retrieved from the NASA Exoplanet Archive \citep{akeson2013nasa}. The dashed line (and associated 1-$\sigma$ error band) shows the estimate for the position of the evaporation valley from \citet{martinez2019spectroscopic}, while the dotted line shows a boundary due to gas-depleted formation derived from cool stars in \kepler\ and K2, converted to insolation using stellar parameters for TOI-836 \citep{cloutiernew}.}
    \label{fig:insolation}
\end{figure}

In Figure~\ref{fig:radiusvalley} we plot a histogram of planets with orbital periods less than 100 days based on data from \citet{Fulton:2018}, along with the positions of \Tplanetb\ and \Tplanetc\ using the modelled values from \texttt{exoplanet} in Table \ref{tab:planet_props}. We also plot a diagram of planetary radius against the insolation fluxes in Figure~\ref{fig:insolation}, alongside a sample of the exoplanet population and the position of the radius valley as estimated by \citet{martinez2019spectroscopic}. \Tplanetb\ can be seen to sit directly within this valley, and \Tplanetc\ can be seen close to the peak on the higher radius side of the valley. \Tplanetb\ is set at a particularly interesting location, and there may be scope for further investigation of the extent and composition of its atmosphere, especially as the host star is suspected to not be young in age (see Section~\ref{sec:stellaranalysis}).

%%%%%%%%%%%%%%%%%%%%%%%%%%%%%%%%%%%%%%%%%%%%%%%%%%%%%%%%%%%%%%%%%%%%%%%%%%%%%%%%%%%%%%%%%%%%%%%%%%%%%%

%\subsection{Transit timing variations} \label{sec:ttvdiscussion}

%%%%%%%%%%%%%%%%%%%%%%%%%%%%%%%%%%%%%%%%%%%%%%%%%%%%%%%%%%%%%%%%%%%%%%%%%%%%%%%%%%%%%%%%%%%%%%%%%%%%%%

In order to evaluate \Tstar\ as a potential target for transmission spectroscopy follow-up in the era of \jwst\ \citep[James Webb Space Telescope;][]{jwst}, we calculate a Transmission Spectroscopy Metric (TSM) for each of the planets based upon equation 1 in \citet{kempton2018}. This value is an estimate of the observed SNR of each planet as would be achieved by the \textit{NIRSPEC} instrument on \jwst. We find a TSM for \Tplanetb\ of \TSMbshort, and a TSM for \Tplanetc\ of \TSMcshort\ (see Table~\ref{tab:planet_props}). We also note that the system has been allocated time on \jwst\ as can be seen in \citet{jwstproposal}, with the intention of further examining the atmospheric characteristics of \Tplanetb\ and \Tplanetc\ through molecular abundances. The precise masses provided in this paper will greatly help in the characterisation of the atmospheres of these planets.

\section{Conclusion} \label{sec:conc}
In this paper, we have presented the \TStar\ system and the discovery of its two planets, \Tplanetb\ and \Tplanetc. We base our discovery upon data from two sectors of \tess\ data (11 and 38 from year 1 and year 3 respectively) at 2-minute cadence, and a further five space-based observations ranging from 2020 to 2021 from \cheops, which are complemented by ground-based photometry from the \ngts, \MEarth, \lco\ and \astep\ facilities, with supporting evidence for a stellar rotation period of \Tstarperiodexoshort\ days supported by data from \waspsouth. We model this photometry data jointly with radial velocity data from \harps\ and \pfs\ using the \texttt{exoplanet} package to constrain short-term trends, and \hires\ data for long-term trends. We are also able to rule out the presence of blended stellar companions that may affect our photometry from an examination of the imaging from \gemini-Zorro. The planets orbit a K-type dwarf star with a mass of \Tstarmassportoshort\ \msun\ and a radius of \Tstarradiusportoshort\ \rsun.

\Tplanetb\ is a super-Earth planet with a mass of \TMassbshort\,\mearth\ and a radius of \TRadiusbshort\,\rearth, on an orbit of \Tperiodbshort\ days. Our internal structure modelling indicates that this planet possesses a relatively small fraction of its mass in the form of gas.

\Tplanetc\ is a sub-Neptune with a mass of \TMasscshort\,\mearth and a radius of \TRadiuscshort\,\rearth, on an orbit of \Tperiodcshort\ days. Our structure modelling indicates that it contains a higher proportion of gas and/or water than \Tplanetb. We also find significant Transit Timing Variations within our observations of this planet, which may indicate the presence of a third non-transiting planet in the system - however we find no transits of a third planet within our current set of photometry data, or any indication of an additional periodic signal in our current radial velocity data.

\Tplanetb\ appears in the centre of the radius valley, and \Tplanetc\ appears to sit close to the peak on the right hand side of the valley, which is an area of interest in terms of the formation and structure of terrestrial planets and the dynamics of atmospheric loss and retention. The planets also contribute to the \tess\ Level 1 Mission requirement, and are particularly amenable to follow-up observations in the era of \jwst.

%%%%%%%%%%%%%%%%%%%%%%%%%%%%%%%%%%%%%%%%%%%%%%%%%%%
%%%%%%%%%%%%%%%% ACKNOWLEDGEMENTS %%%%%%%%%%%%%%%%%
%%%%%%%%%%%%%%%%%%%%%%%%%%%%%%%%%%%%%%%%%%%%%%%%%%%

\section*{Acknowledgements}

%TPFPLOTTER
This work makes use of \texttt{tpfplotter} by J. Lillo-Box (publicly available in www.github.com/jlillo/tpfplotter), which also made use of the python packages \texttt{astropy}, \texttt{lightkurve}, \texttt{matplotlib} and \texttt{numpy}.

%EXOPLANET
This research makes use of \texttt{exoplanet} \citep{exoplanet:exoplanet} and its dependencies \citep{exoplanet:agol20, exoplanet:arviz, exoplanet:astropy13,
exoplanet:astropy18, exoplanet:exoplanet, exoplanet:kipping13,
exoplanet:luger18, exoplanet:pymc3, exoplanet:theano}.

%EXOFAST
This paper makes use of \texttt{EXOFAST} \citep{Exofast, Exofastv2} as provided by the NASA Exoplanet Archive, which is operated by the California Institute of Technology, under contract with the National Aeronautics and Space Administration under the Exoplanet Exploration Program.

%DACE
This publication makes use of The Data \& Analysis Center for Exoplanets (DACE), which is a facility based at the University of Geneva (CH) dedicated to extrasolar planets data visualisation, exchange and analysis. DACE is a platform of the Swiss National Centre of Competence in Research (NCCR) PlanetS, federating the Swiss expertise in Exoplanet research. The DACE platform is available at https://dace.unige.ch.

%GAIA
This work makes use of data from the European Space Agency (ESA) mission {\it Gaia} (\url{https://www.cosmos.esa.int/gaia}), processed by the {\it Gaia} Data Processing and Analysis Consortium (DPAC, \url{https://www.cosmos.esa.int/web/gaia/dpac/consortium}). Funding for the DPAC has been provided by national institutions, in particular the institutions participating in the {\it Gaia} Multilateral Agreement.

%TESS
This paper includes data collected by the \TESS\ mission. Funding for the \TESS\ mission is provided by the NASA Explorer Program. Resources supporting this work were provided by the NASA High-End Computing (HEC) Program through the NASA Advanced Supercomputing (NAS) Division at Ames Research Center for the production of the SPOC data products. The TESS team shall assure that the masses of fifty (50) planets with radii less than 4 REarth are determined.

%TESS SPOC
We acknowledge the use of public \TESS\ Alert data from pipelines at the \TESS\ Science Office and at the \TESS\ Science Processing Operations Center.

%TESS QLP
%We acknowledge the use of TESS High Level Science Products (HLSP) produced by the Quick-Look Pipeline (QLP) at the TESS Science Office at MIT, which are publicly available from the Mikulski Archive for Space Telescopes (MAST). Funding for the TESS mission is provided by NASA's Science Mission directorate.

%EXOFOP
This research makes use of the Exoplanet Follow-up Observation Program website, which is operated by the California Institute of Technology, under contract with the National Aeronautics and Space Administration under the Exoplanet Exploration Program.

%MAST
This paper includes data collected by the TESS mission that are publicly available from the Mikulski Archive for Space Telescopes (MAST).

%CHEOPS
CHEOPS is an ESA mission in partnership with Switzerland with important contributions to the payload and the ground segment from Austria, Belgium, France, Germany, Hungary, Italy, Portugal, Spain, Sweden, and the United Kingdom. The CHEOPS Consortium would like to gratefully acknowledge the support received by all the agencies, offices, universities, and industries involved. Their flexibility and willingness to explore new approaches were essential to the success of this mission.

%NGTS
This paper is in part based on data collected under the NGTS project at the ESO La Silla Paranal Observatory.  The NGTS facility is operated by the consortium institutes with support from the UK Science and Technology Facilities Council (STFC)  projects ST/M001962/1 and  ST/S002642/1.

%MEARTH
The MEarth Team gratefully acknowledges funding from the David and Lucile Packard Fellowship for Science and Engineering (awarded to D.C.). This material is based upon work supported by the National Science Foundation under grants AST-0807690, AST-1109468, AST-1004488 (Alan T. Waterman Award), and AST-1616624, and upon work supported by the National Aeronautics and Space Administration under Grant No. 80NSSC18K0476 issued through the XRP Program. This work is made possible by a grant from the John Templeton Foundation. The opinions expressed in this publication are those of the authors and do not necessarily reflect the views of the John Templeton Foundation.

%LCOGT
This work makes use of observations from the LCOGT network. Part of the LCOGT telescope time was granted by NOIRLab through the Mid-Scale Innovations Program (MSIP). MSIP is funded by NSF.

%ASTEP
The \ASTEP\ project was funded by the Agence Nationale de la Recherche (ANR), the Institut National des Sciences de l'Univers (INSU), the Programme National de Plan\'etologie (PNP), and the Idex UCAJEDI (ANR-15-IDEX-01). The logistics at Concordia is handled by the French Institut Paul-Emile Victor (IPEV) and the Italian Programma Nazionale di Ricerche in Antartide (PNRA). We acknowledge support from the European Space Agency SCI-S Faculty Research Project Programme. This research is supported by the European Research Council (ERC) under the European Union's Horizon 2020 research and innovation programme (grant agreement n$^\circ$ 803193/BEBOP), and from the Science and Technology Facilities Council (STFC; grant n$^\circ$ ST/S00193X/1). 

%WASP
WASP-South is hosted by the South African Astronomical Observatory and we are grateful for their ongoing support and assistance. Funding for WASP comes from consortium universities and from the UK's Science and Technology Facilities Council.

%HARPS
This study is based on observations collected at the European Southern Observatory under ESO programme 1102.C-0249 (PI: Armstrong).

%PFS
This paper includes data gathered with the 6.5-m Magellan Telescopes located at Las Campanas Observatory, Chile.

%MINERVA
\minerva\ is supported by Australian Research Council LIEF Grant LE160100001, Discovery Grants DP180100972 and DP220100365, Mount Cuba Astronomical Foundation, and institutional partners University of Southern Queensland, UNSW Sydney, MIT, Nanjing University, George Mason University, University of Louisville, University of California Riverside, University of Florida, and The University of Texas at Austin.

We respectfully acknowledge the traditional custodians of all lands throughout Australia, and recognise their continued cultural and spiritual connection to the land, waterways, cosmos, and community. We pay our deepest respects to all Elders, ancestors and descendants of the Giabal, Jarowair, and Kambuwal nations, upon whose lands the \minerva\ facility at Mt Kent is situated.

%GEMINI
Supported by the international Gemini Observatory, a program of NSF’s NOIRLab, which is managed by the Association of Universities for Research in Astronomy (AURA) under a cooperative agreement with the National Science Foundation, on behalf of the Gemini partnership of Argentina, Brazil, Canada, Chile, the Republic of Korea, and the United States of America.

%GEMINI
Some of the observations in the paper make use of the High-Resolution Imaging instrument(s) ‘Alopeke and Zorro. ‘Alopeke and Zorro were funded by the NASA Exoplanet Exploration Program and built at the NASA Ames Research Center by Steve B. Howell, Nic Scott, Elliott P. Horch, and Emmett Quigley. ‘Alopeke and Zorro were mounted on the Gemini North and South telescopes of the international Gemini Observatory, a program of NSF’s NOIRLab, which is managed by the Association of Universities for Research in Astronomy (AURA) under a cooperative agreement with the National Science Foundation. on behalf of the Gemini partnership: the National Science Foundation (United States), National Research Council (Canada), Agencia Nacional de Investigación y Desarrollo (Chile), Ministerio de Ciencia, Tecnología e Innovación (Argentina), Ministério da Ciência, Tecnologia, Inovações e Comunicações (Brazil), and Korea Astronomy and Space Science Institute (Republic of Korea).

This work has been carried out within the framework of the NCCR PlanetS supported by the Swiss National Science Foundation.

%PEOPLE
FH is supported by an STFC studentship. The French group acknowledges financial support from the French Programme National de Plan\'etologie (PNP, INSU). AO is supported by an STFC studentship. This work has been carried out within the framework of the NCCR PlanetS supported by the Swiss National Science Foundation. MNG acknowledges support from the European Space Agency (ESA) as an ESA Research Fellow. DJA acknowledges support from the STFC via an Ernest Rutherford Fellowship (ST/R00384X/1). PJW acknowledges support from STFC consolidated grant ST/T000406/1. JSJ greatfully acknowledges support by FONDECYT grant 1201371 and from the ANID BASAL projects ACE210002 and FB210003. JL-B acknowledges financial support received from "la Caixa" Foundation (ID 100010434) with fellowship code LCF/BQ/PI20/11760023, and the Projects No. PID2019-107061GB-C61 and No. MDM-2017-0737. EDM acknowledges the support from Funda\c{c}\~ao para a Ci\^encia e a Tecnologia (FCT) by the Investigador FCT contract IF/00849/2015/CP1273/CT0003. SH acknowledges CNES funding through the grant 837319. We acknowledge the support by FCT – Funda\c{c}\~ao para a Ci\^encia e a Tecnologia through national funds and by FEDER through COMPETE2020 – Programa Operacional Competitividade e Internacionaliza\c{c}\~ao by these grants: UID/FIS/04434/2019; UIDB/04434/2020; UIDP/04434/2020; PTDC/FIS-AST/32113/2017 \& POCI-01-0145-FEDER-032113; PTDC/FISAST/28953/2017 \& POCI-01-0145-FEDER-028953. S.G.S acknowledges the support from FCT through Investigador FCT contract nr. CEECIND/00826/2018 and POPH/FSE (EC). SMO is supported by an STFC studentship. VA acknowledges the support from FCT by the Investigador FCT contract IF/00650/2015/CP1273/CT0001. TGW, ACC, and KH acknowledge support from STFC consolidated grant numbers ST/R000824/1 and ST/V000861/1, and UKSA grant ST/R003203/1. YA and MJH acknowledge the support of the Swiss National Fund under grant 200020\_172746. SCCB acknowledges support from FCT through FCT contracts nr. IF/01312/2014/CP1215/CT0004. XB and SC acknowledge their role as ESA-appointed CHEOPS science team members. ABr was supported by the SNSA. This project was supported by the CNES. The Belgian participation to CHEOPS has been supported by the Belgian Federal Science Policy Office (BELSPO) in the framework of the PRODEX Program, and by the University of Liège through an ARC grant for Concerted Research Actions financed by the Wallonia-Brussels Federation; LD is an F.R.S.-FNRS Postdoctoral Researcher. ODSD is supported in the form of work contract (DL 57/2016/CP1364/CT0004) funded by national funds through FCT. B-OD acknowledges support from the Swiss National Science Foundation (PP00P2-190080). This project has received funding from the European Research Council (ERC) under the European Union’s Horizon 2020 research and innovation programme (project {\sc Four Aces}; grant agreement No 724427). It has also been carried out in the frame of the National Centre for Competence in Research PlanetS supported by the Swiss National Science Foundation (SNSF). DE acknowledges financial support from the Swiss National Science Foundation for project 200021\_200726. MF and CMP gratefully acknowledge the support of the Swedish National Space Agency (DNR 65/19, 174/18). MF acknowledges their role as ESA-appointed CHEOPS science team members. DG gratefully acknowledges financial support from the CRT foundation under Grant No. 2018.2323 ``Gaseousor rocky? Unveiling the nature of small worlds''. DG acknowledges their role as ESA-appointed CHEOPS science team members. MG is an F.R.S.-FNRS Senior Research Associate. This work was granted access to the HPC resources of MesoPSL financed by the Region Ile de France and the project Equip@Meso (reference ANR-10-EQPX-29-01) of the programme Investissements d'Avenir supervised by the Agence Nationale pour la Recherche. JL acknowledges their role as ESA-appointed CHEOPS science team members. ML acknowledges support of the Swiss National Science Foundation under grant number PCEFP2\_194576. PM acknowledges support from STFC research grant number ST/M001040/1. VNa, Ipa, GPi, RRa, and GSc, acknowledge the funding support from Italian Space Agency (ASI) regulated by “Accordo ASI-INAF n. 2013-016-R.0 del 9 luglio 2013 e integrazione del 9 luglio 2015 CHEOPS Fasi A/B/C”. This work was also partially supported by a grant from the Simons Foundation (PI Queloz, grant number 327127). IRI acknowledges support from the Spanish Ministry of Science and Innovation and the European Regional Development Fund through grant PGC2018-098153-B- C33, as well as the support of the Generalitat de Catalunya/CERCA programme. S.S. has received funding from the European
Research Council (ERC) under the European Union’s Horizon 2020 research and innovation programme (grant agreement No 833925, project STAREX). GyMSz acknowledges the support of the Hungarian National Research, Development and Innovation Office (NKFIH) grant K-125015, a PRODEX Institute Agreement between the ELTE E\"otv\"os Lor\'and University and the European Space Agency (ESA-D/SCI-LE-2021-0025), the Lend\"ulet LP2018-7/2021 grant of the Hungarian Academy of Science and the support of the city of Szombathely. VVG is an F.R.S-FNRS Research Associate. DB has been funded by the Spanish State Research Agency (AEI) Projects No. PID2019-107061GB-C61 and No. MDM-2017-0737 Unidad de Excelencia “María de Maeztu”- Centro de Astrobiología (CSIC/INTA).

%%%%%%%%%%%%%%%%%%%%%%%%%%%%%%%%%%%%%%%%%%%%%%%%%%
%%%%%%%%%%%%%% DATA AVAILABILITY %%%%%%%%%%%%%%%%%
%%%%%%%%%%%%%%%%%%%%%%%%%%%%%%%%%%%%%%%%%%%%%%%%%%

\section*{Data Availability}
The \tess\ data is accessible via the MAST (Mikulski Archive for Space Telescopes) portal at \url{https://mast.stsci.edu/portal/Mashup/Clients/Mast/Portal.html}. Photometry and imaging data from \ngts, \MEarth, \lco, \astep\ and \gemini\ are accessible via the ExoFOP-\tess\ archive at \url{https://exofop.ipac.caltech.edu/tess/target.php?id=440887364}. The \texttt{exoplanet} modelling code and associated \texttt{python} scripts for parameter analysis and plotting are available upon reasonable request to the author.
The posterior plots are available online as supplementary material to this publication.

%%%%%%%%%%%%%%%%%%%%%%%%%%%%%%%%%%%%%%%%%%%%%%%%%%
%%%%%%%%%%%%%%%%%% REFERENCES %%%%%%%%%%%%%%%%%%%%
%%%%%%%%%%%%%%%%%%%%%%%%%%%%%%%%%%%%%%%%%%%%%%%%%%

\bibliographystyle{mnras}
\bibliography{bib}

\begin{thebibliography}{}
\makeatletter
\relax
\def\mn@urlcharsother{\let\do\@makeother \do\$\do\&\do\#\do\^\do\_\do\%\do\~}
\def\mn@doi{\begingroup\mn@urlcharsother \@ifnextchar [ {\mn@doi@}
  {\mn@doi@[]}}
\def\mn@doi@[#1]#2{\def\@tempa{#1}\ifx\@tempa\@empty \href
  {http://dx.doi.org/#2} {doi:#2}\else \href {http://dx.doi.org/#2} {#1}\fi
  \endgroup}
\def\mn@eprint#1#2{\mn@eprint@#1:#2::\@nil}
\def\mn@eprint@arXiv#1{\href {http://arxiv.org/abs/#1} {{\tt arXiv:#1}}}
\def\mn@eprint@dblp#1{\href {http://dblp.uni-trier.de/rec/bibtex/#1.xml}
  {dblp:#1}}
\def\mn@eprint@#1:#2:#3:#4\@nil{\def\@tempa {#1}\def\@tempb {#2}\def\@tempc
  {#3}\ifx \@tempc \@empty \let \@tempc \@tempb \let \@tempb \@tempa \fi \ifx
  \@tempb \@empty \def\@tempb {arXiv}\fi \@ifundefined
  {mn@eprint@\@tempb}{\@tempb:\@tempc}{\expandafter \expandafter \csname
  mn@eprint@\@tempb\endcsname \expandafter{\@tempc}}}

\bibitem[\protect\citeauthoryear{{Abe} et~al.,}{{Abe} et~al.}{2013}]{Abe2013}
{Abe} L.,  et~al., 2013, \mn@doi [\aap] {10.1051/0004-6361/201220351}, \href
  {http://adsabs.harvard.edu/abs/2013A\%26A...553A..49A} {553, A49}

\bibitem[\protect\citeauthoryear{{Addison} et~al.,}{{Addison}
  et~al.}{2019}]{addison2019}
{Addison} B.,  et~al., 2019, \mn@doi [\pasp] {10.1088/1538-3873/ab03aa}, \href
  {https://ui.adsabs.harvard.edu/abs/2019PASP..131k5003A} {131, 115003}

\bibitem[\protect\citeauthoryear{{Addison} et~al.,}{{Addison}
  et~al.}{2021}]{addison2021}
{Addison} B.~C.,  et~al., 2021, \mn@doi [\mnras] {10.1093/mnras/staa3960},
  \href {https://ui.adsabs.harvard.edu/abs/2021MNRAS.502.3704A} {502, 3704}

\bibitem[\protect\citeauthoryear{{Adibekyan}, {Sousa}, {Santos}, {Delgado
  Mena}, {Gonz{\'a}lez Hern{\'a}ndez}, {Israelian}, {Mayor}  \&
  {Khachatryan}}{{Adibekyan} et~al.}{2012}]{adibekyan2012}
{Adibekyan} V.~Z.,  {Sousa} S.~G.,  {Santos} N.~C.,  {Delgado Mena} E.,
  {Gonz{\'a}lez Hern{\'a}ndez} J.~I.,  {Israelian} G.,  {Mayor} M.,
  {Khachatryan} G.,  2012, \mn@doi [\aap] {10.1051/0004-6361/201219401}, \href
  {https://ui.adsabs.harvard.edu/abs/2012A&A...545A..32A} {545, A32}

\bibitem[\protect\citeauthoryear{{Adibekyan} et~al.,}{{Adibekyan}
  et~al.}{2015}]{adibekyan2015}
{Adibekyan} V.,  et~al., 2015, \mn@doi [\aap] {10.1051/0004-6361/201527120},
  \href {https://ui.adsabs.harvard.edu/abs/2015A&A...583A..94A} {583, A94}

\bibitem[\protect\citeauthoryear{{Adibekyan} et~al.,}{{Adibekyan}
  et~al.}{2021}]{adibekyan2021}
{Adibekyan} V.,  et~al., 2021, \mn@doi [Science] {10.1126/science.abg8794},
  \href {https://ui.adsabs.harvard.edu/abs/2021Sci...374..330A} {374, 330}

\bibitem[\protect\citeauthoryear{{Agol}, {Luger}  \& {Foreman-Mackey}}{{Agol}
  et~al.}{2020}]{exoplanet:agol20}
{Agol} E.,  {Luger} R.,   {Foreman-Mackey} D.,  2020, \mn@doi [\aj]
  {10.3847/1538-3881/ab4fee}, \href
  {https://ui.adsabs.harvard.edu/abs/2020AJ....159..123A} {159, 123}

\bibitem[\protect\citeauthoryear{Akeson et~al.,}{Akeson
  et~al.}{2013b}]{akeson2013nasa}
Akeson R.,  et~al., 2013b, Publications of the Astronomical Society of the
  Pacific, 125, 989

\bibitem[\protect\citeauthoryear{{Akeson} et~al.,}{{Akeson}
  et~al.}{2013a}]{akeson2013}
{Akeson} R.~L.,  et~al., 2013a, \mn@doi [\pasp] {10.1086/672273}, \href
  {https://ui.adsabs.harvard.edu/abs/2013PASP..125..989A} {125, 989}

\bibitem[\protect\citeauthoryear{{Aller}, {Lillo-Box}, {Jones}, {Miranda}  \&
  {Barcel{\'o} Forteza}}{{Aller} et~al.}{2020}]{tpfplotter}
{Aller} A.,  {Lillo-Box} J.,  {Jones} D.,  {Miranda} L.~F.,   {Barcel{\'o}
  Forteza} S.,  2020, \mn@doi [\aap] {10.1051/0004-6361/201937118}, \href
  {https://ui.adsabs.harvard.edu/abs/2020A&A...635A.128A} {635, A128}

\bibitem[\protect\citeauthoryear{{Astropy Collaboration} et~al.,}{{Astropy
  Collaboration} et~al.}{2013}]{exoplanet:astropy13}
{Astropy Collaboration} et~al., 2013, \mn@doi [\aap]
  {10.1051/0004-6361/201322068}, \href
  {http://adsabs.harvard.edu/abs/2013A%26A...558A..33A} {558, A33}

\bibitem[\protect\citeauthoryear{{Astropy Collaboration} et~al.,}{{Astropy
  Collaboration} et~al.}{2018}]{exoplanet:astropy18}
{Astropy Collaboration} et~al., 2018, \mn@doi [\aj] {10.3847/1538-3881/aabc4f},
  \href {http://adsabs.harvard.edu/abs/2018AJ....156..123A} {156, 123}

\bibitem[\protect\citeauthoryear{{Auvergne} et~al.,}{{Auvergne}
  et~al.}{2009}]{corot2009}
{Auvergne} M.,  et~al., 2009, \mn@doi [\aap] {10.1051/0004-6361/200810860},
  \href {https://ui.adsabs.harvard.edu/abs/2009A&A...506..411A} {506, 411}

\bibitem[\protect\citeauthoryear{{Bakos}, {Noyes}, {Kov{\'a}cs}, {Stanek},
  {Sasselov}  \& {Domsa}}{{Bakos} et~al.}{2004}]{bakos2004}
{Bakos} G.,  {Noyes} R.~W.,  {Kov{\'a}cs} G.,  {Stanek} K.~Z.,  {Sasselov}
  D.~D.,   {Domsa} I.,  2004, \mn@doi [\pasp] {10.1086/382735}, \href
  {https://ui.adsabs.harvard.edu/abs/2004PASP..116..266B} {116, 266}

\bibitem[\protect\citeauthoryear{{Bakos} et~al.,}{{Bakos}
  et~al.}{2013}]{bakos2013}
{Bakos} G.~{\'A}.,  et~al., 2013, \mn@doi [\pasp] {10.1086/669529}, \href
  {https://ui.adsabs.harvard.edu/abs/2013PASP..125..154B} {125, 154}

\bibitem[\protect\citeauthoryear{{Batalha} et~al.,}{{Batalha}
  et~al.}{2021}]{jwstproposal}
{Batalha} N.,  et~al., 2021, {Seeing the Forest and the Trees: Unveiling Small
  Planet Atmospheres with a Population-Level Framework}, JWST Proposal. Cycle 1

\bibitem[\protect\citeauthoryear{{Bensby}, {Feltzing}  \&
  {Lundstr{\"o}m}}{{Bensby} et~al.}{2003}]{bensby2003}
{Bensby} T.,  {Feltzing} S.,   {Lundstr{\"o}m} I.,  2003, \mn@doi [\aap]
  {10.1051/0004-6361:20031213}, \href
  {https://ui.adsabs.harvard.edu/abs/2003A&A...410..527B} {410, 527}

\bibitem[\protect\citeauthoryear{{Bensby}, {Feltzing}  \& {Oey}}{{Bensby}
  et~al.}{2014}]{bensby2014}
{Bensby} T.,  {Feltzing} S.,   {Oey} M.~S.,  2014, \mn@doi [\aap]
  {10.1051/0004-6361/201322631}, \href
  {https://ui.adsabs.harvard.edu/abs/2014A&A...562A..71B} {562, A71}

\bibitem[\protect\citeauthoryear{{Benz} et~al.,}{{Benz}
  et~al.}{2021}]{benz2021}
{Benz} W.,  et~al., 2021, \mn@doi [Experimental Astronomy]
  {10.1007/s10686-020-09679-4}, \href
  {https://ui.adsabs.harvard.edu/abs/2021ExA....51..109B} {51, 109}

\bibitem[\protect\citeauthoryear{{Blackwell} \& {Shallis}}{{Blackwell} \&
  {Shallis}}{1977}]{Blackwell1977}
{Blackwell} D.~E.,  {Shallis} M.~J.,  1977, \mn@doi [\mnras]
  {10.1093/mnras/180.2.177}, \href
  {https://ui.adsabs.harvard.edu/abs/1977MNRAS.180..177B} {180, 177}

\bibitem[\protect\citeauthoryear{{Bonfanti}, {Ortolani}, {Piotto}  \&
  {Nascimbeni}}{{Bonfanti} et~al.}{2015}]{bonfanti15}
{Bonfanti} A.,  {Ortolani} S.,  {Piotto} G.,   {Nascimbeni} V.,  2015, \mn@doi
  [\aap] {10.1051/0004-6361/201424951}, \href
  {http://adsabs.harvard.edu/abs/2015A%26A...575A..18B} {575, A18}

\bibitem[\protect\citeauthoryear{{Bonfanti}, {Ortolani}  \&
  {Nascimbeni}}{{Bonfanti} et~al.}{2016}]{bonfanti16}
{Bonfanti} A.,  {Ortolani} S.,   {Nascimbeni} V.,  2016, \mn@doi [\aap]
  {10.1051/0004-6361/201527297}, \href
  {http://adsabs.harvard.edu/abs/2016A%26A...585A...5B} {585, A5}

\bibitem[\protect\citeauthoryear{{Bonfanti} et~al.,}{{Bonfanti}
  et~al.}{2021}]{Bonfanti2021}
{Bonfanti} A.,  et~al., 2021, \mn@doi [\aap] {10.1051/0004-6361/202039608},
  \href {https://ui.adsabs.harvard.edu/abs/2021A&A...646A.157B} {646, A157}

\bibitem[\protect\citeauthoryear{{Borucki} et~al.,}{{Borucki}
  et~al.}{2010}]{kepler2010}
{Borucki} W.~J.,  et~al., 2010, \mn@doi [Science] {10.1126/science.1185402},
  \href {https://ui.adsabs.harvard.edu/abs/2010Sci...327..977B} {327, 977}

\bibitem[\protect\citeauthoryear{{Brown} et~al.,}{{Brown}
  et~al.}{2013}]{lcogt2013}
{Brown} T.~M.,  et~al., 2013, \mn@doi [\pasp] {10.1086/673168}, \href
  {https://ui.adsabs.harvard.edu/abs/2013PASP..125.1031B} {125, 1031}

\bibitem[\protect\citeauthoryear{{Bryant} et~al.,}{{Bryant}
  et~al.}{2020}]{bryant:wasp166b:2020}
{Bryant} E.~M.,  et~al., 2020, \mn@doi [\mnras] {10.1093/mnras/staa1075}, \href
  {https://ui.adsabs.harvard.edu/abs/2020MNRAS.494.5872B} {494, 5872}

\bibitem[\protect\citeauthoryear{Buchhave et~al.,}{Buchhave
  et~al.}{2012}]{spc2012}
Buchhave L.~A.,  et~al., 2012, Nature, 486, 375

\bibitem[\protect\citeauthoryear{{Buchhave} et~al.,}{{Buchhave}
  et~al.}{2014}]{spc2014}
{Buchhave} L.~A.,  et~al., 2014, \mn@doi [\nat] {10.1038/nature13254}, \href
  {https://ui.adsabs.harvard.edu/abs/2014Natur.509..593B} {509, 593}

\bibitem[\protect\citeauthoryear{{Butler}, {Marcy}, {Williams}, {McCarthy},
  {Dosanjh}  \& {Vogt}}{{Butler} et~al.}{1996}]{butler1996}
{Butler} R.~P.,  {Marcy} G.~W.,  {Williams} E.,  {McCarthy} C.,  {Dosanjh} P.,
   {Vogt} S.~S.,  1996, \mn@doi [\pasp] {10.1086/133755}, \href
  {https://ui.adsabs.harvard.edu/abs/1996PASP..108..500B} {108, 500}

\bibitem[\protect\citeauthoryear{{Butler} et~al.,}{{Butler}
  et~al.}{2017}]{butler2017}
{Butler} R.~P.,  et~al., 2017, \mn@doi [\aj] {10.3847/1538-3881/aa66ca}, \href
  {https://ui.adsabs.harvard.edu/abs/2017AJ....153..208B} {153, 208}

\bibitem[\protect\citeauthoryear{{Cale} et~al.,}{{Cale}
  et~al.}{2021}]{cale2021}
{Cale} B.~L.,  et~al., 2021, \mn@doi [\aj] {10.3847/1538-3881/ac2c80}, \href
  {https://ui.adsabs.harvard.edu/abs/2021AJ....162..295C} {162, 295}

\bibitem[\protect\citeauthoryear{{Castelli} \& {Kurucz}}{{Castelli} \&
  {Kurucz}}{2003}]{Castelli2003}
{Castelli} F.,  {Kurucz} R.~L.,  2003, in {Piskunov} N.,  {Weiss} W.~W.,
  {Gray} D.~F.,  eds,  IAU Symposium Vol. 210, Modelling of Stellar
  Atmospheres. p.~A20 (\mn@eprint {arXiv} {astro-ph/0405087})

\bibitem[\protect\citeauthoryear{{Celisse}}{{Celisse}}{2008}]{loocv}
{Celisse} A.,  2008, arXiv e-prints, \href
  {https://ui.adsabs.harvard.edu/abs/2008arXiv0811.0802C} {p. arXiv:0811.0802}

\bibitem[\protect\citeauthoryear{{Chen} et~al.,}{{Chen}
  et~al.}{2021}]{chen2021}
{Chen} D.-C.,  et~al., 2021, \mn@doi [\apj] {10.3847/1538-4357/abd5be}, \href
  {https://ui.adsabs.harvard.edu/abs/2021ApJ...909..115C} {909, 115}

\bibitem[\protect\citeauthoryear{{Chen} et~al.,}{{Chen}
  et~al.}{2022}]{chen2022}
{Chen} D.-C.,  et~al., 2022, \mn@doi [\aj] {10.3847/1538-3881/ac641f}, \href
  {https://ui.adsabs.harvard.edu/abs/2022AJ....163..249C} {163, 249}

\bibitem[\protect\citeauthoryear{{Ciardi}, {Beichman}, {Horch}  \&
  {Howell}}{{Ciardi} et~al.}{2015}]{ciardi2015}
{Ciardi} D.~R.,  {Beichman} C.~A.,  {Horch} E.~P.,   {Howell} S.~B.,  2015,
  \mn@doi [\apj] {10.1088/0004-637X/805/1/16}, \href
  {https://ui.adsabs.harvard.edu/abs/2015ApJ...805...16C} {805, 16}

\bibitem[\protect\citeauthoryear{{Cloutier} \& {Menou}}{{Cloutier} \&
  {Menou}}{2020}]{cloutiernew}
{Cloutier} R.,  {Menou} K.,  2020, \mn@doi [\aj] {10.3847/1538-3881/ab8237},
  \href {https://ui.adsabs.harvard.edu/abs/2020AJ....159..211C} {159, 211}

\bibitem[\protect\citeauthoryear{{Cloutier} et~al.,}{{Cloutier}
  et~al.}{2020}]{cloutier2020}
{Cloutier} R.,  et~al., 2020, \mn@doi [\aj] {10.3847/1538-3881/ab9534}, \href
  {https://ui.adsabs.harvard.edu/abs/2020AJ....160...22C} {160, 22}

\bibitem[\protect\citeauthoryear{{Collins}, {Kielkopf}, {Stassun}  \&
  {Hessman}}{{Collins} et~al.}{2017}]{Collins:2017}
{Collins} K.~A.,  {Kielkopf} J.~F.,  {Stassun} K.~G.,   {Hessman} F.~V.,  2017,
  \mn@doi [\aj] {10.3847/1538-3881/153/2/77}, \href
  {http://adsabs.harvard.edu/abs/2017AJ....153...77C} {153, 77}

\bibitem[\protect\citeauthoryear{{Crane}, {Shectman}  \& {Butler}}{{Crane}
  et~al.}{2006}]{pfs2006}
{Crane} J.~D.,  {Shectman} S.~A.,   {Butler} R.~P.,  2006, in {McLean} I.~S.,
  {Iye} M.,  eds,  Society of Photo-Optical Instrumentation Engineers (SPIE)
  Conference Series Vol. 6269, Society of Photo-Optical Instrumentation
  Engineers (SPIE) Conference Series. p. 626931, \mn@doi{10.1117/12.672339}

\bibitem[\protect\citeauthoryear{{Crane}, {Shectman}, {Butler}, {Thompson}  \&
  {Burley}}{{Crane} et~al.}{2008}]{crane2008}
{Crane} J.~D.,  {Shectman} S.~A.,  {Butler} R.~P.,  {Thompson} I.~B.,
  {Burley} G.~S.,  2008, in {McLean} I.~S.,  {Casali} M.~M.,  eds,  Society of
  Photo-Optical Instrumentation Engineers (SPIE) Conference Series Vol. 7014,
  Ground-based and Airborne Instrumentation for Astronomy II. p. 701479,
  \mn@doi{10.1117/12.789637}

\bibitem[\protect\citeauthoryear{{Crane}, {Shectman}, {Butler}, {Thompson},
  {Birk}, {Jones}  \& {Burley}}{{Crane} et~al.}{2010}]{crane2010}
{Crane} J.~D.,  {Shectman} S.~A.,  {Butler} R.~P.,  {Thompson} I.~B.,  {Birk}
  C.,  {Jones} P.,   {Burley} G.~S.,  2010, in {McLean} I.~S.,  {Ramsay} S.~K.,
    {Takami} H.,  eds,  Society of Photo-Optical Instrumentation Engineers
  (SPIE) Conference Series Vol. 7735, Ground-based and Airborne Instrumentation
  for Astronomy III. p. 773553, \mn@doi{10.1117/12.857792}

\bibitem[\protect\citeauthoryear{{Crouzet} et~al.,}{{Crouzet}
  et~al.}{2010}]{Crouzet2010}
{Crouzet} N.,  et~al., 2010, \mn@doi [\aap] {10.1051/0004-6361/200913629},
  \href {http://adsabs.harvard.edu/abs/2010A\%26A...511A..36C} {511, A36}

\bibitem[\protect\citeauthoryear{{Crouzet} et~al.,}{{Crouzet}
  et~al.}{2018}]{Crouzet2018}
{Crouzet} N.,  et~al., 2018, \mn@doi [\aap] {10.1051/0004-6361/201732565},
  \href {https://ui.adsabs.harvard.edu/abs/2018A&A...619A.116C} {619, A116}

\bibitem[\protect\citeauthoryear{{Crouzet} et~al.,}{{Crouzet}
  et~al.}{2020}]{Crouzet2020}
{Crouzet} N.,  et~al., 2020, in Society of Photo-Optical Instrumentation
  Engineers (SPIE) Conference Series. p. 114470O, \mn@doi{10.1117/12.2562550}

\bibitem[\protect\citeauthoryear{{Daban} et~al.,}{{Daban}
  et~al.}{2010}]{Daban2010}
{Daban} J.-B.,  et~al., 2010, in Ground-based and Airborne Telescopes III. p.
  77334T, \mn@doi{10.1117/12.854946}

\bibitem[\protect\citeauthoryear{{Delgado Mena}, {Adibekyan}, {Santos},
  {Tsantaki}, {Gonz{\'a}lez Hern{\'a}ndez}, {Sousa}  \& {Bertr{\'a}n de
  Lis}}{{Delgado Mena} et~al.}{2021}]{delgadomena2021}
{Delgado Mena} E.,  {Adibekyan} V.,  {Santos} N.~C.,  {Tsantaki} M.,
  {Gonz{\'a}lez Hern{\'a}ndez} J.~I.,  {Sousa} S.~G.,   {Bertr{\'a}n de Lis}
  S.,  2021, arXiv e-prints, \href
  {https://ui.adsabs.harvard.edu/abs/2021arXiv210904844D} {p. arXiv:2109.04844}

\bibitem[\protect\citeauthoryear{{Delrez} et~al.,}{{Delrez}
  et~al.}{2021}]{Delrez2021}
{Delrez} L.,  et~al., 2021, \mn@doi [Nature Astronomy]
  {10.1038/s41550-021-01381-5}, \href
  {https://ui.adsabs.harvard.edu/abs/2021NatAs...5..775D} {5, 775}

\bibitem[\protect\citeauthoryear{{Eastman}, {Gaudi}  \& {Agol}}{{Eastman}
  et~al.}{2013}]{Exofast}
{Eastman} J.,  {Gaudi} B.~S.,   {Agol} E.,  2013, \mn@doi [\pasp]
  {10.1086/669497}, \href {http://adsabs.harvard.edu/abs/2013PASP..125...83E}
  {125, 83}

\bibitem[\protect\citeauthoryear{{Eastman} et~al.,}{{Eastman}
  et~al.}{2019}]{Exofastv2}
{Eastman} J.~D.,  et~al., 2019, arXiv e-prints, \href
  {https://ui.adsabs.harvard.edu/abs/2019arXiv190709480E} {p. arXiv:1907.09480}

\bibitem[\protect\citeauthoryear{F\H{u}r\'esz}{F\H{u}r\'esz}{2008}]{tres}
F\H{u}r\'esz G.,  2008, PhD thesis, University of Szeged, Hungary

\bibitem[\protect\citeauthoryear{{Fabrycky} et~al.,}{{Fabrycky}
  et~al.}{2014}]{fabrycky2014}
{Fabrycky} D.~C.,  et~al., 2014, \mn@doi [\apj] {10.1088/0004-637X/790/2/146},
  \href {https://ui.adsabs.harvard.edu/abs/2014ApJ...790..146F} {790, 146}

\bibitem[\protect\citeauthoryear{{Fang} \& {Margot}}{{Fang} \&
  {Margot}}{2012}]{fangmargot2012}
{Fang} J.,  {Margot} J.-L.,  2012, \mn@doi [\apj] {10.1088/0004-637X/761/2/92},
  \href {https://ui.adsabs.harvard.edu/abs/2012ApJ...761...92F} {761, 92}

\bibitem[\protect\citeauthoryear{{Figueira} et~al.,}{{Figueira}
  et~al.}{2012}]{figueira2012}
{Figueira} P.,  et~al., 2012, \mn@doi [\aap] {10.1051/0004-6361/201219017},
  \href {https://ui.adsabs.harvard.edu/abs/2012A&A...541A.139F} {541, A139}

\bibitem[\protect\citeauthoryear{{Foreman-Mackey}, {Agol}, {Ambikasaran}  \&
  {Angus}}{{Foreman-Mackey} et~al.}{2017}]{celerite}
{Foreman-Mackey} D.,  {Agol} E.,  {Ambikasaran} S.,   {Angus} R.,  2017,
  {celerite: Scalable 1D Gaussian Processes in C++, Python, and Julia}
  (\mn@eprint {ascl} {1709.008})

\bibitem[\protect\citeauthoryear{Foreman-Mackey et~al.,}{Foreman-Mackey
  et~al.}{2021a}]{exoplanetdanforeman}
Foreman-Mackey D.,  et~al., 2021a, exoplanet-dev/exoplanet: exoplanet v0.4.5,
  \mn@doi{10.5281/zenodo.4604868}, \url
  {https://doi.org/10.5281/zenodo.4604868}

\bibitem[\protect\citeauthoryear{Foreman-Mackey et~al.,}{Foreman-Mackey
  et~al.}{2021b}]{exoplanet:exoplanet}
Foreman-Mackey D.,  et~al., 2021b, exoplanet-dev/exoplanet v0.4.5,
  \mn@doi{10.5281/zenodo.1998447}, \url
  {https://doi.org/10.5281/zenodo.1998447}

\bibitem[\protect\citeauthoryear{{Foreman-Mackey} et~al.,}{{Foreman-Mackey}
  et~al.}{2021c}]{exoplanetnew}
{Foreman-Mackey} D.,  et~al., 2021c, arXiv e-prints, \href
  {https://ui.adsabs.harvard.edu/abs/2021arXiv210501994F} {p. arXiv:2105.01994}

\bibitem[\protect\citeauthoryear{{Fressin} et~al.,}{{Fressin}
  et~al.}{2005}]{Fressin2005a}
{Fressin} F.,  et~al., 2005, in {M.~Giard, F.~Casoli, \& F.~Paletou} ed.,  EAS
  Publications Series Vol. 14, EAS Publications Series. pp 309--312,
  \mn@doi{10.1051/eas:2005049}

\bibitem[\protect\citeauthoryear{{Fulton} \& {Petigura}}{{Fulton} \&
  {Petigura}}{2018}]{Fulton:2018}
{Fulton} B.~J.,  {Petigura} E.~A.,  2018, \mn@doi [\aj]
  {10.3847/1538-3881/aae828}, \href
  {https://ui.adsabs.harvard.edu/abs/2018AJ....156..264F} {156, 264}

\bibitem[\protect\citeauthoryear{{Fulton} et~al.,}{{Fulton}
  et~al.}{2017}]{Fulton:2017}
{Fulton} B.~J.,  et~al., 2017, \mn@doi [\aj] {10.3847/1538-3881/aa80eb}, \href
  {https://ui.adsabs.harvard.edu/abs/2017AJ....154..109F} {154, 109}

\bibitem[\protect\citeauthoryear{{Furlan} et~al.,}{{Furlan}
  et~al.}{2017}]{furlan17}
{Furlan} E.,  et~al., 2017, \mn@doi [\aj] {10.3847/1538-3881/153/2/71}, \href
  {https://ui.adsabs.harvard.edu/abs/2017AJ....153...71F} {153, 71}

\bibitem[\protect\citeauthoryear{{Gaia Collaboration} et~al.,}{{Gaia
  Collaboration} et~al.}{2018}]{GAIA_DR2}
{Gaia Collaboration} et~al., 2018, \mn@doi [\aap]
  {10.1051/0004-6361/201833051}, \href
  {https://ui.adsabs.harvard.edu/abs/2018A&A...616A...1G} {616, A1}

\bibitem[\protect\citeauthoryear{{Gaia Collaboration} et~al.,}{{Gaia
  Collaboration} et~al.}{2021a}]{GAIA_EDR3}
{Gaia Collaboration} et~al., 2021a, \mn@doi [\aap]
  {10.1051/0004-6361/202039657}, \href
  {https://ui.adsabs.harvard.edu/abs/2021A&A...649A...1G} {649, A1}

\bibitem[\protect\citeauthoryear{{Gaia Collaboration} et~al.,}{{Gaia
  Collaboration} et~al.}{2021b}]{GaiaCollaboration2021}
{Gaia Collaboration} et~al., 2021b, \mn@doi [\aap]
  {10.1051/0004-6361/202039657}, \href
  {https://ui.adsabs.harvard.edu/abs/2021A&A...649A...1G} {649, A1}

\bibitem[\protect\citeauthoryear{{Gardner} et~al.,}{{Gardner}
  et~al.}{2006}]{jwst}
{Gardner} J.~P.,  et~al., 2006, \mn@doi [\ssr] {10.1007/s11214-006-8315-7},
  \href {https://ui.adsabs.harvard.edu/abs/2006SSRv..123..485G} {123, 485}

\bibitem[\protect\citeauthoryear{{Giacalone} et~al.,}{{Giacalone}
  et~al.}{2022}]{Giacalone2022}
{Giacalone} S.,  et~al., 2022, arXiv e-prints, \href
  {https://ui.adsabs.harvard.edu/abs/2022arXiv220112661G} {p. arXiv:2201.12661}

\bibitem[\protect\citeauthoryear{{Ginzburg}, {Schlichting}  \&
  {Sari}}{{Ginzburg} et~al.}{2016}]{Ginzburg2016}
{Ginzburg} S.,  {Schlichting} H.~E.,   {Sari} R.,  2016, \mn@doi [\apj]
  {10.3847/0004-637X/825/1/29}, \href
  {https://ui.adsabs.harvard.edu/abs/2016ApJ...825...29G} {825, 29}

\bibitem[\protect\citeauthoryear{{Grunblatt}, {Howard}  \&
  {Haywood}}{{Grunblatt} et~al.}{2015}]{grunblatt2015}
{Grunblatt} S.~K.,  {Howard} A.~W.,   {Haywood} R.~D.,  2015, \mn@doi [\apj]
  {10.1088/0004-637X/808/2/127}, \href
  {https://ui.adsabs.harvard.edu/abs/2015ApJ...808..127G} {808, 127}

\bibitem[\protect\citeauthoryear{{Guerrero} et~al.,}{{Guerrero}
  et~al.}{2021}]{toi2021}
{Guerrero} N.~M.,  et~al., 2021, \mn@doi [\apjs] {10.3847/1538-4365/abefe1},
  \href {https://ui.adsabs.harvard.edu/abs/2021ApJS..254...39G} {254, 39}

\bibitem[\protect\citeauthoryear{{Guillot} et~al.,}{{Guillot}
  et~al.}{2015}]{Guillot2015}
{Guillot} T.,  et~al., 2015, \mn@doi [Astronomische Nachrichten]
  {10.1002/asna.201512174}, \href
  {https://ui.adsabs.harvard.edu/abs/2015AN....336..638G} {336, 638}

\bibitem[\protect\citeauthoryear{{Henden}, {Templeton}, {Terrell}, {Smith},
  {Levine}  \& {Welch}}{{Henden} et~al.}{2016}]{apass}
{Henden} A.~A.,  {Templeton} M.,  {Terrell} D.,  {Smith} T.~C.,  {Levine} S.,
  {Welch} D.,  2016, VizieR Online Data Catalog, \href
  {https://ui.adsabs.harvard.edu/abs/2016yCat.2336....0H} {p. II/336}

\bibitem[\protect\citeauthoryear{{Hoffman} \& {Gelman}}{{Hoffman} \&
  {Gelman}}{2011}]{mcmcnuts}
{Hoffman} M.~D.,  {Gelman} A.,  2011, arXiv e-prints, \href
  {https://ui.adsabs.harvard.edu/abs/2011arXiv1111.4246H} {p. arXiv:1111.4246}

\bibitem[\protect\citeauthoryear{{Howell}, {Everett}, {Sherry}, {Horch}  \&
  {Ciardi}}{{Howell} et~al.}{2011}]{howell11}
{Howell} S.~B.,  {Everett} M.~E.,  {Sherry} W.,  {Horch} E.,   {Ciardi} D.~R.,
  2011, \mn@doi [\aj] {10.1088/0004-6256/142/1/19}, \href
  {http://adsabs.harvard.edu/abs/2011AJ....142...19H} {142, 19}

\bibitem[\protect\citeauthoryear{{Howell} et~al.,}{{Howell}
  et~al.}{2014}]{howell2014}
{Howell} S.~B.,  et~al., 2014, \mn@doi [\pasp] {10.1086/676406}, \href
  {https://ui.adsabs.harvard.edu/abs/2014PASP..126..398H} {126, 398}

\bibitem[\protect\citeauthoryear{{Hoyer}, {Guterman}, {Demangeon}, {Sousa},
  {Deleuil}, {Meunier}  \& {Benz}}{{Hoyer} et~al.}{2020}]{Hoyer2020}
{Hoyer} S.,  {Guterman} P.,  {Demangeon} O.,  {Sousa} S.~G.,  {Deleuil} M.,
  {Meunier} J.~C.,   {Benz} W.,  2020, \mn@doi [\aap]
  {10.1051/0004-6361/201936325}, \href
  {https://ui.adsabs.harvard.edu/abs/2020A&A...635A..24H} {635, A24}

\bibitem[\protect\citeauthoryear{{Irwin}, {Irwin}, {Aigrain}, {Hodgkin}, {Hebb}
   \& {Moraux}}{{Irwin} et~al.}{2007}]{irwin2007}
{Irwin} J.,  {Irwin} M.,  {Aigrain} S.,  {Hodgkin} S.,  {Hebb} L.,   {Moraux}
  E.,  2007, \mn@doi [\mnras] {10.1111/j.1365-2966.2006.11408.x}, \href
  {https://ui.adsabs.harvard.edu/abs/2007MNRAS.375.1449I} {375, 1449}

\bibitem[\protect\citeauthoryear{{Irwin}, {Berta-Thompson}, {Charbonneau},
  {Dittmann}, {Falco}, {Newton}  \& {Nutzman}}{{Irwin}
  et~al.}{2015a}]{irwin2015}
{Irwin} J.~M.,  {Berta-Thompson} Z.~K.,  {Charbonneau} D.,  {Dittmann} J.,
  {Falco} E.~E.,  {Newton} E.~R.,   {Nutzman} P.,  2015a, in 18th Cambridge
  Workshop on Cool Stars, Stellar Systems, and the Sun. pp 767--772 (\mn@eprint
  {arXiv} {1409.0891})

\bibitem[\protect\citeauthoryear{{Irwin}, {Berta-Thompson}, {Charbonneau},
  {Dittmann}  \& {Newton}}{{Irwin} et~al.}{2015b}]{mearth2015}
{Irwin} J.,  {Berta-Thompson} Z.~K.,  {Charbonneau} D.,  {Dittmann} J.,
  {Newton} E.~R.,  2015b, in American Astronomical Society Meeting Abstracts
  \#225. p. 258.01

\bibitem[\protect\citeauthoryear{{Jenkins}}{{Jenkins}}{2002}]{jenkins2002}
{Jenkins} J.~M.,  2002, \mn@doi [\apj] {10.1086/341136}, \href
  {http://adsabs.harvard.edu/abs/2002ApJ...575..493J} {575, 493}

\bibitem[\protect\citeauthoryear{{Jenkins} et~al.,}{{Jenkins}
  et~al.}{2010}]{jenkins2010}
{Jenkins} J.~M.,  et~al., 2010, in {Radziwill} N.~M.,  {Bridger} A.,  eds,
  Society of Photo-Optical Instrumentation Engineers (SPIE) Conference Series
  Vol. 7740, Software and Cyberinfrastructure for Astronomy. p. 77400D,
  \mn@doi{10.1117/12.856764}

\bibitem[\protect\citeauthoryear{{Jenkins} et~al.,}{{Jenkins}
  et~al.}{2016}]{Jenkins:2016}
{Jenkins} J.~M.,  et~al., 2016, in Software and Cyberinfrastructure for
  Astronomy IV. p. 99133E, \mn@doi{10.1117/12.2233418}

\bibitem[\protect\citeauthoryear{{Jenkins}, {Tenenbaum}, {Seader}, {Burke},
  {McCauliff}, {Smith}, {Twicken}  \& {Chandrasekaran}}{{Jenkins}
  et~al.}{2020}]{jenkins2020}
{Jenkins} J.~M.,  {Tenenbaum} P.,  {Seader} S.,  {Burke} C.~J.,  {McCauliff}
  S.~D.,  {Smith} J.~C.,  {Twicken} J.~D.,   {Chandrasekaran} H.,  2020,
  {Kepler Data Processing Handbook: Transiting Planet Search}, Kepler Science
  Document KSCI-19081-003

\bibitem[\protect\citeauthoryear{{Jensen}}{{Jensen}}{2013}]{Jensen:2013}
{Jensen} E.,  2013, {Tapir: A web interface for transit/eclipse observability},
  Astrophysics Source Code Library (\mn@eprint {ascl} {1306.007})

\bibitem[\protect\citeauthoryear{Jin \& Mordasini}{Jin \&
  Mordasini}{2017}]{Jin2018}
Jin S.,  Mordasini C.,  2017, \mn@doi [The Astrophysical Journal]
  {10.3847/1538-4357/aa9f1e}, 853

\bibitem[\protect\citeauthoryear{{Johansen}, {Davies}, {Church}  \&
  {Holmelin}}{{Johansen} et~al.}{2012}]{johansen2012}
{Johansen} A.,  {Davies} M.~B.,  {Church} R.~P.,   {Holmelin} V.,  2012,
  \mn@doi [\apj] {10.1088/0004-637X/758/1/39}, \href
  {https://ui.adsabs.harvard.edu/abs/2012ApJ...758...39J} {758, 39}

\bibitem[\protect\citeauthoryear{{Johnson} \& {Soderblom}}{{Johnson} \&
  {Soderblom}}{1987}]{johnson1987}
{Johnson} D. R.~H.,  {Soderblom} D.~R.,  1987, \mn@doi [\aj] {10.1086/114370},
  \href {https://ui.adsabs.harvard.edu/abs/1987AJ.....93..864J} {93, 864}

\bibitem[\protect\citeauthoryear{{Kempton} et~al.,}{{Kempton}
  et~al.}{2018}]{kempton2018}
{Kempton} E. M.~R.,  et~al., 2018, \mn@doi [\pasp] {10.1088/1538-3873/aadf6f},
  \href {https://ui.adsabs.harvard.edu/abs/2018PASP..130k4401K} {130, 114401}

\bibitem[\protect\citeauthoryear{{Kipping}}{{Kipping}}{2013a}]{exoplanet:kipping13_2}
{Kipping} D.~M.,  2013a, \mn@doi [\mnras] {10.1093/mnrasl/slt075}, \href
  {https://ui.adsabs.harvard.edu/abs/2013MNRAS.434L..51K} {434, L51}

\bibitem[\protect\citeauthoryear{{Kipping}}{{Kipping}}{2013b}]{exoplanet:kipping13}
{Kipping} D.~M.,  2013b, \mn@doi [\mnras] {10.1093/mnras/stt1435}, \href
  {http://adsabs.harvard.edu/abs/2013MNRAS.435.2152K} {435, 2152}

\bibitem[\protect\citeauthoryear{Kumar, Carroll, Hartikainen  \& Martin}{Kumar
  et~al.}{2019}]{exoplanet:arviz}
Kumar R.,  Carroll C.,  Hartikainen A.,   Martin O.~A.,  2019, \mn@doi [The
  Journal of Open Source Software] {10.21105/joss.01143}

\bibitem[\protect\citeauthoryear{{Kurucz}}{{Kurucz}}{1993}]{Kurucz:1993}
{Kurucz} R.~L.,  1993, \mn@doi [Physica Scripta Volume T]
  {10.1088/0031-8949/1993/T47/017}, \href
  {https://ui.adsabs.harvard.edu/abs/1993PhST...47..110K} {47, 110}

\bibitem[\protect\citeauthoryear{{Leleu} et~al.,}{{Leleu}
  et~al.}{2021}]{leleu2021}
{Leleu} A.,  et~al., 2021, \mn@doi [\aap] {10.1051/0004-6361/202039767}, \href
  {https://ui.adsabs.harvard.edu/abs/2021A&A...649A..26L} {649, A26}

\bibitem[\protect\citeauthoryear{{Lendl} et~al.,}{{Lendl}
  et~al.}{2020}]{Lendl2020}
{Lendl} M.,  et~al., 2020, \mn@doi [\aap] {10.1051/0004-6361/202038677}, \href
  {https://ui.adsabs.harvard.edu/abs/2020A&A...643A..94L} {643, A94}

\bibitem[\protect\citeauthoryear{{Li}, {Tenenbaum}, {Twicken}, {Burke},
  {Jenkins}, {Quintana}, {Rowe}  \& {Seader}}{{Li} et~al.}{2019}]{li2019}
{Li} J.,  {Tenenbaum} P.,  {Twicken} J.~D.,  {Burke} C.~J.,  {Jenkins} J.~M.,
  {Quintana} E.~V.,  {Rowe} J.~F.,   {Seader} S.~E.,  2019, \mn@doi [\pasp]
  {10.1088/1538-3873/aaf44d}, \href
  {https://ui.adsabs.harvard.edu/abs/2019PASP..131b4506L} {131, 024506}

\bibitem[\protect\citeauthoryear{{Lillo-Box}, {Barrado}  \& {Bouy}}{{Lillo-Box}
  et~al.}{2012}]{lillobox12}
{Lillo-Box} J.,  {Barrado} D.,   {Bouy} H.,  2012, \mn@doi [\aap]
  {10.1051/0004-6361/201219631}, \href
  {http://ads.nao.ac.jp/abs/2012A%26A...546A..10L} {546, A10}

\bibitem[\protect\citeauthoryear{{Lillo-Box}, {Barrado}  \& {Bouy}}{{Lillo-Box}
  et~al.}{2014}]{lillobox14}
{Lillo-Box} J.,  {Barrado} D.,   {Bouy} H.,  2014, \mn@doi [\aap]
  {10.1051/0004-6361/201423497}, \href
  {http://adsabs.harvard.edu/abs/2014A%26A...566A.103L} {566, A103}

\bibitem[\protect\citeauthoryear{{Lindegren} et~al.,}{{Lindegren}
  et~al.}{2021}]{Lindegren2021}
{Lindegren} L.,  et~al., 2021, \mn@doi [\aap] {10.1051/0004-6361/202039653},
  \href {https://ui.adsabs.harvard.edu/abs/2021A&A...649A...4L} {649, A4}

\bibitem[\protect\citeauthoryear{{Lissauer} et~al.,}{{Lissauer}
  et~al.}{2011}]{lissauer2011kepler}
{Lissauer} J.~J.,  et~al., 2011, \mn@doi [\apjs] {10.1088/0067-0049/197/1/8},
  \href {https://ui.adsabs.harvard.edu/abs/2011ApJS..197....8L} {197, 8}

\bibitem[\protect\citeauthoryear{{Lissauer} et~al.,}{{Lissauer}
  et~al.}{2012}]{Lissauer2012}
{Lissauer} J.~J.,  et~al., 2012, \mn@doi [\apj] {10.1088/0004-637X/750/2/112},
  \href {https://ui.adsabs.harvard.edu/abs/2012ApJ...750..112L} {750, 112}

\bibitem[\protect\citeauthoryear{{Lovis} \& {Pepe}}{{Lovis} \&
  {Pepe}}{2007}]{lovis2007}
{Lovis} C.,  {Pepe} F.,  2007, \mn@doi [\aap] {10.1051/0004-6361:20077249},
  \href {https://ui.adsabs.harvard.edu/abs/2007A&A...468.1115L} {468, 1115}

\bibitem[\protect\citeauthoryear{{Luger}, {Agol}, {Foreman-Mackey}, {Fleming},
  {Lustig-Yaeger}  \& {Deitrick}}{{Luger} et~al.}{2019a}]{starry}
{Luger} R.,  {Agol} E.,  {Foreman-Mackey} D.,  {Fleming} D.~P.,
  {Lustig-Yaeger} J.,   {Deitrick} R.,  2019a, \mn@doi [aj]
  {10.3847/1538-3881/aae8e5}, \href
  {http://adsabs.harvard.edu/abs/2019AJ....157...64L} {157, 64}

\bibitem[\protect\citeauthoryear{{Luger}, {Agol}, {Foreman-Mackey}, {Fleming},
  {Lustig-Yaeger}  \& {Deitrick}}{{Luger} et~al.}{2019b}]{exoplanet:luger18}
{Luger} R.,  {Agol} E.,  {Foreman-Mackey} D.,  {Fleming} D.~P.,
  {Lustig-Yaeger} J.,   {Deitrick} R.,  2019b, \mn@doi [\aj]
  {10.3847/1538-3881/aae8e5}, \href
  {http://adsabs.harvard.edu/abs/2019AJ....157...64L} {157, 64}

\bibitem[\protect\citeauthoryear{{Marigo} et~al.,}{{Marigo}
  et~al.}{2017}]{marigo17}
{Marigo} P.,  et~al., 2017, \mn@doi [\apj] {10.3847/1538-4357/835/1/77}, \href
  {http://adsabs.harvard.edu/abs/2017ApJ...835...77M} {835, 77}

\bibitem[\protect\citeauthoryear{Martinez, Cunha, Ghezzi  \& Smith}{Martinez
  et~al.}{2019}]{martinez2019spectroscopic}
Martinez C.~F.,  Cunha K.,  Ghezzi L.,   Smith V.~V.,  2019, The Astrophysical
  Journal, 875, 29

\bibitem[\protect\citeauthoryear{{Maxted} et~al.,}{{Maxted}
  et~al.}{2011}]{2011PASP..123..547M}
{Maxted} P.~F.~L.,  et~al., 2011, \mn@doi [\pasp] {10.1086/660007}, \href
  {https://ui.adsabs.harvard.edu/abs/2011PASP..123..547M} {123, 547}

\bibitem[\protect\citeauthoryear{{Mayor} \& {Queloz}}{{Mayor} \&
  {Queloz}}{1995}]{MayorQueloz1995}
{Mayor} M.,  {Queloz} D.,  1995, \mn@doi [\nat] {10.1038/378355a0}, \href
  {https://ui.adsabs.harvard.edu/abs/1995Natur.378..355M} {378, 355}

\bibitem[\protect\citeauthoryear{{Mayor} et~al.,}{{Mayor} et~al.}{2003}]{HARPS}
{Mayor} M.,  et~al., 2003, The Messenger, \href
  {http://adsabs.harvard.edu/abs/2003Msngr.114...20M} {114, 20}

\bibitem[\protect\citeauthoryear{{McCully}, {Volgenau}, {Harbeck}, {Lister},
  {Saunders}, {Turner}, {Siiverd}  \& {Bowman}}{{McCully}
  et~al.}{2018}]{McCully:2018}
{McCully} C.,  {Volgenau} N.~H.,  {Harbeck} D.-R.,  {Lister} T.~A.,  {Saunders}
  E.~S.,  {Turner} M.~L.,  {Siiverd} R.~J.,   {Bowman} M.,  2018, in \procspie.
  p. 107070K (\mn@eprint {arXiv} {1811.04163}), \mn@doi{10.1117/12.2314340}

\bibitem[\protect\citeauthoryear{{M{\'e}karnia} et~al.,}{{M{\'e}karnia}
  et~al.}{2016}]{Mekarnia2016}
{M{\'e}karnia} D.,  et~al., 2016, \mn@doi [\mnras] {10.1093/mnras/stw1934},
  \href {http://adsabs.harvard.edu/abs/2016MNRAS.463...45M} {463, 45}

\bibitem[\protect\citeauthoryear{{Millholland} \& {Winn}}{{Millholland} \&
  {Winn}}{2021}]{milholland2021}
{Millholland} S.~C.,  {Winn} J.~N.,  2021, \mn@doi [\apjl]
  {10.3847/2041-8213/ac2c77}, \href
  {https://ui.adsabs.harvard.edu/abs/2021ApJ...920L..34M} {920, L34}

\bibitem[\protect\citeauthoryear{{Millholland}, {Wang}  \&
  {Laughlin}}{{Millholland} et~al.}{2017}]{milholland2017}
{Millholland} S.,  {Wang} S.,   {Laughlin} G.,  2017, \mn@doi [\apjl]
  {10.3847/2041-8213/aa9714}, \href
  {https://ui.adsabs.harvard.edu/abs/2017ApJ...849L..33M} {849, L33}

\bibitem[\protect\citeauthoryear{{Nesvorn{\'y}}, {Kipping}, {Terrell}  \&
  {Feroz}}{{Nesvorn{\'y}} et~al.}{2014}]{nesvorny2014}
{Nesvorn{\'y}} D.,  {Kipping} D.,  {Terrell} D.,   {Feroz} F.,  2014, \mn@doi
  [\apj] {10.1088/0004-637X/790/1/31}, \href
  {https://ui.adsabs.harvard.edu/abs/2014ApJ...790...31N} {790, 31}

\bibitem[\protect\citeauthoryear{{Osborn} et~al.,}{{Osborn}
  et~al.}{2021}]{ares}
{Osborn} A.,  et~al., 2021, \mn@doi [\mnras] {10.1093/mnras/stab2313}, \href
  {https://ui.adsabs.harvard.edu/abs/2021MNRAS.507.2782O} {507, 2782}

\bibitem[\protect\citeauthoryear{{Owen} \& {Wu}}{{Owen} \&
  {Wu}}{2013}]{Owen2013}
{Owen} J.~E.,  {Wu} Y.,  2013, \mn@doi [\apj] {10.1088/0004-637X/775/2/105},
  \href {https://ui.adsabs.harvard.edu/abs/2013ApJ...775..105O} {775, 105}

\bibitem[\protect\citeauthoryear{{Pepper} et~al.,}{{Pepper}
  et~al.}{2007}]{pepper2007}
{Pepper} J.,  et~al., 2007, \mn@doi [\pasp] {10.1086/521836}, \href
  {https://ui.adsabs.harvard.edu/abs/2007PASP..119..923P} {119, 923}

\bibitem[\protect\citeauthoryear{{Pollacco} et~al.,}{{Pollacco}
  et~al.}{2006}]{pollacco2006}
{Pollacco} D.~L.,  et~al., 2006, \mn@doi [\pasp] {10.1086/508556}, \href
  {https://ui.adsabs.harvard.edu/abs/2006PASP..118.1407P} {118, 1407}

\bibitem[\protect\citeauthoryear{{Rayner}, {Bond}, {Bonnet}, {Jaffe}, {Muller}
  \& {Tokunaga}}{{Rayner} et~al.}{2012}]{ishell2012}
{Rayner} J.,  {Bond} T.,  {Bonnet} M.,  {Jaffe} D.,  {Muller} G.,   {Tokunaga}
  A.,  2012, in {McLean} I.~S.,  {Ramsay} S.~K.,   {Takami} H.,  eds,  Society
  of Photo-Optical Instrumentation Engineers (SPIE) Conference Series Vol.
  8446, Ground-based and Airborne Instrumentation for Astronomy IV. p. 84462C,
  \mn@doi{10.1117/12.925511}

\bibitem[\protect\citeauthoryear{{Reddy}, {Lambert}  \& {Allende
  Prieto}}{{Reddy} et~al.}{2006}]{reddy2006}
{Reddy} B.~E.,  {Lambert} D.~L.,   {Allende Prieto} C.,  2006, \mn@doi [\mnras]
  {10.1111/j.1365-2966.2006.10148.x}, \href
  {https://ui.adsabs.harvard.edu/abs/2006MNRAS.367.1329R} {367, 1329}

\bibitem[\protect\citeauthoryear{{Ricker} et~al.,}{{Ricker}
  et~al.}{2015}]{Ricker:2015}
{Ricker} G.~R.,  et~al., 2015, \mn@doi [Journal of Astronomical Telescopes,
  Instruments, and Systems] {10.1117/1.JATIS.1.1.014003}, \href
  {http://adsabs.harvard.edu/abs/2015JATIS...1a4003R} {1, 014003}

\bibitem[\protect\citeauthoryear{{Rousset} et~al.,}{{Rousset}
  et~al.}{2003}]{rousset2003}
{Rousset} G.,  et~al., 2003, in {Wizinowich} P.~L.,  {Bonaccini} D.,  eds,
  Society of Photo-Optical Instrumentation Engineers (SPIE) Conference Series
  Vol. 4839, Adaptive Optical System Technologies II. pp 140--149,
  \mn@doi{10.1117/12.459332}

\bibitem[\protect\citeauthoryear{{Salmon}, {Van Grootel}, {Buldgen}, {Dupret}
  \& {Eggenberger}}{{Salmon} et~al.}{2021}]{salmon21}
{Salmon} S.~J.~A.~J.,  {Van Grootel} V.,  {Buldgen} G.,  {Dupret} M.~A.,
  {Eggenberger} P.,  2021, \mn@doi [\aap] {10.1051/0004-6361/201937174}, \href
  {https://ui.adsabs.harvard.edu/abs/2021A&A...646A...7S} {646, A7}

\bibitem[\protect\citeauthoryear{Salvatier, Wiecki  \& Fonnesbeck}{Salvatier
  et~al.}{2016a}]{exoplanet:pymc3}
Salvatier J.,  Wiecki T.~V.,   Fonnesbeck C.,  2016a, PeerJ Computer Science,
  2, e55

\bibitem[\protect\citeauthoryear{Salvatier, Wiecki  \& Fonnesbeck}{Salvatier
  et~al.}{2016b}]{pymc3}
Salvatier J.,  Wiecki T.~V.,   Fonnesbeck C.,  2016b, PeerJ Computer Science,
  2, e55

\bibitem[\protect\citeauthoryear{{Santerne} et~al.,}{{Santerne}
  et~al.}{2018}]{santerne2018}
{Santerne} A.,  et~al., 2018, \mn@doi [Nature Astronomy]
  {10.1038/s41550-018-0420-5}, \href
  {https://ui.adsabs.harvard.edu/abs/2018NatAs...2..393S} {2, 393}

\bibitem[\protect\citeauthoryear{{Santos} et~al.,}{{Santos}
  et~al.}{2013}]{Santos:2013}
{Santos} N.~C.,  et~al., 2013, \mn@doi [\aap] {10.1051/0004-6361/201321286},
  \href {https://ui.adsabs.harvard.edu/abs/2013A&A...556A.150S} {556, A150}

\bibitem[\protect\citeauthoryear{{Schanche} et~al.,}{{Schanche}
  et~al.}{2020}]{Schanche2020}
{Schanche} N.,  et~al., 2020, \mn@doi [\mnras] {10.1093/mnras/staa2848}, \href
  {https://ui.adsabs.harvard.edu/abs/2020MNRAS.499..428S} {499, 428}

\bibitem[\protect\citeauthoryear{{Schlichting}, {Sari}  \&
  {Yalinewich}}{{Schlichting} et~al.}{2015}]{Schlichting2015}
{Schlichting} H.~E.,  {Sari} R.,   {Yalinewich} A.,  2015, \mn@doi [\icarus]
  {10.1016/j.icarus.2014.09.053}, \href
  {https://ui.adsabs.harvard.edu/abs/2015Icar..247...81S} {247, 81}

\bibitem[\protect\citeauthoryear{{Scott} et~al.,}{{Scott}
  et~al.}{2021}]{scott2021}
{Scott} N.~J.,  et~al., 2021, \mn@doi [Frontiers in Astronomy and Space
  Sciences] {10.3389/fspas.2021.716560}, \href
  {https://ui.adsabs.harvard.edu/abs/2021FrASS...8..138S} {8, 138}

\bibitem[\protect\citeauthoryear{{Scuflaire}, {Th{\'e}ado}, {Montalb{\'a}n},
  {Miglio}, {Bourge}, {Godart}, {Thoul}  \& {Noels}}{{Scuflaire}
  et~al.}{2008}]{scuflaire08}
{Scuflaire} R.,  {Th{\'e}ado} S.,  {Montalb{\'a}n} J.,  {Miglio} A.,  {Bourge}
  P.-O.,  {Godart} M.,  {Thoul} A.,   {Noels} A.,  2008, \mn@doi [\apss]
  {10.1007/s10509-007-9650-1}, \href
  {http://adsabs.harvard.edu/abs/2008Ap%26SS.316...83S} {316, 83}

\bibitem[\protect\citeauthoryear{{Skrutskie} et~al.,}{{Skrutskie}
  et~al.}{2006}]{Skrutskie2006}
{Skrutskie} M.~F.,  et~al., 2006, \mn@doi [\aj] {10.1086/498708}, \href
  {https://ui.adsabs.harvard.edu/abs/2006AJ....131.1163S} {131, 1163}

\bibitem[\protect\citeauthoryear{{Smith} et~al.,}{{Smith}
  et~al.}{2012}]{Smith2012}
{Smith} J.~C.,  et~al., 2012, \mn@doi [\pasp] {10.1086/667697}, \href
  {https://ui.adsabs.harvard.edu/abs/2012PASP..124.1000S} {124, 1000}

\bibitem[\protect\citeauthoryear{{Smith} et~al.,}{{Smith}
  et~al.}{2020}]{smith2020}
{Smith} A. M.~S.,  et~al., 2020, \mn@doi [Astronomische Nachrichten]
  {10.1002/asna.202013768}, \href
  {https://ui.adsabs.harvard.edu/abs/2020AN....341..273S} {341, 273}

\bibitem[\protect\citeauthoryear{{Sneden}}{{Sneden}}{1973}]{Sneden:1973}
{Sneden} C.~A.,  1973, PhD thesis, THE UNIVERSITY OF TEXAS AT AUSTIN.

\bibitem[\protect\citeauthoryear{Sousa}{Sousa}{2014}]{sousa2014}
Sousa S.~G.,  2014, \mn@doi [GeoPlanet: Earth and Planetary Sciences]
  {10.1007/978-3-319-06956-2_26}, p. 297–310

\bibitem[\protect\citeauthoryear{{Sousa}, {Santos}, {Adibekyan}, {Delgado-Mena}
   \& {Israelian}}{{Sousa} et~al.}{2015}]{sousa2015}
{Sousa} S.~G.,  {Santos} N.~C.,  {Adibekyan} V.,  {Delgado-Mena} E.,
  {Israelian} G.,  2015, \mn@doi [\aap] {10.1051/0004-6361/201425463}, \href
  {https://ui.adsabs.harvard.edu/abs/2015A&A...577A..67S} {577, A67}

\bibitem[\protect\citeauthoryear{{Sousa} et~al.,}{{Sousa}
  et~al.}{2021}]{sousa2021}
{Sousa} S.~G.,  et~al., 2021, arXiv e-prints, \href
  {https://ui.adsabs.harvard.edu/abs/2021arXiv210904781S} {p. arXiv:2109.04781}

\bibitem[\protect\citeauthoryear{{Southworth}}{{Southworth}}{2011}]{tepcat}
{Southworth} J.,  2011, \mn@doi [\mnras] {10.1111/j.1365-2966.2011.19399.x},
  \href {https://ui.adsabs.harvard.edu/abs/2011MNRAS.417.2166S} {417, 2166}

\bibitem[\protect\citeauthoryear{{Stassun} et~al.,}{{Stassun}
  et~al.}{2019}]{Stassun2019}
{Stassun} K.~G.,  et~al., 2019, \mn@doi [\aj] {10.3847/1538-3881/ab3467}, \href
  {https://ui.adsabs.harvard.edu/abs/2019AJ....158..138S} {158, 138}

\bibitem[\protect\citeauthoryear{{Stumpe} et~al.,}{{Stumpe}
  et~al.}{2012}]{Stumpe2012}
{Stumpe} M.~C.,  et~al., 2012, \mn@doi [\pasp] {10.1086/667698}, \href
  {https://ui.adsabs.harvard.edu/abs/2012PASP..124..985S} {124, 985}

\bibitem[\protect\citeauthoryear{{Stumpe}, {Smith}, {Catanzarite}, {Van Cleve},
  {Jenkins}, {Twicken}  \& {Girouard}}{{Stumpe} et~al.}{2014}]{Stumpe2014}
{Stumpe} M.~C.,  {Smith} J.~C.,  {Catanzarite} J.~H.,  {Van Cleve} J.~E.,
  {Jenkins} J.~M.,  {Twicken} J.~D.,   {Girouard} F.~R.,  2014, \mn@doi [\pasp]
  {10.1086/674989}, \href
  {https://ui.adsabs.harvard.edu/abs/2014PASP..126..100S} {126, 100}

\bibitem[\protect\citeauthoryear{{Teske} et~al.,}{{Teske}
  et~al.}{2021}]{magellanteske}
{Teske} J.,  et~al., 2021, \mn@doi [\apjs] {10.3847/1538-4365/ac0f0a}, \href
  {https://ui.adsabs.harvard.edu/abs/2021ApJS..256...33T} {256, 33}

\bibitem[\protect\citeauthoryear{{Theano Development Team}}{{Theano Development
  Team}}{2016}]{exoplanet:theano}
{Theano Development Team} 2016, arXiv e-prints, abs/1605.02688

\bibitem[\protect\citeauthoryear{{Tremaine} \& {Dong}}{{Tremaine} \&
  {Dong}}{2012}]{tremainedong}
{Tremaine} S.,  {Dong} S.,  2012, \mn@doi [\aj] {10.1088/0004-6256/143/4/94},
  \href {https://ui.adsabs.harvard.edu/abs/2012AJ....143...94T} {143, 94}

\bibitem[\protect\citeauthoryear{{Twicken} et~al.,}{{Twicken}
  et~al.}{2018}]{twicken2018}
{Twicken} J.~D.,  et~al., 2018, \mn@doi [\pasp] {10.1088/1538-3873/aab694},
  \href {http://adsabs.harvard.edu/abs/2018PASP..130f4502T} {130, 064502}

\bibitem[\protect\citeauthoryear{{Van Eylen}, {Agentoft}, {Lundkvist},
  {Kjeldsen}, {Owen}, {Fulton}, {Petigura}  \& {Snellen}}{{Van Eylen}
  et~al.}{2018}]{vaneylen2017}
{Van Eylen} V.,  {Agentoft} C.,  {Lundkvist} M.~S.,  {Kjeldsen} H.,  {Owen}
  J.~E.,  {Fulton} B.~J.,  {Petigura} E.,   {Snellen} I.,  2018, \mn@doi
  [\mnras] {10.1093/mnras/sty1783}, \href
  {https://ui.adsabs.harvard.edu/abs/2018MNRAS.479.4786V} {479, 4786}

\bibitem[\protect\citeauthoryear{{Van Eylen} et~al.,}{{Van Eylen}
  et~al.}{2021}]{VanEylen2021}
{Van Eylen} V.,  et~al., 2021, \mn@doi [\mnras] {10.1093/mnras/stab2143}, \href
  {https://ui.adsabs.harvard.edu/abs/2021MNRAS.507.2154V} {507, 2154}

\bibitem[\protect\citeauthoryear{{Vogt} \& {Penrod}}{{Vogt} \&
  {Penrod}}{1988}]{hires}
{Vogt} S.~S.,  {Penrod} G.~D.,  1988, in Instrumentation for Ground-Based
  Optical Astronomy. p.~68

\bibitem[\protect\citeauthoryear{{Wheatley} et~al.,}{{Wheatley}
  et~al.}{2018}]{Wheatley2018}
{Wheatley} P.~J.,  et~al., 2018, \mn@doi [\mnras] {10.1093/mnras/stx2836},
  \href {https://ui.adsabs.harvard.edu/abs/2018MNRAS.475.4476W} {475, 4476}

\bibitem[\protect\citeauthoryear{{Wilson} et~al.,}{{Wilson}
  et~al.}{2022}]{Wilson2022}
{Wilson} T.~G.,  et~al., 2022, \mn@doi [\mnras] {10.1093/mnras/stab3799}, \href
  {https://ui.adsabs.harvard.edu/abs/2022MNRAS.tmp..107W} {}

\bibitem[\protect\citeauthoryear{{Wittenmyer} et~al.,}{{Wittenmyer}
  et~al.}{2018}]{minerva2018}
{Wittenmyer} R.,  et~al., 2018, in American Astronomical Society Meeting
  Abstracts \#231. p. 128.01

\bibitem[\protect\citeauthoryear{{Wright} et~al.}{{Wright} et~al.}{2010}]{wise}
{Wright} E.~L.,  et~al., 2010, \mn@doi [\aj] {10.1088/0004-6256/140/6/1868},
  \href {http://adsabs.harvard.edu/abs/2010AJ....140.1868W} {140, 1868}

\bibitem[\protect\citeauthoryear{{Zeng}, {Sasselov}  \& {Jacobsen}}{{Zeng}
  et~al.}{2016}]{zeng}
{Zeng} L.,  {Sasselov} D.~D.,   {Jacobsen} S.~B.,  2016, \mn@doi [\apj]
  {10.3847/0004-637X/819/2/127}, \href
  {https://ui.adsabs.harvard.edu/abs/2016ApJ...819..127Z} {819, 127}

\bibitem[\protect\citeauthoryear{{Ziegler}, {Tokovinin}, {Brice{\~n}o}, {Mang},
  {Law}  \& {Mann}}{{Ziegler} et~al.}{2020}]{ziegler2020}
{Ziegler} C.,  {Tokovinin} A.,  {Brice{\~n}o} C.,  {Mang} J.,  {Law} N.,
  {Mann} A.~W.,  2020, \mn@doi [\aj] {10.3847/1538-3881/ab55e9}, \href
  {https://ui.adsabs.harvard.edu/abs/2020AJ....159...19Z} {159, 19}

\makeatother
\end{thebibliography}

%%%%%%%%%%%%%%%%%%%%%%%%%%%%%%%%%%%%%%%%%%%%%%%%
%%%%%%%%%%%%%%%%% APPENDIX %%%%%%%%%%%%%%%%%%%%%
%%%%%%%%%%%%%%%%%%%%%%%%%%%%%%%%%%%%%%%%%%%%%%%%

\appendix

\section{Priors} \label{sec:priors}

\begin{center}
\begin{table}
    \centering
    \caption{Global fit parameter prior function type and prior limits for TOI-836.}
    \label{tab:priorsstar}
    \begin{tabularx}{0.5\textwidth}{ l l l }
    \toprule
    \textbf{Parameter} &\textbf{Prior}  &\textbf{Value} \\
    \hline
    Baseline flux &$\mathcal{N}$(0, 1) & \\
    \mstar\ (\msun) &$\mathcal{N}$(0.678, 0.049, 0.65) & Table~\ref{tab:star_props_results}\\
    \rstar\ (\rsun) &$\mathcal{N}$(0.666, 0.010, 0.56) & Table~\ref{tab:star_props_results} \\
    Period (days) &$\mathcal{N}$(22, 0.1) & Table~\ref{tab:star_props_results} \\
    LD coefficient \textit{u\textsubscript{1}} &\citet{exoplanet:kipping13} & Table~\ref{tab:star_props_results} \\
    LD coefficient \textit{u\textsubscript{2}} &\citet{exoplanet:kipping13} & Table~\ref{tab:star_props_results} \\ \toprule
    \textbf{TESS GP} & \\
    \hline
    \textbf{Sector 11} & \\
    Mean &$\mathcal{N}$(0, 1)   & 0.00006 $\pm$ 0.00021 \\
    log(\textit{s}2) &$\mathcal{N}$(-14.704$^*$, 0.1)  & -14.98064 $\pm$ 0.01205 \\
    log(\textit{w}0) &$\mathcal{N}$(0, 0.1)     & 0.10400 $\pm$ 0.09697 \\
    log(\textit{Sw}4) &$\mathcal{N}$(-14.704$^*$, 0.1)   & -14.12245 $\pm$ 0.09004 \\
    \hline
    \textbf{Sector 38} & \\
    Mean &$\mathcal{N}$(0, 1) & 0.00008 $\pm$ 0.00031 \\
    log(\textit{s}2) &$\mathcal{N}$(-13.903$^*$, 0.1) & -14.86420 $\pm$ 0.01063 \\
    log(\textit{w}0) &$\mathcal{N}$(0, 0.05) & 0.00736 $\pm$ 0.04815 \\
    log(\textit{Sw}4) &$\mathcal{N}$(-13.903$^*$, 0.1)   & -13.47408 $\pm$ 0.04995 \\
    \toprule
    \textbf{RV GP} & \\
    \hline
    Amplitude &$\mathcal{C}$(5) & 7.13782 $\pm$ 1.05463 \\
    \textit{l}\textsubscript{\textit{E}} &$\mathcal{T}$(22, 20, 20) & 31.59616 $\pm$ 5.63098 \\
    \textit{l}\textsubscript{\textit{P}} &$\mathcal{T}$(0.1, 10, 0, 1)  & 0.21018 $\pm$ 0.02573 \\
    \harps\ offset  & $\mathcal{N}$(-26274.131$^{\dagger}$,10) & -26144.6 $\pm$ 2622.4  \\
    log(Jitter)\textsubscript{\harps}  & $\mathcal{N}$(0.247$^{\ddagger}$,5)  & -3.01818 $\pm$ 3.12178   \\
    \pfs\ offset    & $\mathcal{N}$(0.403$^{\dagger}$,10)   & -0.75678 $\pm$ 1.72435  \\
    log(Jitter)\textsubscript{\pfs}  & $\mathcal{N}$(-1.270$^{\ddagger}$,5)  & -1.51981 $\pm$ 3.07024   \\
    \bottomrule
    \end{tabularx}
    \begin{tablenotes}
    \item \textbf{Prior distributions:}
    \item (lower limit \textit{x}, upper limit \textit{y}) for uniform distribution $\mathcal{U}$(\textit{x},\textit{y})
    \item (mean $\mu$, standard deviation $\sigma$, test value $\alpha$) for normal distribution $\mathcal{N}$($\mu$,$\sigma$,$\alpha$)
    \item (mean $\mu$, standard deviation $\sigma$, lower limit \textit{x}, upper limit \textit{y}) for truncated normal distribution $\mathcal{T}$($\mu$,$\sigma$,\textit{x},\textit{y})
    \item (scale parameter $\beta$) for half-Cauchy distribution $\mathcal{C}$($\beta$)
    \item \textbf{Prior values:}
    \item $^*$ Equivalent to the log of the variance of the \TESS\ flux from the corresponding sector
    \item $^{\dagger}$ Equivalent to the mean of the radial velocity from the corresponding spectrographs
    \item $^{\ddagger}$ Equivalent to 2 times the log of the minimum radial velocity error from the corresponding spectrographs
    \end{tablenotes}
\end{table}
\end{center}

\begin{center}
\begin{table}
    \centering
    \caption{Global fit parameter prior function type and prior limits for TOI-836 b.}
     \label{tab:priorsb}
    \begin{tabular}{l l}
    \toprule
    \textbf{Parameter} &\textbf{Prior} \\
    \hline
    \textbf{TOI-836 b} &  \\
    Period (days) &$\mathcal{U}$(3.7, 3.9) \\
    Transit ephemeris (TBJD) &$\mathcal{U}$(2458599.98, 2458600.03) \\
    \textit{K}\textsubscript{RV} (\ms) &$\mathcal{U}$(0, 10) \\
    log(\rpl) &$\mathcal{N}$(-4.062\S, 1) \\
    \textit{b} &$\mathcal{U}$(0, 1) \\
    \textit{e} & \citet{exoplanet:kipping13_2}, $\mathcal{B}$(e, 0.867, 3.03) \\
    \textit{$\omega$} (rad) & $\mathcal{U}$($-\pi,\pi$) \\
    \bottomrule
    \end{tabular}
    \begin{tablenotes}
    \item Numbers in brackets represent:
    \item (lower limit \textit{x}, upper limit \textit{y}) for uniform distribution $\mathcal{U}$(\textit{x},\textit{y})
    \item (mean $\mu$, standard deviation $\sigma$, test value $\alpha$) for normal distribution $\mathcal{N}$($\mu$,$\sigma$,$\alpha$)
    \item Distributions for eccentricity \textit{e} are built into the \texttt{exoplanet} package and based on \citet{exoplanet:kipping13_2} which includes the Beta distribution $\mathcal{B}$(e,a,b) (exponential e, shape parameter a, shape parameter b)
    \item \S\ Equivalent to 0.5$\times$log($\delta$)+log(\rstar), $\delta$ represents transit depth (based on ExoFOP catalog values)
    \end{tablenotes}
\end{table}
\end{center}

\begin{center}
\begin{table}
    \centering
    \caption{Global fit parameter prior function type and prior limits for TOI-836 c.}
    \label{tab:priorsc}
    \begin{tabular}{l l}
    \toprule
    \textbf{Parameter} &\textbf{Prior} \\
    \hline
    \textbf{TOI-836 c} &  \\
    Period (days) &$\mathcal{U}$(8.5, 8.7) \\
    Transit ephemeris (TBJD) &$\mathcal{U}$(2458599.74, 2458599.79) \\
    \textit{K}\textsubscript{RV} (\ms) &$\mathcal{U}$(0, 10) \\
    log(\rpl) &$\mathcal{N}$(-3.701\S, 1) \\
    \textit{b} &$\mathcal{U}$(0, 1) \\
    \textit{e} & \citet{exoplanet:kipping13_2}, $\mathcal{B}$(e, 0.867, 3.03)  \\
    \textit{$\omega$} (rad) & $\mathcal{U}$($-\pi,\pi$) \\
    \bottomrule
    \end{tabular}
    \begin{tablenotes}
    \item Numbers in brackets represent:
    \item (lower limit \textit{x}, upper limit \textit{y}) for uniform distribution $\mathcal{U}$(\textit{x},\textit{y})
    \item (mean $\mu$, standard deviation $\sigma$, test value $\alpha$) for normal distribution $\mathcal{N}$($\mu$,$\sigma$,$\alpha$)
    \item Distributions for eccentricity \textit{e} are built into the \texttt{exoplanet} package and based on \citet{exoplanet:kipping13_2} which includes the Beta distribution $\mathcal{B}$(e,a,b) (exponential e, shape parameter a, shape parameter b)
    \item \S\ Equivalent to 0.5$\times$log($\delta$)+log(\rstar), $\delta$ represents transit depth (based on ExoFOP catalog values)
    \end{tablenotes}
\end{table}
\end{center}

%%%%%%%%%%%%%%%%%%%%%%%%%%%%%%%%%%%%%%%%%%%%%%%%%%%%%%%%%%%%%%%

% \section{Posterior plots}

% \begin{figure}
%     \centering
%     \includegraphics[width=0.48\textwidth]{3.Results/TOI-836_corner_plots_3.pdf}
%     \caption{\textcolor{red}{posterior plot 1}}
%     \label{fig:corner3}
% \end{figure}

% \begin{figure}
%     \centering
%     \includegraphics[width=0.48\textwidth]{3.Results/TOI-836_corner_plots_4.pdf}
%     \caption{\textcolor{red}{posterior plot 2}}
%     \label{fig:corner4}
% \end{figure}

% \begin{figure}
%     \centering
%     \includegraphics[width=0.48\textwidth]{3.Results/TOI-836_corner_plots_5.pdf}
%     \caption{\textcolor{red}{posterior plot 3}}
%     \label{fig:corner5}
% \end{figure}

%%%%%%%%%%%%%%%%%%%%%%%%%%%%%%%%%%%%%%%%%%%%%%%%%
%%%%%%%%%%%% AUTHOR AFFILIATIONS %%%%%%%%%%%%%%%%
%%%%%%%%%%%%%%%%%%%%%%%%%%%%%%%%%%%%%%%%%%%%%%%%%

\section{Author affiliations} \label{sec:affiliations}

% List of institutions

 $^{1}$ Department of Physics, University of Warwick, Gibbet Hill Road, Coventry CV4 7AL, UK\\
$^{2}$ Centre for Exoplanets and Habitability, University of Warwick, Gibbet Hill Road, Coventry CV4 7AL, UK\\
$^{3}$ Centre for Exoplanet Science, SUPA School of Physics and Astronomy, University of St Andrews, North Haugh, St Andrews KY16 9SS, UK \\
$^{4}$ Space Research Institute, Austrian Academy of Sciences, Schmiedlstrasse 6, A-8042 Graz, Austria \\
$^{5}$ Instituto de Astrof\'isica e Ci\^encias do Espa\c{c}o, Universidade do Porto, CAUP, Rua das Estrelas, 4150-762 Porto, Portugal \\
$^{6}$ Departamento de F\'isica e Astronomia, Faculdade de Ci\^encias, Universidade do Porto, Rua do Campo Alegre, 4169-007 Porto, Portugal \\
$^{7}$ Physikalisches Institut, University of Bern, Gesellsschaftstrasse 6, 3012 Bern, Switzerland \\
$^{8}$ Center for Astrophysics \textbar\ Harvard \& Smithsonian, 60 Garden St., Cambridge, MA 02138, USA \\
$^{9}$ Universit\'e C\^ote d'Azur, Observatoire de la C\^ote d'Azur, CNRS, Laboratoire Lagrange, CS 34229, F-06304 Nice Cedex 4, France \\
$^{10}$ School of Physics and Astronomy, University of Leicester, Leicester LE1 7RH, UK \\
$^{11}$ University of Southern Queensland, Centre for Astrophysics, USQ Toowoomba, West Street, QLD 4350 Australia \\
$^{12}$ Instituto de Astrofisica de Canarias, 38200 La Laguna, Tenerife, Spain \\
$^{13}$ Departamento de Astrofisica, Universidad de La Laguna, 38206 La Laguna, Tenerife, Spain \\
$^{14}$ Departamento de Astronomia, Universidad de Chile, Casilla 36-D, Santiago, Chile \\
$^{15}$ Institut de Ciencies de l'Espai (ICE, CSIC), Campus UAB, Can Magrans s/n, 08193 Bellaterra, Spain \\
$^{16}$ Institut d'Estudis Espacials de Catalunya (IEEC), 08034 Barcelona, Spain \\
$^{17}$ Admatis, 5. Kand\'o K\'alm\'an Street, 3534 Miskolc, Hungary \\
$^{18}$ NASA Goddard Space Flight Center, 8800 Greenbelt Road, Greenbelt, MD 20771, USA \\
$^{19}$ University of Maryland, Baltimore County, 1000 Hilltop Circle, Baltimore, MD 21250, USA \\
$^{20}$ Depto. de Astrof\'{\i}sica, Centro de Astrobiolog\'{\i}a (CSIC-INTA), ESAC campus, 28692 Villanueva de la Ca\~nada (Madrid), Spain \\
$^{21}$ Center for Space and Habitability, Gesellsschaftstrasse 6, 3012 Bern, Switzerland \\
$^{22}$ Universit\'e Grenoble Alpes, CNRS, IPAG, 38000 Grenoble, France \\
$^{23}$ Observatoire de Genève, Université de Genève, Chemin de Pegasi, 51, 1290 Versoix, Switzerland \\
$^{24}$ Department of Astronomy, Stockholm University, AlbaNova University Center, 10691 Stockholm, Sweden \\
$^{25}$ Programma Nazionale di Ricerche in Antartide (PNRA), Concordia station, Antarctica \\
$^{26}$ Institute of Planetary Research, German Aerospace Center (DLR), Rutherfordstrasse 2, 12489 Berlin, Germany \\
$^{27}$ SETI Institute/NASA Ames Research Center, Moffett Field, CA 94035 \\
$^{28}$ Université de Paris, Institut de physique du globe de Paris, CNRS, F-75005 Paris, France \\
$^{29}$ George Mason University, 4400 University Drive, Fairfax, VA 22030, USA \\
$^{30}$ American Association of Variable Star Observers, 49 Bay State Road, Cambridge, MA 02138, USA \\
$^{31}$ European Space Agency (ESA), European Space Research and Technology Centre (ESTEC), Keplerlaan 1, 2201 AZ Noordwijk, The Netherlands \\
$^{32}$ Centre for Mathematical Sciences, Lund University, Box 118, 221 00 Lund, Sweden \\
$^{33}$ Aix Marseille Univ, CNRS, CNES, LAM, 38 rue Fr\'ed\'eric Joliot-Curie, 13388 Marseille, France \\
$^{34}$ Astrobiology Research Unit, Universit\'e de Li\`ege, All\'ee du 6 Ao\^ut 19C, B-4000 Li\`ege, Belgium \\
$^{35}$ Space sciences, Technologies and Astrophysics Research (STAR) Institute, Universit\'e de Li\`ege, All\'ee du 6 Ao\^ut 19C, 4000 Li\`ege, Belgium \\
$^{36}$ School of Physics \& Astronomy, University of Birmingham, Edgbaston, Birmingham B15 2TT, United Kingdom \\
$^{37}$ Department of Earth, Atmospheric and Planetary Sciences, Massachusetts Institute of Technology, Cambridge, MA 02139, USA \\
$^{38}$ Department of Physics and Kavli Institute for Astrophysics and Space Research, Massachusetts Institute of Technology, Cambridge, MA 02139, USA \\
$^{39}$ Leiden Observatory, University of Leiden, PO Box 9513, 2300 RA Leiden, The Netherlands \\
$^{40}$ Department of Space, Earth and Environment, Chalmers University of Technology, Onsala Space Observatory, 43992 Onsala, Sweden \\
$^{41}$ University of Vienna, Department of Astrophysics, T\"urkenschanzstrasse 17, 1180 Vienna, Austria \\
$^{42}$ Dipartimento di Fisica, Universita degli Studi di Torino, via Pietro Giuria 1, I-10125, Torino, Italy \\
$^{43}$ NASA Ames Research Center, Moffett Field, CA 94035, USA \\
$^{44}$ NASA Exoplanet Science Institute, Caltech/IPAC, Mail Code 100-22, 1200 E. California Blvd., Pasadena, CA 91125, USA \\
$^{45}$ Astrophysics Group, Keele University, Keele ST5 5BG, UK \\
$^{46}$ Cavendish Laboratory, JJ Thomson Avenue, Cambridge CB3 0HE, UK \\
$^{47}$ N\'ucleo de Astronom\'ia, Facultad de Ingenier\'ia y Ciencias, Universidad Diego Portales, Av. Ej\'ercito 441, Santiago, Chile \\
$^{48}$ Centro de Astrof\'isica y Tecnolog\'ias Afines (CATA), Casilla 36-D, Santiago, Chile \\
$^{49}$ Department of Physics \& Astronomy, Swarthmore College, Swarthmore PA 19081, USA \\
$^{50}$ Department of Earth and Planetary Sciences, University of California, Riverside, CA 92521, USA
$^{51}$ Department of Physics and Astronomy, University of Louisville, Louisville, KY 40292, USA
$^{52}$ Konkoly Observatory, Research Centre for Astronomy and Earth Sciences, 1121 Budapest, Konkoly Thege Miklós út 15-17, Hungary \\
$^{53}$ ELTE E\"otv\"os Lor\'and University, Institute of Physics, P\'azm\'any P\'eter s\'et\'any 1/A, 1117 Budapest, Hungary
$^{54}$ INAF, Osservatorio Astronomico di Padova, Vicolo dell'Osservatorio 5, 35122 Padova, Italy \\
$^{55}$ IMCCE, UMR8028 CNRS, Observatoire de Paris, PSL Univ., Sorbonne Univ., 77 av. Denfert-Rochereau, 75014 Paris, France \\
$^{56}$ Institut d'astrophysique de Paris, UMR7095 CNRS, Universit\'e Pierre \& Marie Curie, 98bis blvd. Arago, 75014 Paris, France \\
$^{57}$ Villa '39 Observatory, Landers, CA 92285, USA \\
$^{58}$ European Southern Observatory, Karl-Schwarzschildstrasse 2, D-85748 Garching bei M\"unchen, Germany \\
$^{59}$ Astrophysics Research Centre, School of Mathematics and Physics, Queen's University Belfast, Belfast, BT7 1NN, UK \\
$^{60}$ INAF, Osservatorio Astrofisico di Catania, Via S. Sofia 78, 95123 Catania, Italy \\
$^{61}$ Chalmers University of Technology, Chalmersplatsen 4, 412 96 Göteborg, Sweden \\
$^{62}$ Dipartimento di Fisica e Astronomia "Galileo Galilei", Universita degli Studi di Padova, Vicolo dell'Osservatorio 3, 35122 Padova, Italy \\
$^{63}$ ETH Zurich, Department of Physics, Wolfgang-Pauli-Strasse 2, CH-8093 Zurich, Switzerland \\
$^{64}$ Cavendish Laboratory, JJ Thomson Avenue, Cambridge CB3 0HE, UK \\
$^{65}$ Center for Astronomy and Astrophysics, Technical University Berlin, Hardenberstrasse 36, 10623 Berlin, Germany \\
$^{66}$ Institut f\"ur Geologische Wissenschaften, Freie Universit\~at Berlin, 12249 Berlin, Germany \\
$^{67}$ Patashnick Voorheesville Observatory, Voorheesville, NY 12186, USA \\
$^{68}$ Kotizarovci Observatory, Sarsoni 90, 51216 Viskovo, Croatia \\
$^{69}$ ELTE E\"otv\"os Lor\'and University, Gothard Astrophysical Observatory, 9700 Szombathely, Szent Imre h. u. 112, Hungary \\
$^{70}$ MTA-ELTE Exoplanet Research Group, 9700 Szombathely, Szent Imre h. u. 112, Hungary \\
$^{71}$ Earth and Planets Laboratory, Carnegie Institution for Science, 5241 Broad Branch Road NW, Washington, DC 20015 \\
$^{72}$ Institute of Astronomy, University of Cambridge, Madingley Road, Cambridge CB3 0HA, UK \\
$^{73}$ Department of Astronomy, Tsinghua University, Beijing 100084, People's Republic of China \\
$^{74}$ Department of Astrophysical Sciences, Princeton University, Princeton, NJ 08544, USA \\
$^{75}$ Shanghai Astronomical Observatory, 80 Nandan Road, Shanghai, 200030, China

%%%%%%%%%%%%%%%%%%%%%%%%%%%%%%%%%%%%%%%%
%%%%%%%%%%%%% FINISHING %%%%%%%%%%%%%%%%
%%%%%%%%%%%%%%%%%%%%%%%%%%%%%%%%%%%%%%%%

% Don't change these lines
\bsp	% typesetting comment
\label{lastpage}
\end{document}